%% file: dsphs-gnn.tex
\documentclass[prd,aps,10pt,nofootinbib,twocolumn,superscriptaddress,preprintnumbers,balancelastpage,longbibliography,floatfix]{revtex4-2}

\usepackage{amsmath,amssymb}	
\usepackage{mathtools}
\usepackage{fontawesome}
\usepackage[dvipsnames]{xcolor}
\usepackage{hyperref}
\usepackage{xspace}
\usepackage{dsfont}
\usepackage{graphicx}
\usepackage{siunitx}
\usepackage{bbm}
\usepackage{float}
\usepackage{tabularx}
\usepackage{multirow}
\usepackage{dcolumn}
\usepackage{makecell}
\usepackage{afterpage}
\usepackage{longtable}

\usepackage[section]{placeins}

\usepackage{aas_macros}

\definecolor{linkcolor}{rgb}{0.7752941176470588, 0.22078431372549023, 0.2262745098039215}

\hypersetup{colorlinks=true,
linkcolor=linkcolor,
citecolor=linkcolor,
urlcolor=linkcolor,
linktocpage=true,
pdfproducer=medialab,
}

\newcommand\Tstrut{\rule{0pt}{2.6ex}}         
\newcommand\Bstrut{\rule[-1.6ex]{0pt}{0pt}}   

\newcolumntype{C}[1]{>{\centering\let\newline\\\arraybackslash\hspace{0pt}}m{#1}}

\makeatletter 

\begin{document}

\preprint{MIT/CTP-5459}

\title{Uncovering dark matter density profiles in dwarf galaxies with graph neural networks}

\author{Tri Nguyen}
\email{tnguy@mit.edu}
\affiliation{The NSF AI Institute for Artificial Intelligence and Fundamental Interactions}
\affiliation{Department of Physics and Kavli Institute for Astrophysics and Space Research, Massachusetts Institute of Technology, 77 Massachusetts Ave, Cambridge MA
02139, USA}

\author{Siddharth Mishra-Sharma}
\affiliation{The NSF AI Institute for Artificial Intelligence and Fundamental Interactions}
\affiliation{Center for Theoretical Physics, Massachusetts Institute of Technology, Cambridge, MA 02139, USA}
\affiliation{Department of Physics, Harvard University, Cambridge, MA 02138, USA}

\author{Reuel Williams}
\affiliation{Department of Mathematics, Princeton University, Princeton, NJ 08544, USA}

\author{Lina Necib}
\affiliation{The NSF AI Institute for Artificial Intelligence and Fundamental Interactions}
\affiliation{Department of Physics and Kavli Institute for Astrophysics and Space Research, Massachusetts Institute of Technology, 77 Massachusetts Ave, Cambridge MA
02139, USA}

\date{\today}

\begin{abstract}
Dwarf galaxies are small, dark matter-dominated galaxies, some of which are embedded within the Milky Way. Their lack of baryonic matter (e.g., stars and gas) makes them perfect test beds for probing the properties of dark matter---understanding the spatial dark matter distribution in these systems can be used to constrain microphysical dark matter interactions that influence the formation and evolution of structures in our Universe. 
We introduce a new method that leverages simulation-based inference and graph-based machine learning in order to infer the dark matter density profiles of dwarf galaxies from observable kinematics of stars gravitationally bound to these systems. Our approach aims to address some of the limitations of established methods based on dynamical Jeans modeling. 
We show that this novel method can place stronger constraints on dark matter profiles and, consequently, has the potential to weigh in on some of the ongoing puzzles associated with the small-scale structure of dark matter halos, such as the core-cusp discrepancy. 
\end{abstract}

\maketitle

\section{Introduction}

Cosmological structure formation is known to proceed hierarchically---smaller structures seed the formation of larger structures \citep{White&Rees_1978}. Dark matter (DM) plays an outsized role in this process, acting as a ``scaffolding'' on which structure evolution plays out. At the same time, the precise mechanism of structure formation is keenly sensitive to the microphysical properties of DM e.g., the nature of its self-interactions. Deviations from the canonical $\Lambda$ Cold Dark Matter ($\Lambda$CDM) paradigm of cosmology would be imprinted in the properties of DM clumps (known as \emph{halos}) on smaller spatial scales.
Robustly characterizing the distribution of small-scale structures in our Universe may therefore hold the key to answering one of the major unsolved questions in particle physics and cosmology---the particle nature of DM.

\emph{Dwarf galaxies} are small galaxies, some of which are embedded within larger galaxies like the Milky Way.
They are dominated by DM~\cite{1990AJ....100..127P}, making them versatile astrophysical laboratories for DM studies. 
A major goal in cosmology and particle physics is to detect non-gravitational interactions of DM. 
For the canonical Weakly-Interacting Massive Particle (WIMP) DM paradigm, one of the main avenues to do so is DM indirect detection: WIMPs could annihilate into Standard Model (SM) particles, producing striking signatures from DM-overdense regions in $\gamma$-ray observations~\cite{Cirelli_2011, PhysRevLett.115.231301, 10.1093/mnras/stv942, Geringer_Sameth_2015, PhysRevLett.107.241303, PhysRevLett.107.241302, MAZZIOTTA201226, PhysRevD.91.083535, PhysRevD.89.042001, https://doi.org/10.48550/arxiv.1502.03081}.
Being deficient in baryonic matter, dwarf galaxies act as ideal targets for indirect detection, with a relatively large predicted ratio of DM signal to astrophysical background. 

A pervasive puzzle in cosmology is the so-called core-cusp discrepancy, referring to the question of whether the inner DM density profiles of dwarf galaxies are cuspy (steeply rising) or cored (flattened) \cite{1996MNRAS.283L..72N,2005AJ....129.2119S}.
$N$-body simulations using $\Lambda$CDM cosmology suggest that in the absence of baryonic physics, cold DM halos follow the cuspy Navarro-Frenk-White density (NFW) profile \cite{nfw1997}, which is characterized by a steep rise in the density $\rho \propto r^{-1}$ at small halo-centric radii $r$.
However, recent measurements of stellar dynamics suggest that these systems could instead have a flattened density profile at their center, also known as a \emph{core} \cite{2011AJ....142...24O,2015AJ....149..180O}; see Ref.~\cite{Bullock:2017xww} for a review.
Potential solutions to the core-cusp discrepancy range from stellar feedback which ejects baryons and flattens the DM central density profile~\cite{10.1093/mnras/283.3.L72, 2005MNRAS.356..107R, 2006Natur.442..539M, 2012MNRAS.421.3464P} to alternative DM models like self-interactions \cite{10.1111/j.1365-2966.2011.20200.x, 10.1093/mnras/stv1470,2022MNRAS.513.3458B}.

DM density profiles in dwarf galaxies are traditionally inferred using spectroscopic observations of line-of-sight velocities and angular positions of stars gravitationally bound to these systems. In particular, integral moments of the Jeans equation can be used to relate the velocity dispersions of tracer stars to the gravitational potential of the system \cite{1915MNRAS..76...70J,2015MNRAS.446.3002B}.
Although Jeans modeling has proven highly successful for modeling DM distributions in dwarf galaxies, there are several caveats and limitations associated with this approach (see e.g. Refs.~\cite{2017ApJ...835..193E,read2020,chang2021}). For example, Jeans modeling assumes that the system is in dynamical equilibrium, which may not be a robust assumption given the active merger history of the Milky Way 
(see Ref.~\cite{2020ARA&A..58..205H} for a review). Assumptions such as isotropy of the gravitating system are also often necessary in order to enable a tractable analysis. Finally, by relying on a simplified description of the data through second moments of the stellar velocity distribution, inference based on Jeans modeling is likely to lose some of the salient information available in observations. 
In the absence of additional assumptions, this is expressed as a degeneracy between the mass profile of the system and the anisotropy structure of stellar orbits, known as the mass-anisotropy degeneracy. 
Exploiting additional information about the modeled stellar phase-space distribution can further inform the latent DM density profile, helping break this degeneracy and better constrain the density profile. Several methods have been proposed in the literature to this end~\cite{2021MNRAS.501..978R}, including using higher-order moments of the Jeans equation~\cite{2017MNRAS.471.4541R,Read:2018pft,read2020}, leveraging multiple distinct tracer populations~\cite{2011ApJ...742...20W,2012MNRAS.419..184A,2016MNRAS.463.1117Z}, including measured proper motions when these are available~\cite{Strigari:2007vn,Strigari:2018bcn}, and alternative strategies to solving the Jeans equation~\cite{2013MNRAS.429.3079M}.

In this paper, we introduce a new machine learning-based approach for linking observed stellar properties to the DM density profiles of dwarf galaxies. 
Our method is based on forward modeling simulated dwarf galaxy systems and corresponding observations, learning to extract representative features from these datasets using graph neural networks, and performing simulation-based inference via density estimation to simultaneously extract the spatial profiles associated with the DM and stellar components of the dwarf galaxy. 
As a proof-of-principle and in order to enable a direct comparison with established Jeans analysis methods, in this work we focus on simulated, spherical dwarf galaxies in equilibrium with small velocity measurement errors.
In these idealized systems, we demonstrate the advantages of our method in terms of speed, constraining power, as well as flexibility.

\section{Methodology}

We describe, in turn, the forward model used in this study and its realization via simulations, the representation of stellar kinematic data as a graph, and finally the feature-extractor graph neural network and simulation-based inference procedure.

\subsection{Datasets and the forward model}

In this proof-of-principle exposition, we consider idealized simulations of spherical and dynamically-equilibrated dwarf galaxies.
Our forward model is fully specified by the joint distribution function (DF) $f(\vec x, \vec v)$ of positions and velocities of stars following a certain (a-priori unknown) spatial distribution (known as the \emph{light profile}). 
These tracer stars are gravitationally bound to a DM halo with a density profile that we wish to infer. We use the public code \texttt{StarSampler}\footnote{\url{https://github.com/maoshenl/StarSampler}} to generate simulated realizations of stellar kinematics (6-D position and velocity phase-space components) from the forward model. \texttt{StarSampler} uses importance sampling \cite{gelman2003bayesian, doi:10.1080/00031305.1992.10475856,  rubin1988using} to sample the DF of positions and velocities of tracer stars in a given DM potential. 

We model the DM profile using the generalized Navarro–Frenk–White (gNFW) profile \cite{nfw1997}:
\begin{equation}
    \label{eq:gNFW}
    \rho_\mathrm{DM}^\mathrm{gNFW}(r) = \rho_0 \left(\frac{r}{r_s}\right) ^{-\gamma}\left(1 + \frac{r}{r_s}\right)^{-(3-\gamma)},
\end{equation}
which depends on three free parameters: the density normalization $\rho_0$, the scale radius $r_s$, and the asymptotic inner slope $\gamma$. $\gamma = 1$ corresponds to a cuspy NFW profile, 
while $\gamma=0$ corresponds to a pure DM core. We consider these two parameter points as benchmarks in our study, since the ability to robustly distinguish between the two possibilities would offer a path towards resolution of the core-cusp discrepancy.

\begin{figure*}[!htbp]
    \centering
    \includegraphics[width=0.98\textwidth]{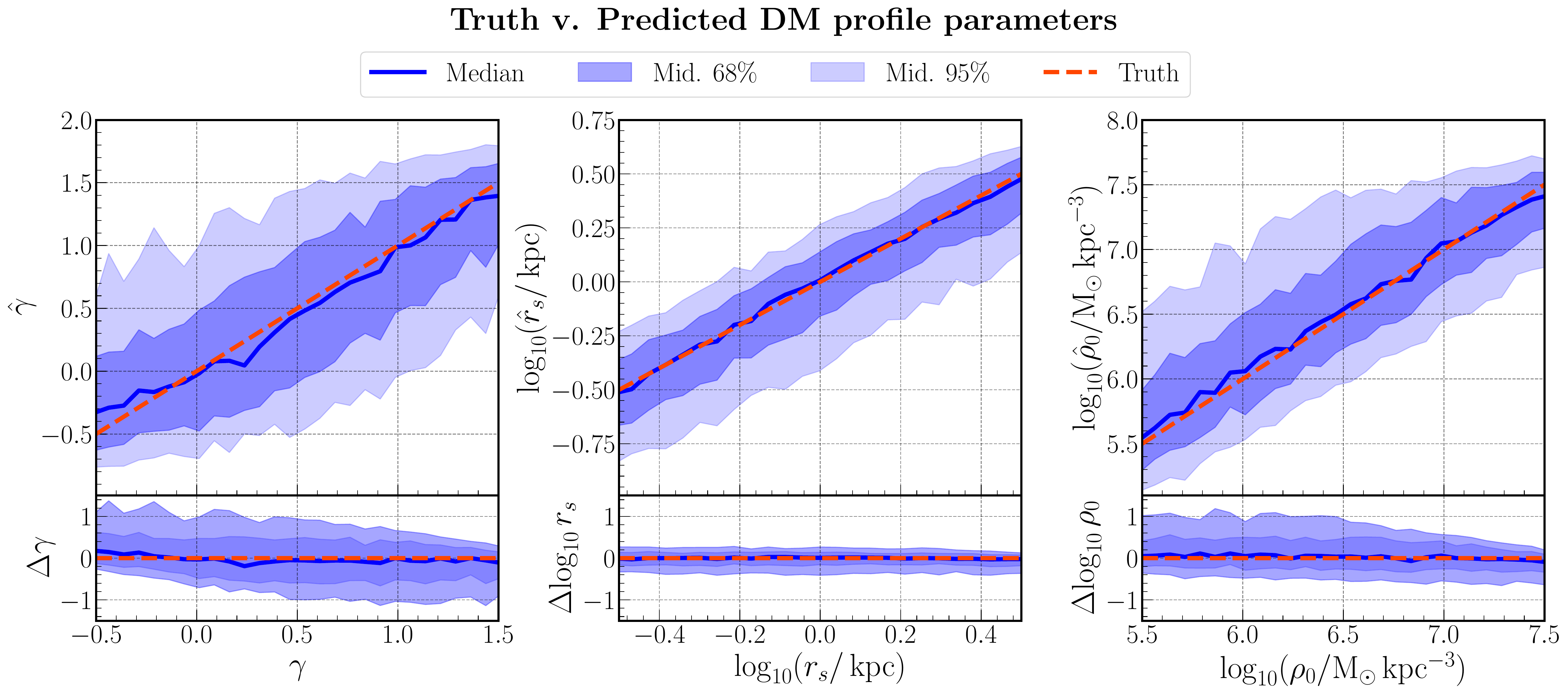}
    \caption{A comparison between the predicted and the truth values of the DM parameters on 10,000 test galaxies for our baseline case containing $\sim 100$ stars with measurement error $0.1 \, \si{km/s}$.
    For each galaxy, the predicted parameters are taken to be the marginal medians of the joint posterior and then sorted into bins based on their truth values.
    The median (solid blue line), middle-68\% percentile (dark blue band), and middle-95\% (light blue band) containment regions of each bin are shown. 
    The diagonal dashed red line denotes where the predicted and truth values are equal.
    The bottom row shows the prediction error on the median $\Delta\theta \equiv \hat{\theta} - \theta_\mathrm{truth}$.
    }
    \label{fig:percentile}
\end{figure*}

We assume a stellar density distribution $\nu(r)$ that follows a 3-D Plummer profile \cite{1911MNRAS..71..460P}:
\begin{equation}
    \label{eq:plummer}
    \nu(r) = \frac{3L}{4\pi r_\star^3} \left(1 + \frac{r^2}{r_\star^2}\right)^{-5/2}
\end{equation}
where $L$ is the total luminosity and $r_\star$ is the scale length.
We also introduce a velocity anisotropy profile $\beta(r)$ in order to model deviations from circular orbits; $\beta(r)$ is defined similarly to Refs.~\cite{1979PAZh....5...77O,1985AJ.....90.1027M} as
\begin{equation}
    \label{eq:beta}
    \beta(r) = \frac{r^2}{r^2 + r_a^2},
\end{equation}
which has an additional parameter $r_a$ describing the radius of transition from an isotropic velocity orbits at small radii to a radially-biased orbits at larger radii. 

In total, our model has three DM parameters $(\rho_0, r_s, \gamma)$ and two stellar parameters $(r_\star, r_a)$.
We assume that the gravitational potential of the system is dominated by DM, and therefore the model is independent of the total luminosity $L$ in Eq.~\ref{eq:plummer}.
We sample the density normalization $\rho_0$ and scale radius $r_s$ $\log_{10}$-uniformly distribution from $[10^5, 10^8] \, \si{\mathrm M_\odot / kpc^3}$ and $[0.1, 5] \, \si{kpc}$ respectively, and central slope $\gamma$ uniformly from $[-1, 2]$.
This implicitly sets the prior distributions of the parameters of interest in our simulation-based inference pipeline.
Because the DM and stellar parameters are correlated, we uniformly sample $r_\star$ from $[0.2, 1] \, r_s$, and $r_a$ from $[0.5, 2] \, r_\star$. 
A summary of the prior specification may be found in App.~\ref{app:priors}, and we provide further details of the forward model and phase-space distribution function in App.~\ref{app:forward_model}.

We generate 80,000 training samples, 10,000 validation samples, and 10,000 test samples using the prior parameter distributions.
Each sample contains the 3-D positions and 3-D velocities of tracer stars with respect to the center of a dwarf galaxy.
The number of stars in each galaxy is sampled from a Poisson distribution $n_\mathrm{stars}\sim\mathrm{Pois}(\mu_\mathrm{stars})$. We set $\mu_\mathrm{stars}=100$ stars in our baseline benchmark, roughly corresponding in order of magnitude to the number of stars typically observed in dwarf galaxies of interest~\cite{2007ApJ...670..313S, 2008ApJ...675..201M, 2009AJ....137.3100W, 2009ApJ...704.1274W}.
For example, the number of observed stars in Segue 1, Leo II, and Draco are 70, 126 and 292, respectively \cite{2007ApJ...670..313S, 2009ApJ...704.1274W, 2017AJ....153..254S}.
We explore variations on this choice in App.~\ref{app:variation}. 

\subsection{Data pre-processing and graph construction}

We pre-process our dataset by adding projection effects and measurement errors reflecting typical observations of dwarf galaxy tracer stars.
For each kinematic sample, we randomly draw a line-of-sight axis and project the galaxy onto the 2-D plane perpendicular to it.
We then derive the 2-D projected spatial coordinates with respect to the center of the galaxy $(X, Y)$ and line-of-sight velocities $v_\mathrm{los}$ for each star in these coordinates.
To study the validity of the method before the inclusion of large measurement errors, we assume a Gaussian velocity noise model of $0.1 \, \si{km/s}$. We show the effect of larger measurement errors in App.~\ref{app:variation}.
For simplicity and consistency with Jeans-based analysis, we do not include positional uncertainty in the angular position measurement.

We can represent the stellar kinematic data in the form of a potentially weighted, undirected graph $\mathcal G = (\mathcal V, \mathcal E, A)$, where $\mathcal V$ is a set of nodes representing $|V| = N_\mathrm{stars}$ individual stars, $\mathcal E$ is a set of edges, and $A \in \mathbb R^{N_\mathrm{stars} \times N_\mathrm{stars}}$ is an adjacency matrix describing the weights of connections between vertices. This representation is well-suited for our purposes since the stars in a dwarf galaxy have no intrinsic ordering, and the graph structure can efficiently capture the phase-space correlation structure containing information about the underlying DM density distribution, including higher-order moments~\cite{2022arXiv220705202M}.

In our analysis, each node represents a star, with the node features being its 
line-of-sight velocity $\tilde{v}_\mathrm{los}$ and the projected radius $R=\sqrt{X^2 + Y^2}$.
We choose to use $R$ instead of the full $(X, Y)$ coordinates in order to incorporate projective rotational invariance into the graph representation, which was found to enhance the simulation-efficiency of our method.

To determine the graph edges $\mathcal E$, we calculate pair-wise distances between all stars using $(X, Y)$, then connect each star to its $k$-nearest stars including itself (i.e. self-loops).
Since the edges are assumed to be undirected, each star can be connected to more than $k$ other stars.
A higher value of $k$ increases the number of edges, which provides more neighboring information at computational and memory cost.
We found $k=20$ to provide a good trade-off between model performance and computational overhead.
Finally, we do not include edge weights in our graph, but note that we have experimented with a variety of weighting schemes, including attention-based learned weights~\cite{2017arXiv171010903V} as well as weights exponentially decaying with inter-star distance, and found them to perform similarly in downstream inference to the unweighted case.

\subsection{Neural network architecture and optimization}

We use a graph neural network (GNN) $g_\varphi: \mathcal G  \rightarrow \mathbb R^{N_\mathrm{feat}}$ in order to extract $N_\mathrm{feat}$ summary features from the constructed graph representation $x\in \mathcal{G}$ of mock dwarf galaxy stellar kinematic data. Here $\varphi$ represent the parameters of the graph neural network. 
The feature-extraction network consists of 5 graph-convolutional layers, each with 128 channels, based on convolutions in the Fourier domain using a basis of Chebyshev polynomials of order $K=4$ as filters~\cite{2016arXiv160609375D}.
This is followed by a global mean pooling layer which aggregates the permutation-equivariant features into a permutation-invariant representation, and a fully-connected layer which projects the output onto a set of $N_\mathrm{feat}=128$ summaries $g_\varphi (x)$.
We show variations on the graph-convolution scheme in App.~\ref{app:variation}.

\begin{figure}[!t]
    \centering
    \includegraphics[width=0.45\textwidth]{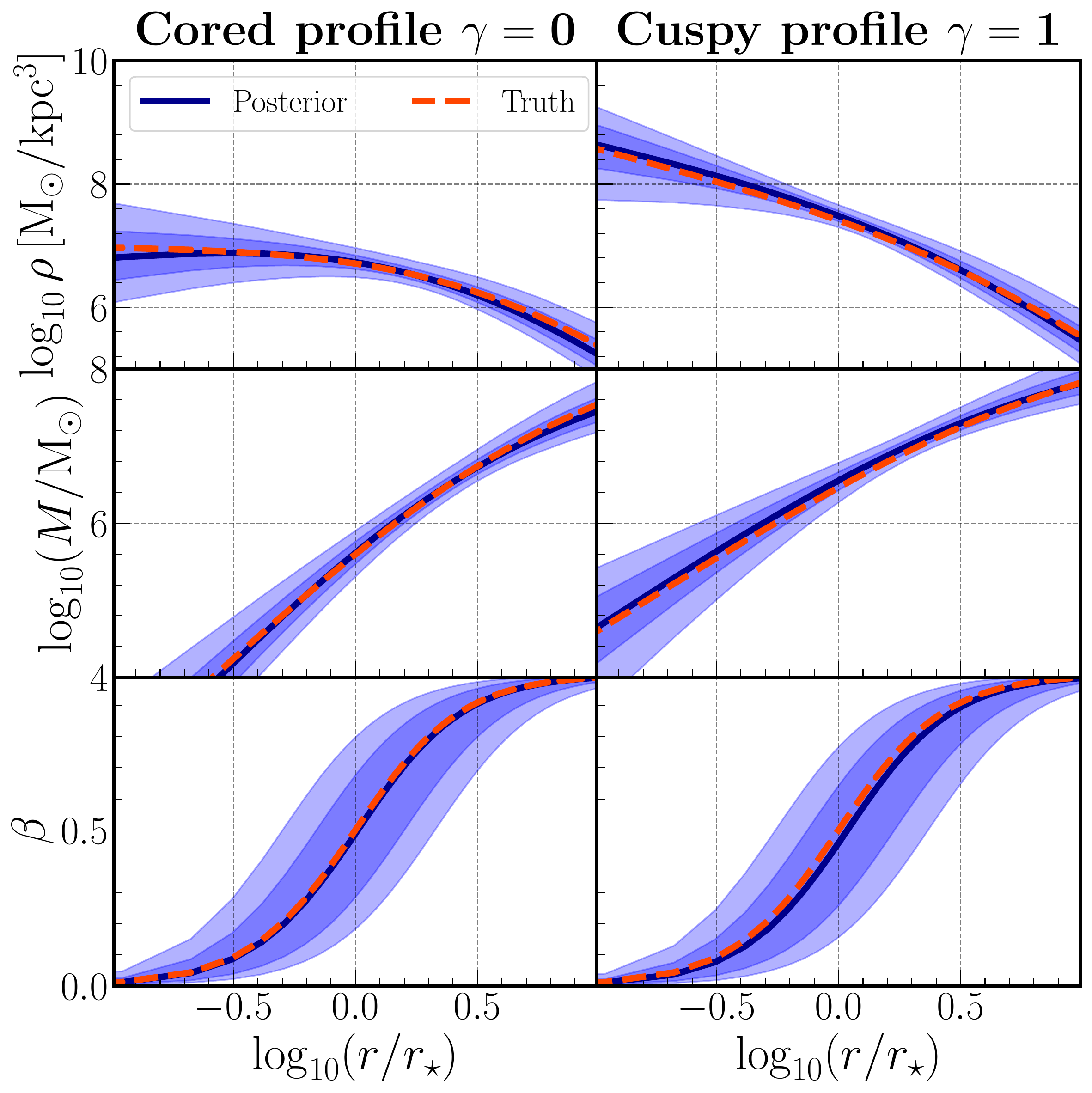}
    \caption{Example inferred posteriors of the density profile (top row), enclosed mass profile (middle row), and velocity anisotropy profile (bottom row) for dwarf galaxies with a cored DM profile (left) and a cuspy DM profile (right).
    The dashed red line is the truth profile, while the blue line and bands represent the median, middle-68\%, and 95\% credible intervals respectively.
    }
    \label{fig:profiles}
\end{figure}

The joint posterior $\hat p_\phi(\theta\mid g_\varphi(x))$ of the five parameters of interest $\theta$ characterizing the DM and stellar profiles is modeled using a normalizing flow ~\cite{papamakarios2019normalizing, DBLP:conf/icml/RezendeM15,10.5555/3294771.3294994}---a class of flexible generative models that allow for efficient density estimation as well as sampling.  
The flow transformation (with parameters $\phi$) is conditioned on the summary features extracted by the graph neural network and its negative log-density $-\log\hat p_\phi\left(\theta\mid g_\varphi(x)\right)$ is used as the optimization loss. 
Our flow model consists of 4 Masked Autoregressive Flow (MAF) transformations, each using a 2-layer Masked Autoencoder for Distribution Estimation (MADE) with hidden dimension 128~\cite{DBLP:conf/icml/GermainGML15,10.5555/3294771.3294994}. This method falls under the class of approaches known as simulation-based inference (see Ref.~\cite{Cranmer:2019eaq} for a review), specifically neural conditional density estimation~\cite{cranmer_kyle_2016_198541,10.5555/3157096.3157212}. 

The GNN and normalizing flow parameters $\{\varphi, \phi\}$ are optimized simultaneously on the 80,000 simulated samples using the AdamW optimizer~\cite{kingma2014adam,adamw2019} with a learning rate of $5 \times 10^{-4}$ and a weight decay of $10^{-2}$ using a batch size of 64.
At the end of each epoch, we evaluate the loss on the 10,000 held out validation samples and reduce the learning rate by a factor of $10$ if no improvement is seen after 4 epochs.
We stop training if the validation loss has not improved after 10 epochs. 
Model training typically terminates after $\sim 30$--$40$ epochs, which takes $\sim 30$ minutes on a single NVIDIA Tesla V100 GPU.

\section{Results and conclusions}

We apply our pipeline to 10,000 test dwarf galaxies and summarize our results in Fig.~\ref{fig:percentile}.
For each galaxy, we condition the trained normalizing flow on features extracted using the trained GNN feature extractor and draw 10,000 samples from the joint DM and stellar posterior.
We then compute the marginal medians as the predicted parameters and sort them into bins based on their truth values. 
Fig.~\ref{fig:percentile} shows the median (solid blue line), middle-68\% (blue bands), and middle-95\% (light blue bands) credible intervals for each bin of the DM parameters.
In general, our method successfully recovers individual DM parameters consistent with the underlying truth.

\begin{figure*}[!htbp]
    \centering
    \includegraphics[width=0.45\textwidth]{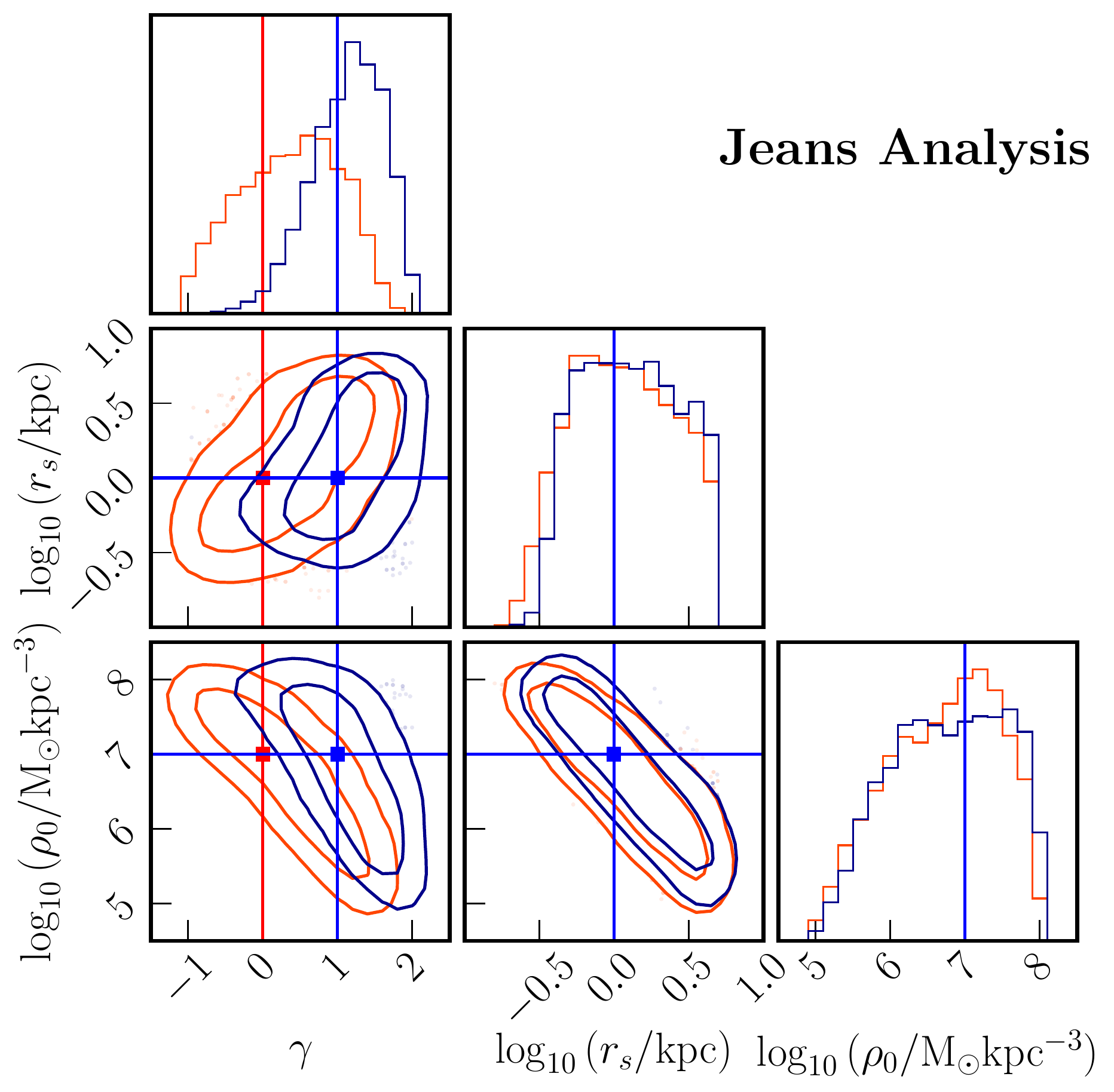}
    \includegraphics[width=0.45\textwidth]{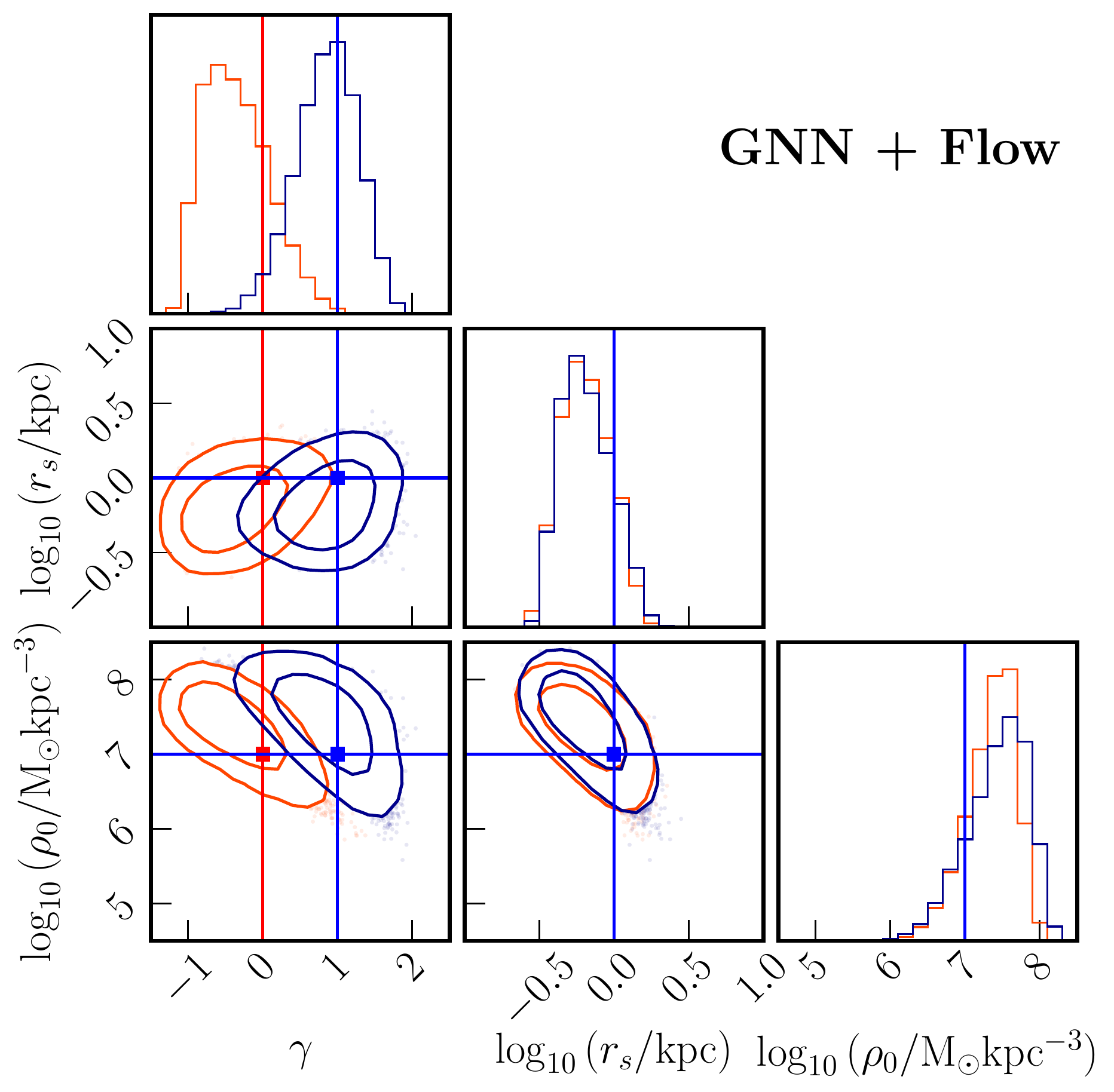}
    \caption{
    Example corner plots of the posterior DM parameters from Jeans dynamical modeling (left) and our method (right) on two test galaxies with cored DM profile (red) and cuspy DM profile (blue).
    Both galaxies have the same central slope $\rho_0$ and scale radius $r_s$.
    The contour lines show the 68\% and the 95\% credible intervals.
    It can be seen that our method provides a stronger constraint on the DM parameters and is able to distinguish more cleanly between a cored and cuspy profile.
    }
    \label{fig:corner}
\end{figure*}

To demonstrate how well our method can distinguish between a cored ($\gamma=0$) and a cuspy ($\gamma=1$) DM profile, in Fig.~\ref{fig:profiles} and Fig.~\ref{fig:corner}, we show the inferred posteriors on two test dwarf galaxies with the same DM density normalization ($\rho_0 = 10^7 \, \si{\mathrm M_\odot / kpc^3}$), scale radius ($r_s = 1 \, \si{kpc}$), and stellar profile, but with different inner density slopes ($\gamma=0$ and $\gamma=1$).

Fig.~\ref{fig:profiles} shows the posteriors on the recovered density profile (top row), enclosed mass profile (middle row), and orbital anisotropy profile (bottom row) as a function of halo-centric radius of a cored (left) and a cupsy (right) profile.
The middle-68\% and 95\% credible intervals are shown as blue bands, and the truth profiles are shown with the dashed red lines.
We find that our method is able to successfully reconstruct the density, mass, and orbital anisotropy profiles at small and large radii.

We apply Jeans analysis on these two test galaxies using a procedure similar to that used in Ref.~\cite{chang2021}.
We approximate the posterior with nested sampling~\cite{2004AIPC..735..395S,skilling2006} using the module \textsc{dynesty}~\cite{dynesty} with $n_\mathrm{live} = 500$ live points and a convergence tolerance of $\Delta\ln\mathcal Z=0.1$ on the estimated evidence. 
We provide details of our Jeans analysis in the App.~\ref{app:jeans}.
In Fig.~\ref{fig:corner}, we show the corner plot of the joint and marginal DM posteriors from Jeans modeling (left panel) and our method (right panel) for $\gamma=0$ (red) and $\gamma=1$ (blue), with the middle-68\% and 95\% credible intervals as the contour lines.
In the Jeans analysis, we see significant overlap between the $\gamma$ posteriors of the cored and cuspy profiles. 
In addition, $\gamma=0$ posterior appears to be peaking close to $\gamma=1$.
On the other hand, the two $\gamma$ posteriors inferred by our method have substantially smaller overlap and maximum a-posteriori values closer to their corresponding truths.

In order to check how well this generalizes to a larger galaxy sample and to quantitatively compare the inferred $\gamma$ posteriors of cored and cuspy profiles, we compute and summarize the Jensen-Shannon (JS)-divergences~\cite{e21050485, manning99foundations}, constrained to lie in the range $[0, 1]$ with our definition, between 100 pairs of cored and cuspy $\gamma$ posteriors using our method as well as Jeans analysis.
We quote the medians and the middle-68\% containment regions of JS divergence $D_\mathrm{JS}^\mathrm{Jeans}=0.417^{+0.288}_{-0.280}$ and $D_\mathrm{JS}^\mathrm{GNN}= 0.629^{+0.196}_{-0.278}$.
A larger value of $D_\mathrm{JS}$ means our method can better distinguish between galaxies with cored and cuspy profiles compared to Jeans analysis. We defer further details of this test and show the distribution of obtained JS-divergences, in App.~\ref{app:slope_results}.

\begin{figure}[!tbp]
    \centering
    \includegraphics[width=0.45\textwidth]{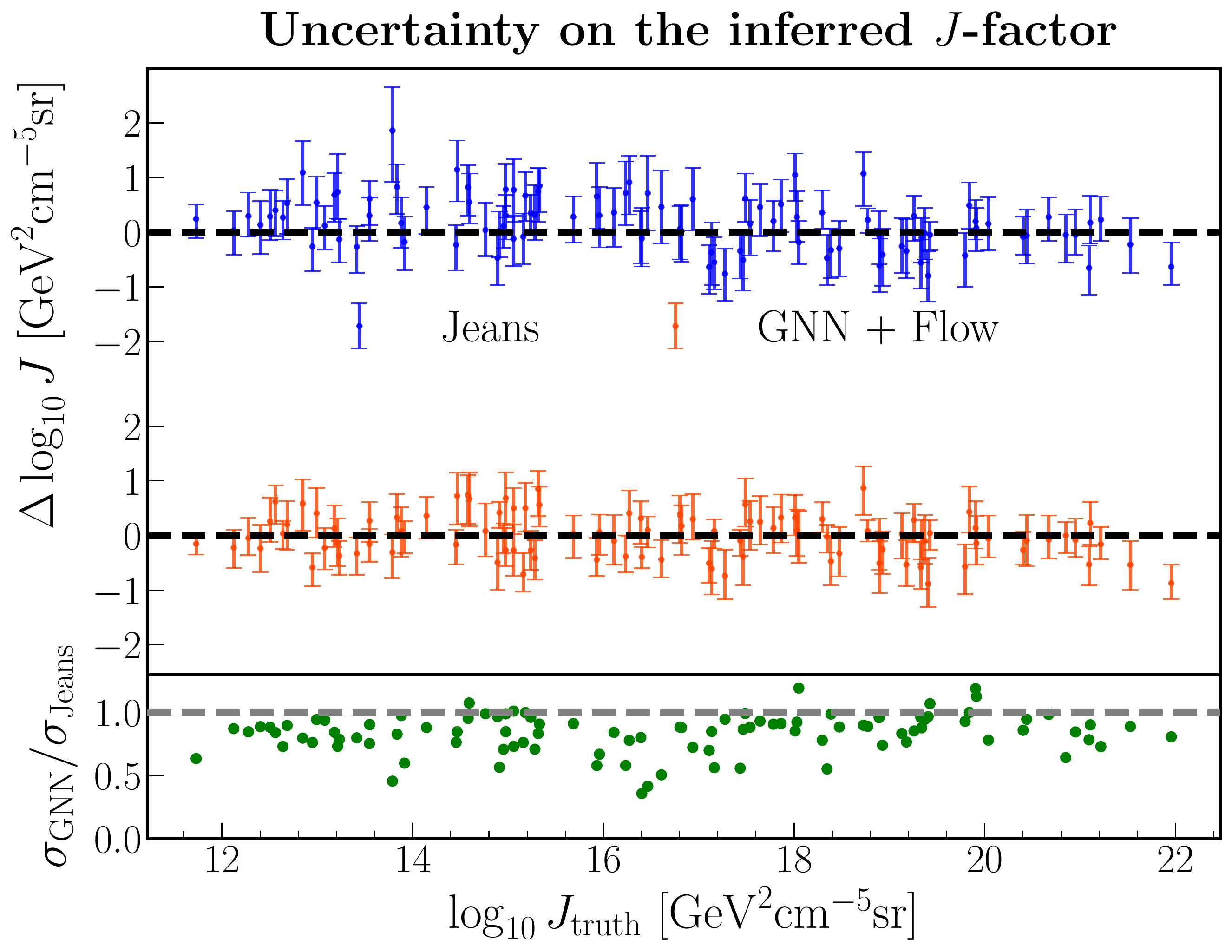}
    \caption{
    Comparison between the $J$-factors predicted by the Jeans analysis and our method.
    (Top) Differences between the predicted $\log_{10} J$-factors and the truth $\log_{10} J$-factors (i.e. $\Delta \log_{10} J = \log_{10} J_\mathrm{predict
    } - \log_{10} J_\mathrm{truth}$) for the Jeans analysis (blue) and our method (red).
    (Bottom) ratios of the symmetrized errors $\sigma_{\log_{10} J}$.
    The $J$-factors are normalized such that the host galaxies are at $100 \, \si{kpc}$.
    Our method generically provides tighter constraints on the $J$-factors than the traditional Jeans analysis.
    }
    \label{fig:j_factor}
\end{figure}

An important quantity in indirect searches for DM is the astrophysical $J$-factor---the integral along the line-of-sight $s$ and over solid angle $\Omega$ of the squared DM density corresponding to a source target,
\begin{equation}
    \label{eq:j_factor}
    J = \int \mathrm ds \int \mathrm d\Omega \, \rho^2(s, \Omega).
\end{equation}
The expected flux of photons sourced by annihilation processes is proportional to the $J$-factor; 
accurately determining it is therefore important for robustly interpreting results of DM indirect detection experiments using dwarf galaxies targets~\cite{Cirelli_2011, PhysRevLett.115.231301, 10.1093/mnras/stv942, Geringer_Sameth_2015, PhysRevLett.107.241303, PhysRevLett.107.241302, MAZZIOTTA201226, PhysRevD.91.083535, PhysRevD.89.042001, https://doi.org/10.48550/arxiv.1502.03081}. 
We compare the $J$-factors as predicted by the Jeans analysis and our method. In Fig.~\ref{fig:j_factor}, we show the inferred $J$-factors, normalized to the truth values, for 100 dwarf galaxies randomly sampled from our test dataset using our method (red error bars) and compare these to the corresponding $J$-factors obtained using Jeans analysis (green error bars).
The $J$-factors are normalized such that the host galaxies lie at a distance of 100 kpc.
We restrict the central slope $\gamma$ to lie between $[0, 1]$.
The bottom panel of Fig.~\ref{fig:j_factor} shows the ratio of the symmetrized errors obtained by our method to those obtained by the Jeans analysis. We see that, in the majority of cases tested, our method provides a better constraint on the $J$-factors than the Jeans analysis, as expected from the fact that the individual DM density parameters tend to be better constrained.
For the cases studied, the uncertainty on the $J$-factor using our method is on average $\sim 20\%$ and up to a factor of $\sim 2$ smaller than that obtained using Jeans analysis.
Details on the $J$-factors calculation can be found in App.~\ref{app:j_factor} and additional results comparing $J$-factors inferred between the two methods in App.~\ref{app:j_factor_results}. \\

To conclude, in this paper we introduced a novel method to reconstruct the DM density profiles of Milky Way dwarf galaxies from measured kinematics of tracer stars based on graph neural networks and simulation-based inference. We showed that the method compares favorably with and can outperform established approaches based on Jeans dynamical modeling in speed as well as constraining power.
The latter is due to the fact that our method incorporates more information about the phase-space structure of bound stars, contrasted with Jeans-based methods which typically rely on the second moments of the velocity distribution. Additionally, the method simultaneously models the stellar light profile and does not require that a fit to it be performed in advance.
While we used simulations of orbitally-anisotropic spherical systems as a proof-of-concept, in future work we plan to incorporate non-equilibrium dynamics using cosmologically-realistic simulations of isolated dwarfs as well as satellites of Milky Way-like systems, which would take into account baryonic effects like tidal disruption~\cite{2019MNRAS.487.1380G} and supernova feedback~\cite{2022MNRAS.513.3458B}. 

Code used to reproduce the results of this paper is available at \url{https://github.com/trivnguyen/dsph_gnn}. \\

\begin{acknowledgments}

We thank Laura Chang, Mariangela Lisanti, and Linda Xu for helpful conversations.
This work was performed in part at the Aspen Center for Physics, which is supported by National Science Foundation grant PHY-1607611.
This work is supported by the National Science Foundation under Cooperative Agreement PHY-2019786 (The NSF AI Institute for Artificial Intelligence and Fundamental Interactions, \url{http://iaifi.org/}).
This material is based upon work supported by the U.S. Department of Energy, Office of Science, Office of High Energy Physics of U.S. Department of Energy under grant Contract Number DE-SC0012567.
RW acknowledges support from the Office of Undergraduate Research at Princeton University and is grateful to Mariangela Lisanti for guidance.
This work used the Extreme Science and Engineering Discovery Environment (XSEDE), which is supported by National Science Foundation grant number ACI-1548562. Specifically, it used the Bridges-2 system, which is supported by NSF award number ACI-1928147, at the Pittsburgh Supercomputing Center (PSC) \cite{bridges2, xsede}.

This research made use of the  
\textsc{Bilby}~\cite{bilby},
\textsc{dynesty}~\cite{dynesty},
\textsc{IPython}~\cite{PER-GRA:2007}, 
\textsc{Jupyter}~\cite{Kluyver2016JupyterN},
\textsc{Matplotlib}~\cite{Hunter:2007},
\textsc{nflows}~\cite{nflows},
\textsc{NumPy}~\cite{harris2020array},
\textsc{PyTorch}~\cite{NEURIPS2019_9015}, 
\textsc{PyTorch Geometric}~\cite{Fey/Lenssen/2019}, 
\textsc{PyTorch Lightning}~\cite{william_falcon_2020_3828935}, and
\textsc{SciPy}~\cite{2020SciPy-NMeth} software packages.
\end{acknowledgments}

\bibliography{dsphs-gnn}

\clearpage

\input{supp.tex}

\end{document}

%% file: supp.tex
\onecolumngrid
\appendix
\makeatletter

\begin{center}
\textbf{\large Uncovering dark matter density profiles in dwarf galaxies with graph neural networks} \\ 
\vspace{0.05in}
{\it \large Appendix}\\ 
\vspace{0.05in}
{Tri Nguyen, Siddharth Mishra-Sharma, Reuel Williams, and Lina Necib}
\end{center}

\setcounter{equation}{0}
\setcounter{figure}{0}
\setcounter{table}{0}
\setcounter{section}{0}
\renewcommand{\thefigure}{S\arabic{figure}}
\renewcommand{\theHfigure}{S\arabic{figure}}
\renewcommand{\thetable}{S\arabic{table}}

\onecolumngrid

This Appendix is organized as follows.
In App.~\ref{app:analysis_details} we describe additional details of our analyses, including the assumed stellar phase-space distribution function in App.~\ref{app:forward_model}, the Jeans dynamical modeling method in App.~\ref{app:jeans}, computation of the annihilation $J$-factor in App.~\ref{app:j_factor}, and the assumed prior distributions in App.~\ref{app:priors}.
In App.~\ref{app:additional_results} we show additional results of our analysis, including systematic variations on the baseline analysis in App.~\ref{app:variation}, a quantitative comparison of the inner density slope and $J$-factor posteriors between our method and the Jeans analysis in Apps.~\ref{app:slope_results} and~\ref{app:j_factor_results} respectively, tests of statistical coverage in App.~\ref{app:exp_cov_prob}, and sensitivity of our method to observational projection in App.~\ref{app:projection} 

\section{Additional details on the analysis}
\label{app:analysis_details}

\begin{figure}[!htbp]
    \centering
    \includegraphics[width=0.98\textwidth]{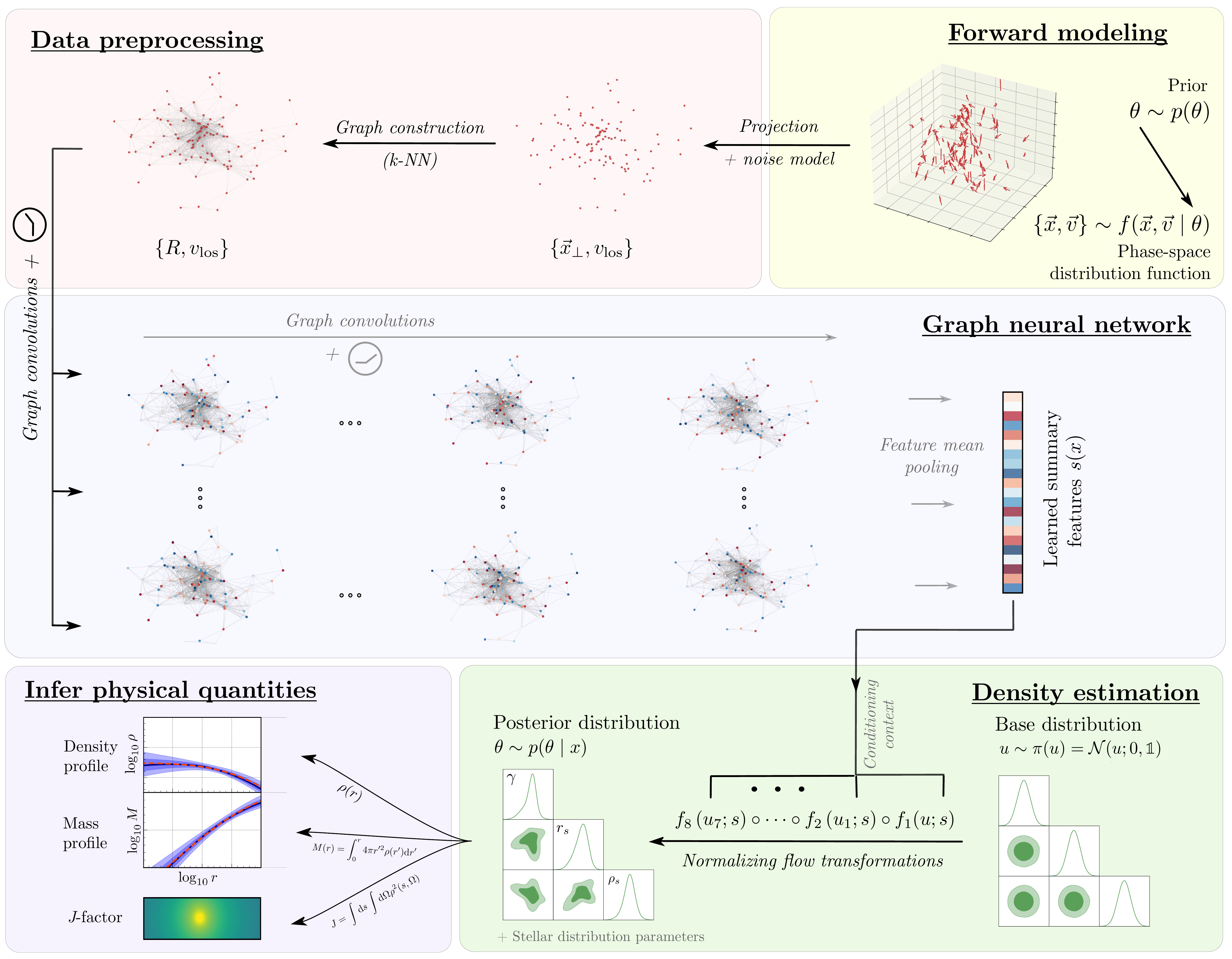}
    \caption{A schematic illustration of our method for inferring DM density profiles of dwarf galaxies from observed stellar kinematics. 
    }
    \label{fig:flow_chart}
\end{figure}

We elaborate here on several details of the analysis presented in the paper. We show a schematic illustration of our method, including a rough breakdown of the different steps of the pipeline, in Fig.~\ref{fig:flow_chart}.

\subsection{Details on the forward model and phase-space distribution function}
\label{app:forward_model}

In this section, we describe in detail the forward model used to generate the dataset.
We model the distribution function (DF) using the Osipkov-Merritt (OM) model as proposed by Refs.~\cite{1979PAZh....5...77O,1985AJ.....90.1027M}.
The OM model depends on the angular momentum $J$ and relative energy per unit mass $\mathcal{E} = \phi - v^2/2$, where $\phi$ is the gravitational potential and $v$ is the velocity, through the variable $Q \equiv \mathcal{E} - J^2/(2 r_a)$.
Here, $r_a$ is the scale radius of the velocity anisotropy profile $\beta(r)$ in Eq.~\eqref{eq:beta} that defines the transition from an isotropic velocity dispersion at small radii to a radially-biased dispersion at larger radii. 
To solve for the DF $f(\mathcal{E}, J) = f(Q)$, we first integrate $f(Q)$ over the velocity space to obtain the stellar mass-density profile
\begin{equation}
    \rho_\star(r) = \frac{4\pi}{1 + r^2/r_a^2} \int_0^\phi \mathrm dQ \, f(Q) \sqrt{2(\phi - Q)}.
\end{equation}
Note that $f(Q)=0$ for $Q < 0$.
The final DF $f(Q)$ can be obtained by Abel transforming $\rho_\star$,
\begin{equation}
    f(Q) = \frac{1}{2\pi \sqrt{2}} \frac{\mathrm d G(Q)}{\mathrm d Q}, 
    \quad \mathrm{where} \quad 
    G(Q) = - \int_0^Q \frac{\mathrm d \rho_Q}{\mathrm d \phi} \frac{\mathrm d\phi }{\sqrt{Q - \phi}}
    \quad \mathrm{and} \quad
    \rho(Q) = \left(1 + \frac{r^2}{r_a^2}\right) \rho_\star(r).
\end{equation}
Assuming that the system is dominated by DM, the gravitational potential $\phi$ depends on the DM density profile via Poisson's equation $\nabla^2 \phi = - 4\pi G \rho_\mathrm{DM}(r)$.
The DM density profile is parameterized through the generalized Navarro-Frenk-White model in Eq.~\eqref{eq:gNFW}.
The stellar density profile is
\begin{equation}
    \rho_\star(r) = \rho_{0, \star}
    \left[1 + \frac{r^2}{r_\star^2}\right]^{-5/ 2}
\end{equation}
where $r_\star$ is the scale radius and $\rho_{0, \star}$ is the density normalization of the stellar profile.
For $\rho_{0, \star}=3M/ 4 \pi r_\star^3$ and $M \propto L$, where $M$ and $L$ is the enclosed mass and luminosity, this is proportional to the 3-D Plummer profile in Eq.~\eqref{eq:plummer}.
The DF $f(Q)$ may be written in terms of the radius $r$ and the tangential and radial velocity $v_r$ and $v_t$ (i.e. $f(Q) = f(r, v_r, v_t)$) since $J=r v_t$ and $\mathcal{E}=\phi(r) - (v_t^2 + v_r^2) / 2$. \texttt{StarSampler} samples the coordinates $r, v_r, v_t$ of each star  from $f(r, v_r, v_t)$ using importance sampling \cite{gelman2003bayesian, doi:10.1080/00031305.1992.10475856,  rubin1988using}. 
The 6-D coordinates $(x, y, z, v_x, v_y, v_z)$ are then calculated by assuming a random projection direction and spherical symmetry.

\subsection{Details on Jeans analysis procedure}
\label{app:jeans}

In this section, we briefly outline the Jeans analysis procedure used in this work, closely following Ref.~\cite{chang2021}.
Following the derivation from Refs.~\cite{1982MNRAS.200..361B,BT2}, we first assume the system follows the collisionless Boltzmann equation
\begin{equation}
    \label{eq:boltzmann}
    \frac{\partial f}{\partial t} + \vec{v} \frac{\partial f}{\partial \vec{x}} - \frac{\partial \Phi}{\partial \vec{x}} \cdot  \frac{\partial{f}}{\partial{\vec{v}}} = 0, 
\end{equation}
where $\phi$ is the gravitational potential, $f=f(\vec{x}, \vec{v})$ is the phase-space distribution function, and ($\vec{x}$, $\vec{v}$) are the phase space coordinates of tracer stars. 
\begin{figure}[!t]
    \centering
    \includegraphics[height=0.32\linewidth]{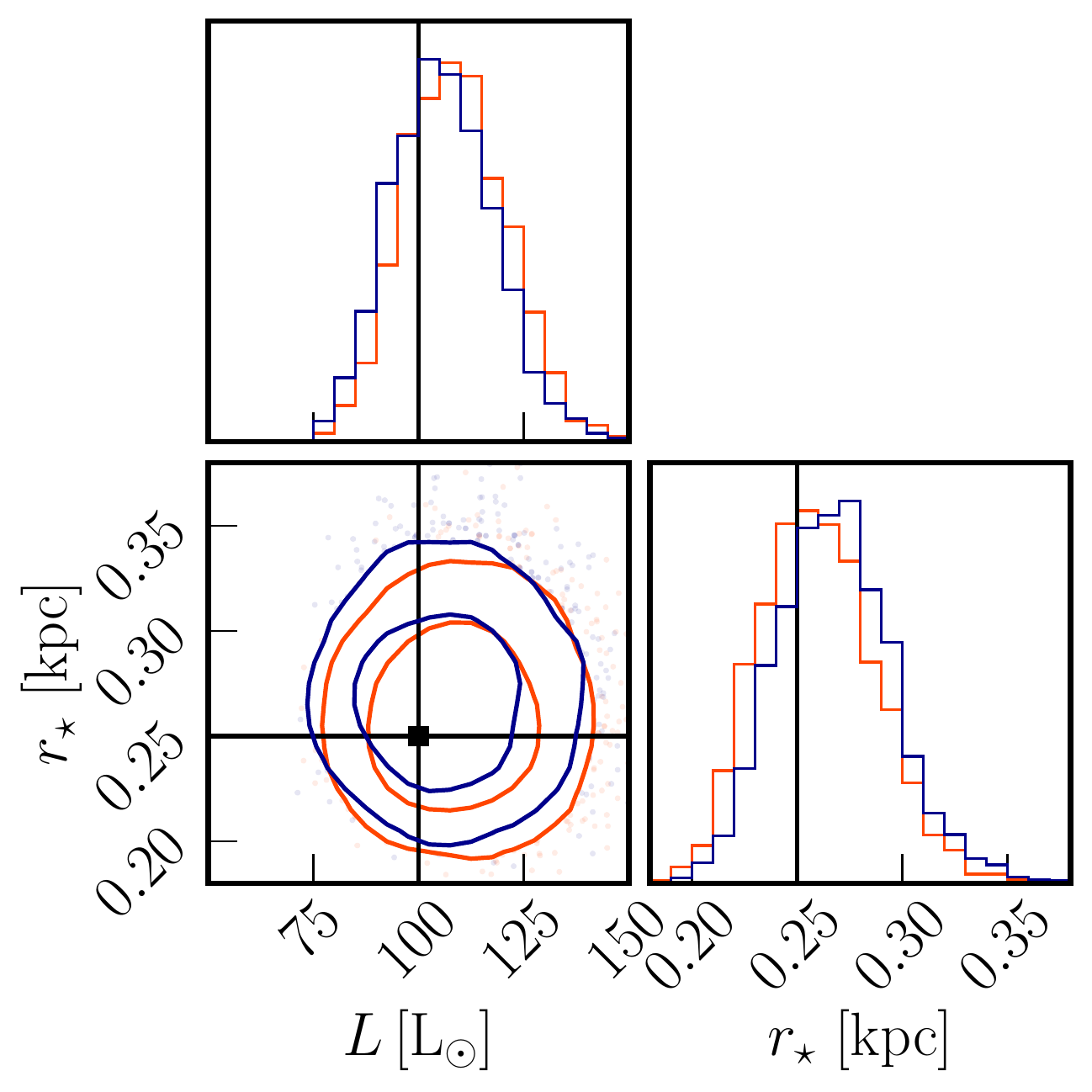}
    \includegraphics[height=0.32\linewidth]{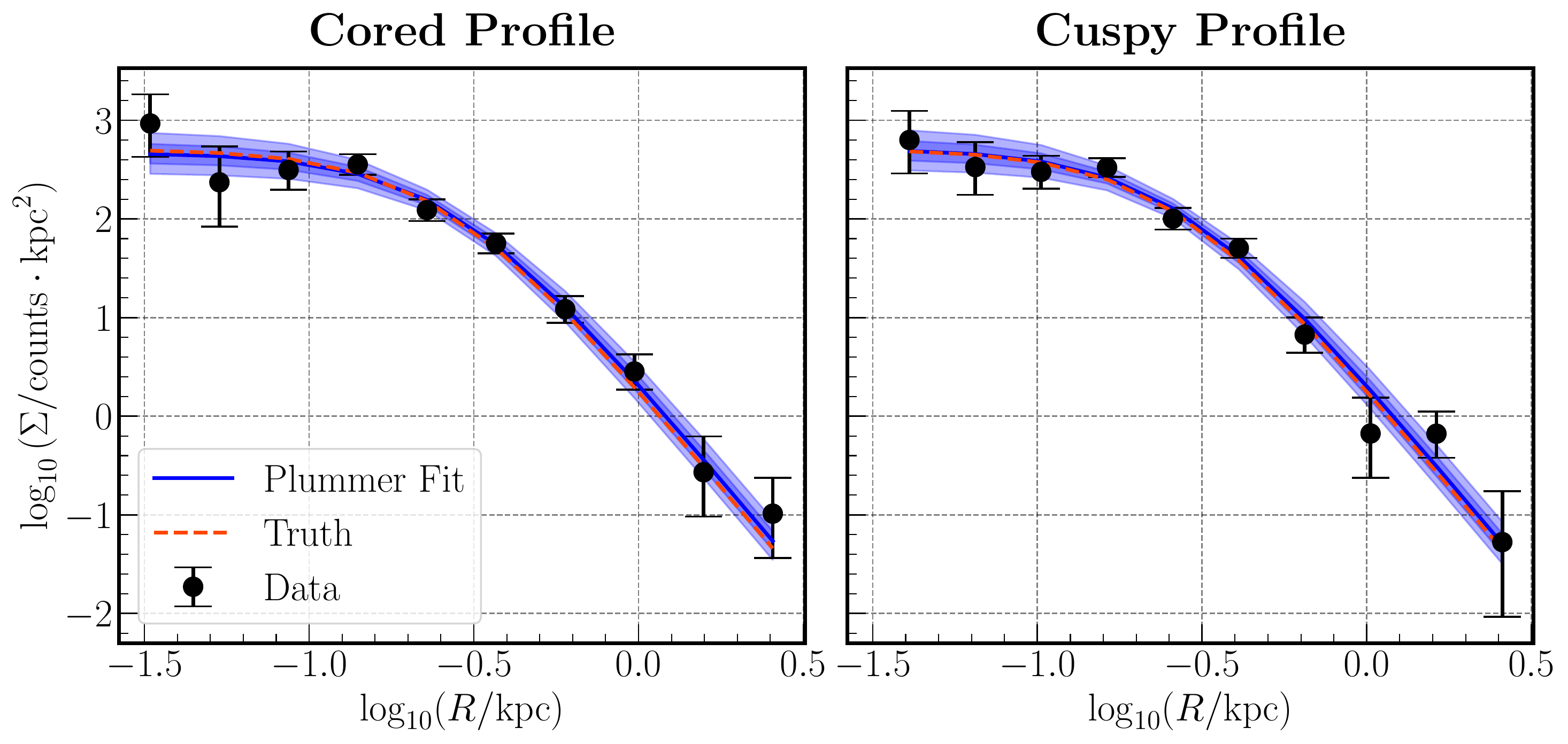}
    \caption{
    The initial Plummer fit for the two test galaxies presented in Fig.~\ref{fig:corner}.
    The left panel shows the 68\% and 95\% credible intervals of the posterior distributions of $L$ and $r_\star$.
    Both galaxies have the same light profile ($L = 100 \, \mathrm L_\odot$ and $r_\star=0.25 \; \si{kpc}$) but different DM profiles: cored DM profile (red) versus cuspy DM profile (blue).
    The right panel shows the best fit, the truth Plummer profile, and the binned data with Poisson uncertainties for both galaxies. 
    The dark blue and light blue bands show the middle-68\% and 95\% credible intervals of the reconstructed profile.
    }
    \label{fig:initial_plummer_post}
\end{figure}

Working in the dwarf galaxy's spherical coordinate system $(r, \theta, \phi)$ and assuming spherical symmetry and steady state, we multiply Eq. \eqref{eq:boltzmann} by the radial velocity $v_r$ and integrate over all velocity components to obtain the spherical Jeans equation
\begin{equation}
    \label{eq:jeans}
    \frac{1}{\nu} \left[\frac{\partial}{\partial r} (\nu \sigma_r^2) + \frac{2 \beta(r)}{r} (\nu \sigma_r^2)\right] = -\frac{\partial \phi}{\partial r} =  -\frac{GM(< r)}{r^2},
\end{equation}
where $\nu=\int \mathrm d^3\vec{v} \, f(\vec{x}, \vec{v})$ is the number density of the tracer stars, $\sigma_i$ is the velocity dispersion $\sigma_i = \sqrt{\langle v_i^2 \rangle - \langle v_i \rangle}$ for $i \in (r, \theta, \phi)$, and $\beta(r)=1-(\sigma^2_\theta + \sigma^2_\phi) / (2\sigma_r^2)$ is the velocity anisotropy profile.
The gravitational potential $\phi$ is assumed to be dominated by DM and may be written as $\phi = - G M(<r) / r$, where $G$ is the gravitational constant and $M(<r)$ is the enclosed mass of DM.
The Jeans equation \eqref{eq:jeans} has therefore the solution
\begin{equation}
    \label{eq:jeans_solution}
    \sigma^2(r)\nu(r)  = \frac{1}{g(r)} \int_r^\infty \frac{G M( < r') \nu(r')}{r'^2}g(r') \mathrm dr'
    \quad \mathrm{where} \quad
    g(r) = \exp\left(2 \int_0^r \frac{\beta(r')}{r'} \mathrm dr' \right).
\end{equation}
Projecting Eq.~\eqref{eq:jeans_solution} along the line of sight using the Abel transformation $s(r) \rightarrow S(R)$ for the spherically symmetric function $s(r)$,
\begin{equation}
    S(R) = 2 \int_R^\infty \frac{s(r) r \, \mathrm dr}{\sqrt{r^2 - R^2}},
\end{equation}
we obtain
\begin{equation}
    \label{eq:jeans_solution_proj}
    \sigma^2_p(R) I(R) = 2 \int_R^\infty 
    \left(1 - \beta(r) \frac{R^2}{r^2}\right)
    \frac{\nu(r) \sigma^2_r(r) r}{\sqrt{r^2 - R^2}} \mathrm dr,
\end{equation}
where $\sigma_p$ is the projected velocity dispersion profile and $I(R)$ is the projected number density of tracer stars, also known as the surface brightness or the light profile.
In our analysis, we parameterize the DM density profile using the gNFW profile in Eq.~\eqref{eq:gNFW}.
The light profile $I(R)$ is the projection along the line-of-sight of the 3-D Plummer profile in Eq.~\eqref{eq:plummer} and is given by
\begin{equation}
    I(R) = \frac{L}{\pi r_\star^2}\left(1 + \frac{R^2}{r_\star^2}\right)^{-2},
\end{equation}
which has two free parameters: the total luminosity $L$ and the scale length $r_\star$.

The Jeans analysis requires two separate fits: fitting the light profile and subsequently fitting the DM profile.
We first fit $L$ and $r_\star$ using the priors in Table~\ref{tab:priors}.
We approximate the posteriors as Gaussian distributions and use them as the priors for $L$ and $r_\star$ in the second fit. 
To compare the performance between Jeans analysis and our machine learning framework, the DM priors are set to be the same as those used to generate the training datasets for the GNN and normalizing flow method, and are summarized in Table~\ref{tab:priors}.

For the initial fit of the light profile, we bin the data in $\log_{10}$-spaced bins in the projected radius $R$.
The number of bins is chosen to be $\sim \sqrt{N_\mathrm{stars}}$ where $N_\mathrm{stars}$ is the number of stars. 
Similarly to Ref.~\cite{chang2021}, we assume Poisson uncertainties on the number of stars in each bin.
Let $n_i(\theta)$ and $\hat{n_i}$ be the predicted and mean number of stars (with $\theta$ is the parameters of the light profile model) in $i^\mathrm{th}$ bins.
We construct the binned likelihood~\cite{Barlow:2004wg}
\begin{equation}
    \ln \mathcal{L}_\mathrm{Plummer} = - \frac{1}{2} \sum_{i=1}^{N_\mathrm{stars}} \frac{(\hat{n}_i - n_i(\theta))^2}{V_i - V'_i(\hat{n_i} - n_i(\theta))},
\end{equation}
where $V = \sigma_\mathrm{lo} \sigma_\mathrm{hi}$ and $V'=\sigma_\mathrm{hi} - \sigma_\mathrm{lo}$; here, $\sigma_\mathrm{lo}$ and $\sigma_\mathrm{hi}$ are the asymmetric Poisson uncertainties. 
We refer to Ref.~\cite{chang2021} for further details. 
As described in the main text, we sample the posterior with nested sampling~~\cite{2004AIPC..735..395S,skilling2006} using the \textsc{dynesty} module~\cite{dynesty}.
We use $n_\mathrm{live} = 500$ live points and a convergence tolerance
of $\Delta \ln \mathcal{Z} = 0.1$ on the estimated remaining contribution to the log-evidence.
In Fig.~\ref{fig:initial_plummer_post}, we show the initial Plummer fit for the two test galaxies presented in Fig.~\ref{fig:corner}. 
The posteriors for $L$ and $r_\star$ are well-constrained and agree well with the true Plummer profile.

As mentioned, we approximate the posteriors of $L$ and $r_\star$ as Gaussian distributions and use them as priors in the second fit of the DM profile.
Unlike the initial Plummer fit, for the Jeans analysis, we construct an unbinned Gaussian likelihood.
The likelihood is given by~\cite{2008ApJ...678..614S},
\begin{equation}
    \mathcal{L}_\mathrm{Jeans} = 
    \prod_{i=1}^{N_\mathrm{stars}} \frac{(2\pi)^{-1/2}}{\sqrt{\sigma_p^2(R_i) + \Delta{v_i}^2}}
    \exp\left[-\frac{1}{2} \left(\frac{(v_i - \bar{v})^2}{\sigma_p^2(R_i) + \Delta{v_i}^2}\right) \right],
\end{equation}
where $\sigma_p (R)$ is the projected velocity dispersion profile in Eq.~\eqref{eq:jeans_solution_proj}, $\bar{v}$ is the mean velocity of tracer stars, and $v_i$ and $\Delta{v_i}$ is the line-of-sight velocity and its measurement error for star $i$. 
We treat the mean velocity $\bar{v}$ as a nuisance parameters and fit it together with the DM and light profile, using the prior distribution in Table~\ref{tab:priors}.
We sample the joint DM and light profile posteriors using \textsc{dynesty}~\cite{dynesty} with the same sampler configuration as in the initial Plummer fit.

\subsection{Details on the $J$-factor calculation}
\label{app:j_factor}

In this section, we outline the details on the annihilation $J$-factor calculation.
We transform the coordinate system in Eq.~\eqref{eq:j_factor} from a spherical coordinate system centered at the Earth's location to a spherical coordinate system centered at the dwarf galaxy considered.
Let $\vec{r'}$ be the position from the center of the galaxy; the $J$-factor in this coordinate system is given by
\begin{equation}
    \label{eq:j_factor_dsph_coord}
    J = \int \mathrm ds \int \mathrm d\Omega \, \rho^2(s, \Omega) = \int \mathrm dV \frac{\rho^2(s, \Omega)}{s^2} = \int \mathrm dV' \frac{\rho^2 (r', \Omega')}{r'^2 - 2 d_c r' \cos\theta' + d_c^2},
\end{equation}
where $d_c = |\vec{r}|$ is the comoving distance and $\vec{r} \cdot \vec{r'} = d_c r' \cos\theta'$.
Note that the volume elements between the two coordinate systems are equal, $\mathrm dV = \mathrm dV'$, since the transformation is only a translation.
We integrate Eq.~\eqref{eq:j_factor_dsph_coord} up to the virial radius $r_\mathrm{vir}$, defined as the radius within which the mean density of the halo is equal to a specified overdensity factor times the critical density of the Universe, $\overline{\rho}(r<r_\mathrm{vir}) = \Delta_\mathrm{vir}\rho_c$.
As per common convention, we take $\Delta_\mathrm{vir} = 200$. We assume the distant-source approximation $d_c \gg r_\mathrm{vir} > r'$, justified for most Milky Way dwarf galaxies, which allows us to write $J$ as a volume integral in spherical coordinates~\cite{Lisanti:2017qoz}
\begin{equation}
    J = \int \mathrm dV' \frac{\rho^2 (r', \Omega')}{r'^2 - 2 d_c r' \cos\theta' + d_c^2}
    \approx \frac{1}{d_c^2} \int \mathrm dV' \, \rho^2(r') .
\end{equation}
Plugging the gNFW density profile Eq.~\eqref{eq:gNFW} into the above expression, we find
\begin{align}
    J &= \frac{1}{d_c^2} \int \mathrm dV' \, (\rho_\mathrm{DM}^\mathrm{gNFW})^2(r') 
    = \frac{4 \pi \rho^2_s r_s^2}{d_c^2} \int_0^{r_\mathrm{vir}} \mathrm dr' \left(\frac{r}{r_s}\right)^{2-2\gamma}\left(1 + \frac{r}{r_s}\right)^{2\gamma - 6} \\
    &= \frac{4 \pi \rho^2_s r_s^3}{d_c^2} 
    \left[
    \frac{c_\mathrm{vir}^{3-2\gamma} (1 + c_\mathrm{vir})^{2\gamma - 5} \left(20 + 4\gamma^2 + 2 c_\mathrm{vir} ( 5 + c_\mathrm{vir}) - 2\gamma ( 9 + 2 c_\mathrm{vir})\right)}{(5-2\gamma)(4-2\gamma)(3-2\gamma)}
    \right]
\end{align}
where $c_\mathrm{vir}\equiv r_\mathrm{vir}/r_s$ is the virial concentration of the dwarf galaxy. 
Note that the integration only converges when $\gamma < 1.5$, which is the case for the DM profiles shown in Fig.~\ref{fig:j_factor}.

\subsection{Prior distributions on the parameters of interest}
\label{app:priors}

\input{tables/priors}

We show the prior distribution for both our method and the Jeans analysis in Table \ref{tab:priors}.
The priors on the DM density profile parameters are the same between the Jeans analysis and our method.
As described in App.~\ref{app:jeans}, the Jeans analysis consists of an initial fit of the light profile parameters $L$ and $r_\star$, and a subsequent joint fit of both the DM and light profile parameters.

\section{Additional results}
\label{app:additional_results}

\subsection{Systematic variations on the analysis}
\label{app:variation}

In this Appendix we explore variations on the assumptions made in our baseline analysis, including the effect of varying the mean number of stars $\mu_\mathrm{stars}$ and line-of-sight velocity measurement error $\Delta v$. 
We also examine the performance of different graph convolution schemes. \\

\begin{figure}[t]
    \centering
    \includegraphics[width=0.95\linewidth]{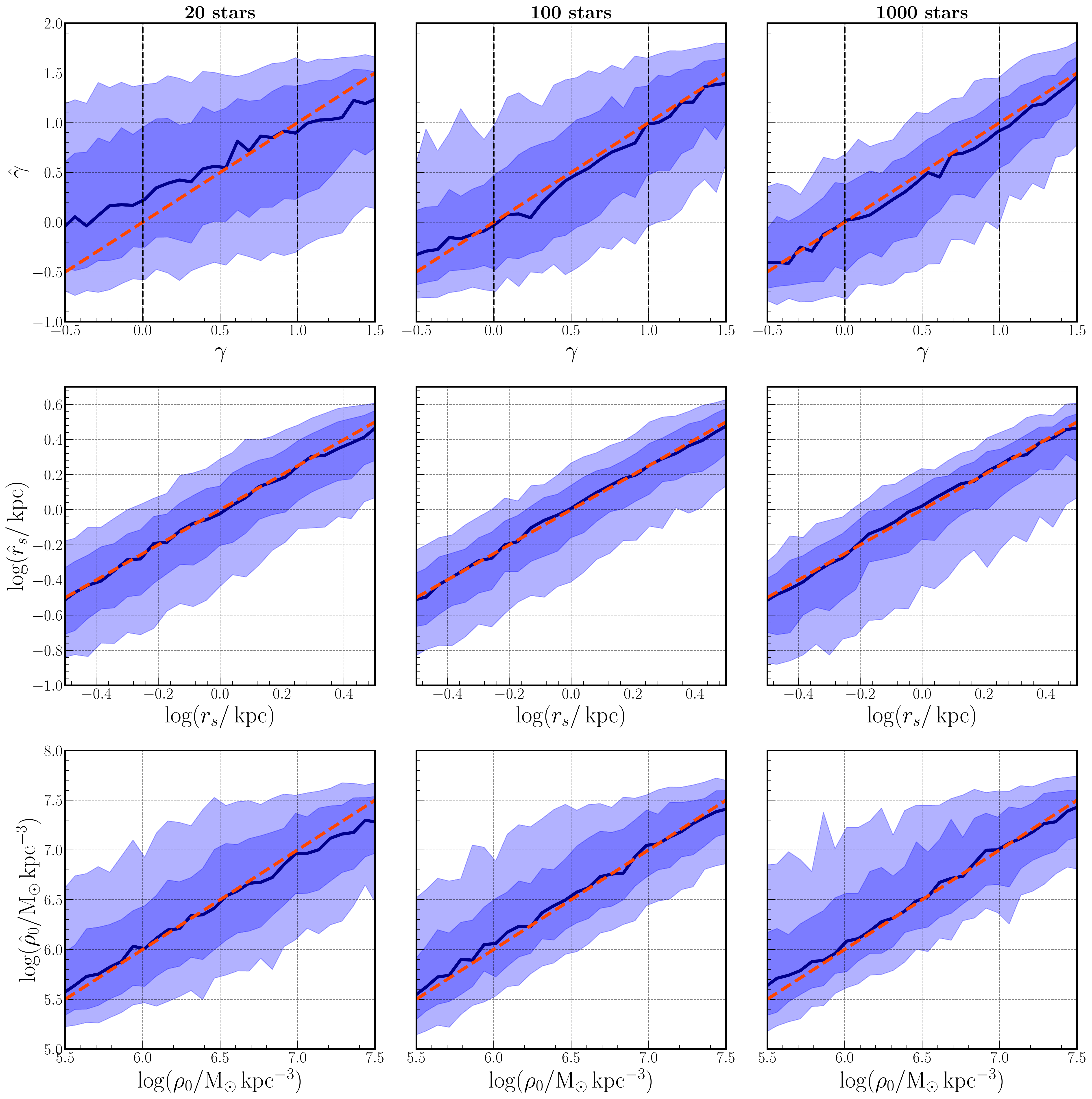}
    \caption{The predicted DM parameters versus truth DM parameters for three different runs.
    In each run, a GNN and normalizing is trained and tested on galaxies with a mean number of stars $\mu_\mathrm{stars}=20 \, (\mathrm{left}), 100 \, (\mathrm{center}), 1000 \, (\mathrm{right})$ stars. 
    The line-of-sight velocity measurement error is set to be $\Delta v = 0.1 \, \si{km/s}$.
    The $\mu_\mathrm{stars}=100$ case (middle column) is the same as in  Fig.~\ref{fig:percentile}. 
    The median (solid blue line), middle-68\% percentile (blue band), and middle-95\% (light blue band) containment regions of each bin are shown. The dashed red line denotes where the predicted values are equal to the true values.
    }
    \label{fig:percentile_nstars}
\end{figure}

\noindent 
\textbf{Variations on the number of stars:} We generate three datasets (including the one presented in the main text) with the same velocity measurement error $\Delta=0.1 \, \si{km/s}$ and different mean number of stars: $\mu_\mathrm{stars}=20, 100, 1000$.
As in the baseline case, each dataset contain 80,000 training samples, 10,000 validation samples, and 10,000 test samples.
For each dataset, we train a GNN and normalizing flow with the same hyperparameters as described in the main text and plot the results on the test samples in Fig.~\ref{fig:percentile_nstars}.
Similarly to Fig.~\ref{fig:percentile}, we take the marginal medians as the predicted DM parameters, bin them based on their truth values, and show the median (solid blue line), middle-68\% (blue bands), and middle-95\% (light blue bands) containment regions.
The baseline $\mu_\mathrm{stars}=100$ stars (middle column) case is the same as that in Fig.~\ref{fig:percentile} and is shown here for comparison. 

The performance on the scale radius $r_s$ (middle row) and density normalization $\rho_s$ (bottom row) is similar for all three cases.
For the inner slope $\gamma$ (top row), we see that increasing the number of stars helps increase the accuracy of the marginal medians.
This is expected because the underlying phase-space distribution is more completely sampled for a larger number of observed tracer stars.
We observed a similar increase in prediction accuracy with increasing sample size in Ref.~\cite{chang2021}.
The marginal medians for the $\mu_\mathrm{stars}=20$ case are slightly biased towards the tails of the $\gamma$ prior distribution.
This is potentially due to clipping caused by a finite prior range that prevents the marginal median from being centered around the true value when constraining power is low. 
\newline

\begin{figure}[t]
    \centering
    \includegraphics[width=0.95\linewidth]{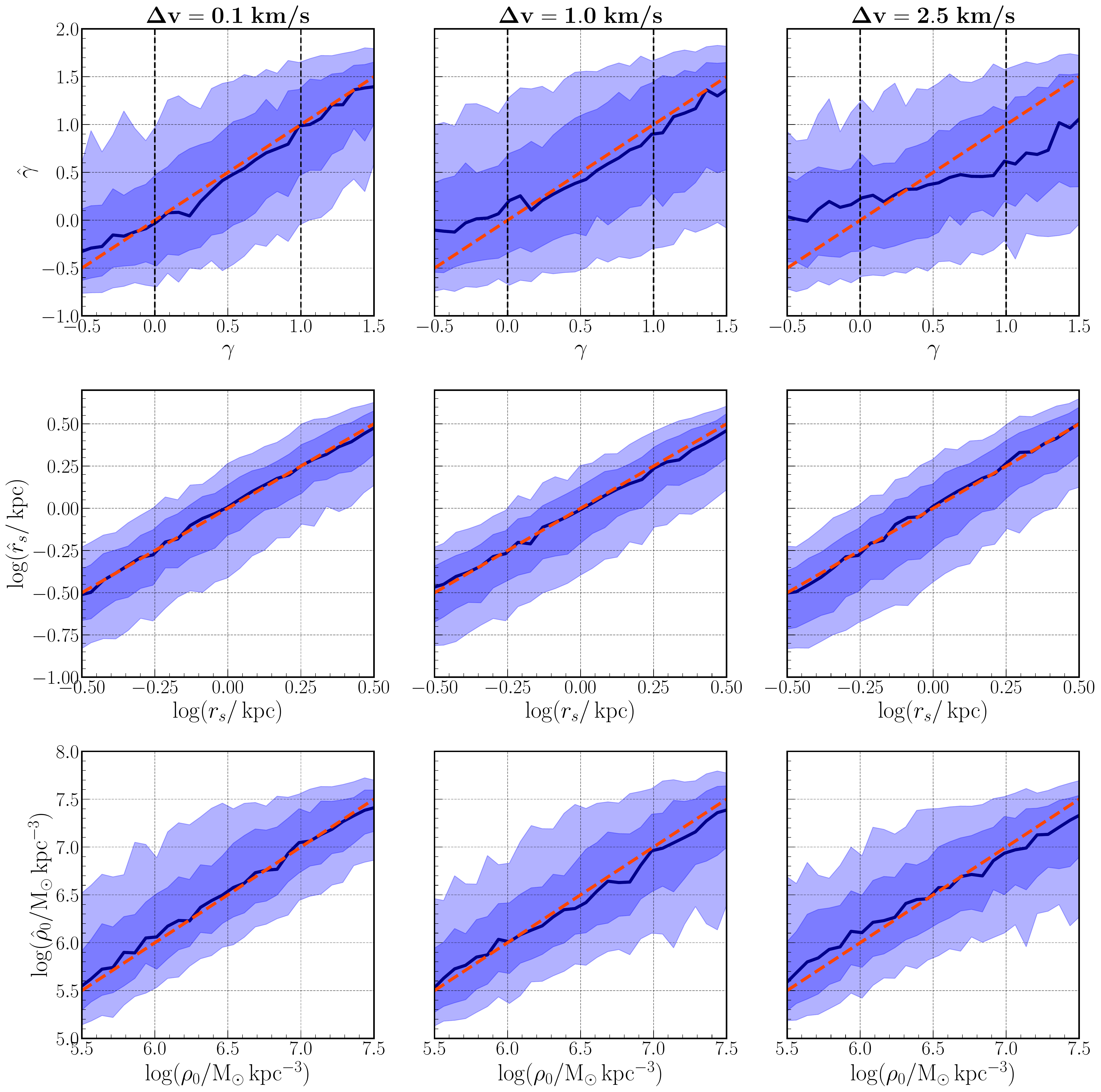}
    \caption{Same as Fig.~\ref{fig:percentile_nstars}, but with different line-of-sight velocity errors $\Delta v$.
    The mean number of stars $\mu_\mathrm{stars}$ is 100 stars.
    The $\Delta v = 0.1 \, \si{km/s}$ case (left column) is the same as in  Fig.~\ref{fig:percentile}. 
    }
    \label{fig:percentile_verr}
\end{figure}

\noindent 
\textbf{Variations on measurement uncertainty:} We generate three datasets with the same mean number of stars $\mu_\mathrm{stars}$ and different velocity measurement errors $\Delta v=0.1, \, 1.0, 2.5 \, \si{km/s}$.
Similar to our baseline case of $\Delta v=0.1\, \si{km/s}$, each dataset has 80,000 training samples, 10,000 validation samples, and 10,000 test samples, and we train our pipeline on each dataset separately.
The results of this variation are shown in Fig.~\ref{fig:percentile_verr} in the same format as Fig.~\ref{fig:percentile_nstars} and Fig.~\ref{fig:percentile} (with the baseline $\Delta v=0.1 \, \si{km/s}$ case the same as in Fig.~\ref{fig:percentile}).

Again, we observe that for the scale radius $r_s$ (middle row) and the density normalization $\rho_0$ (bottom row), the performance is approximately constant across all variations. 
Similar to Ref.~\cite{chang2021}, as the measurement uncertainties increase we observe a decrease in constraining power of the central slope $\gamma$.
This is to be expected because we did not explicitly account for the uncertainties in the neural network architecture---sampling from the noise model is treated as a form of data augmentation, with larger error magnitudes leading to increased sample variance and reduced sample efficiency.
We defer the explicit inclusion of observational uncertainties in the neural network construction to future work.
\newline

\begin{figure}
    \centering
    \includegraphics[width=0.95\linewidth]{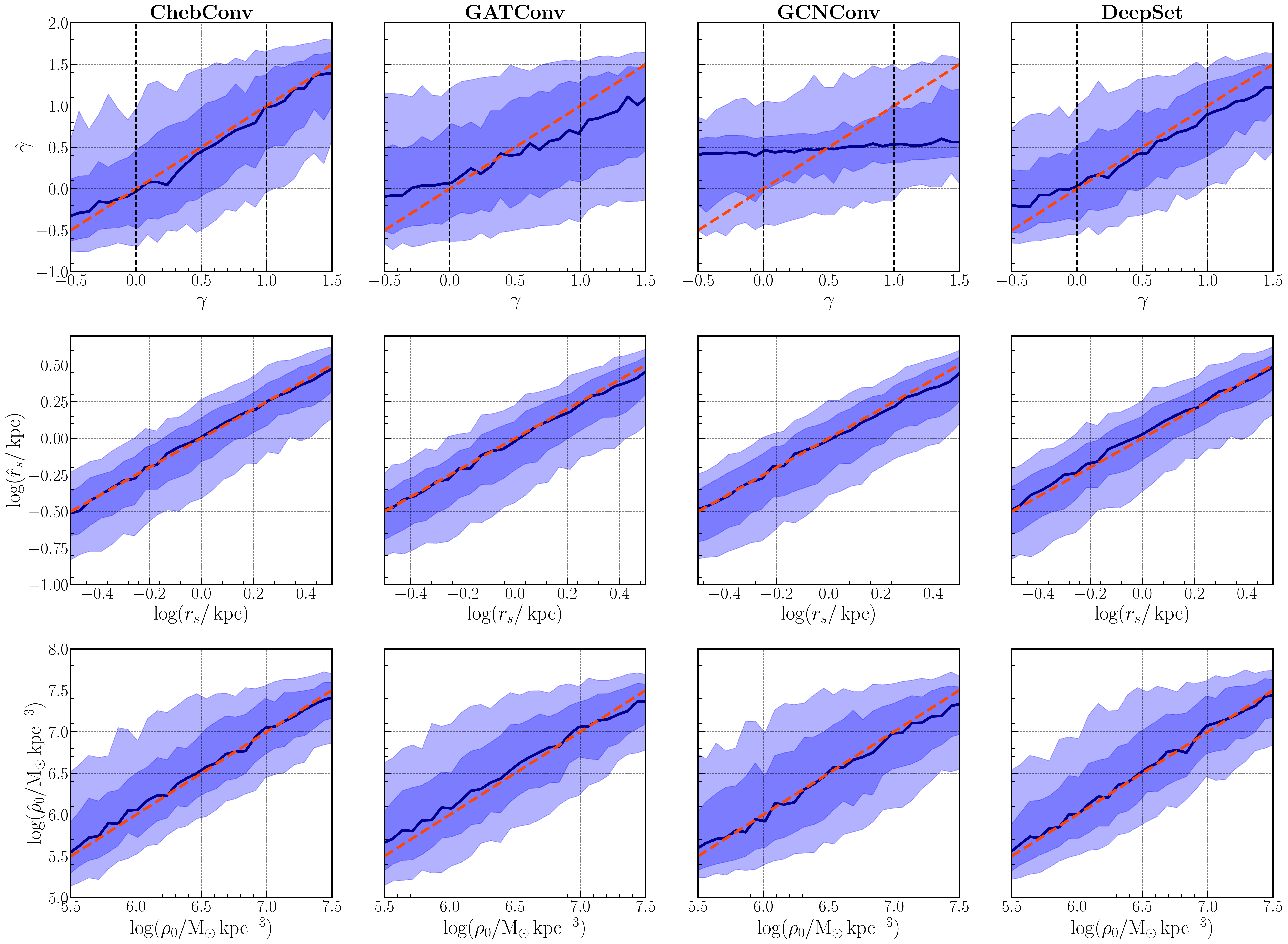}
    \caption{Same as Fig.~\ref{fig:percentile_nstars}, but with variation in the type of graph convolutional layer.
    Each network is trained and tested on the dataset presented in the main text, i.e. $\mu_\mathrm{stars}=100$ stars and $\Delta v = 0.1 \, \si{km/s}$.
    The ChebConv case (left column) refers to the neural network presented in the main text and in Fig.~\ref{fig:percentile}.
    }
    \label{fig:percentile_archs}
\end{figure}

\noindent 
\textbf{Variations on the graph convolution scheme:} In Fig.~\ref{fig:percentile_archs} we show variations on the type of graph convolutional layer used, otherwise keeping all hyperparameters (e.g., channel dimension) the same as the baseline case. This case, labeled `ChebConv' and shown in the left-most column, uses the graph convolution prescription from Ref.~\cite{2016arXiv160609375D}.

The second column shows results using the graph attention layer from Ref.~\cite{velivckovic2017graph}, which uses the attention mechanism to implicitly weigh neighboring nodes. The final column shows results using the deep set architecture, where node-wise features are obtained using a dense network and aggregated through averaging, before finally being passed through a dense layer as in the baseline case. Relatively good recovery of all parameters can be seen in these cases. Together, these results point to the fact that aggregation of neighborhood information may not play a key role in the success of our method on the examples tested. We expect this fact to not hold on more realistic test cases---in particular, the systems in this paper were chosen to be relatively simple (spherical and dynamically equilibrated) in order to enable a direct comparison with the conventional Jeans analysis method. With these assumptions and our choice of stellar and DM profiles, the node-wise features $\{R, v_\mathrm{los}\}$ can be assumed to be independent. 

Finally, the third column of Fig.~\ref{fig:percentile_archs} shows results using the graph convolutional layer from Ref.~\cite{kipf2016semi} with the default configuration in \textsc{PyTorch Geometric}, showing poor recovery of the inner slope $\gamma$.

\subsection{Comparison of inner density slopes obtained using Jeans analysis and GNN + Flow}
\label{app:slope_results}

Given two probability distribution functions $P(x)$ and $Q(x)$, 
defined over the same probability space $\mathcal{X}\ni x$,
the Jensen-Shannon (JS)-divergence is defined as
\begin{equation}
    \label{eq:jsd}
    D_\mathrm{JS}(P || Q) \equiv \frac{1}{2}\left[ D_\mathrm{KL}(P || M) + D_\mathrm{KL}(Q || M)\right],
\end{equation}
where $M(x) = P(x) + Q(x)$ and $D_\mathrm{KL}$ is the Kullback–Leibler (KL) divergence, for which we use the definition
\begin{equation}
    \label{eq:kld}
    D_\mathrm{KL}(P || Q) = \sum_{x \in \mathcal{X}} P(x) \log_{2} \left(\frac{P(x)}{Q(x)}\right).
\end{equation}
The KL divergence of $P$ from $Q$ represents the expected entropy gain from using $Q$ as an approximation for truth distribution $P$~\cite{kld}.
The JS-divergence is based on the KL-divergence but has more desirable properties for the present case, specifically being symmetric in the two posterior distributions.
Note that using a $\log_2$ definition of the constituent KL-divergences, the JS-divergence is constrained to lie within the range $D_\mathrm{JS}\in[0, 1]$, with $0$ corresponding to two identical distributions and $1$ two non-overlapping distributions.

\begin{figure}[t]
    \centering
    \includegraphics[width=0.5\linewidth]{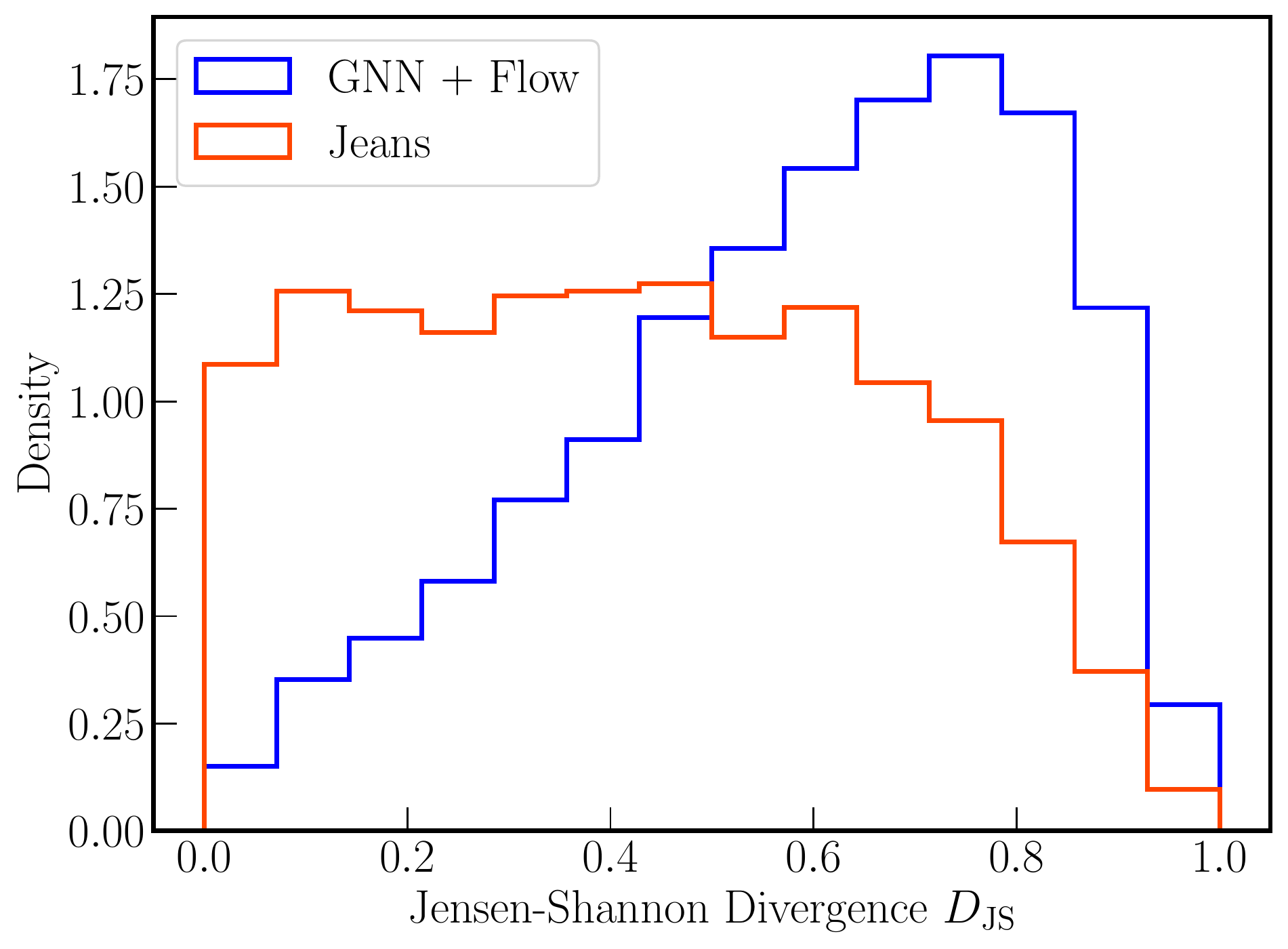}
    \caption{The JS divergence $D_\mathrm{JS}$ between the $\gamma=0$ and $\gamma=1$ posteriors for the Jeans analysis (red) and the GNN and normalizing flow model (blue).
    As evidenced from the higher values of the $D_\mathrm{JS}$, compared to the Jeans analysis, the GNN and normalizing flow produces $\gamma=0$ and $\gamma=1$ more distinct from each other.
    }
    \label{fig:jsd}
\end{figure}

We generate 100 galaxies with cored DM profiles ($\gamma = 0$) and 100 galaxies with cuspy profiles ($\gamma = 1$).
These galaxies have the same DM scale radius $r_s$, density normalization $\rho_0$, and light profile parameters as the two galaxies presented in Figs.~\ref{fig:profiles} and~\ref{fig:corner} of the main text.
We obtain samples the DM parameter posteriors for each galaxy and calculate $D_\mathrm{JS}$ using Eq.~\eqref{eq:jsd} for each pair of cored and cuspy galaxies (in total, there are 10,000 pairs).
We show the distribution of the JS-divergences $D_\mathrm{JS}$ between the $\gamma=0$ posterior and $\gamma=1$ posterior for the Jeans analysis and for our method in Fig.~\ref{fig:jsd}.
As mentioned, in the main text, the median and middle-68\% containment region values of $D_\mathrm{JS}$ are $D_\mathrm{JS}^\mathrm{Jeans}=0.417^{+0.288}_{-0.280}$ and $D_\mathrm{JS}^\mathrm{GNN}= 0.629^{+0.196}_{-0.278}$.
Our method generically produces higher values of $D_\mathrm{JS}$ compared to the Jeans analysis, corresponding to a larger contrast between the cored and cuspy $\gamma$ posteriors and thus increased ability to distinguish between the two scenarios given a set of observations.

\subsection{Comparison of $J$-factors inferred using Jeans analysis and GNN + Flow}
\label{app:j_factor_results}

\begin{figure}[hbt!]
    \centering
    \includegraphics[width=0.9\linewidth]{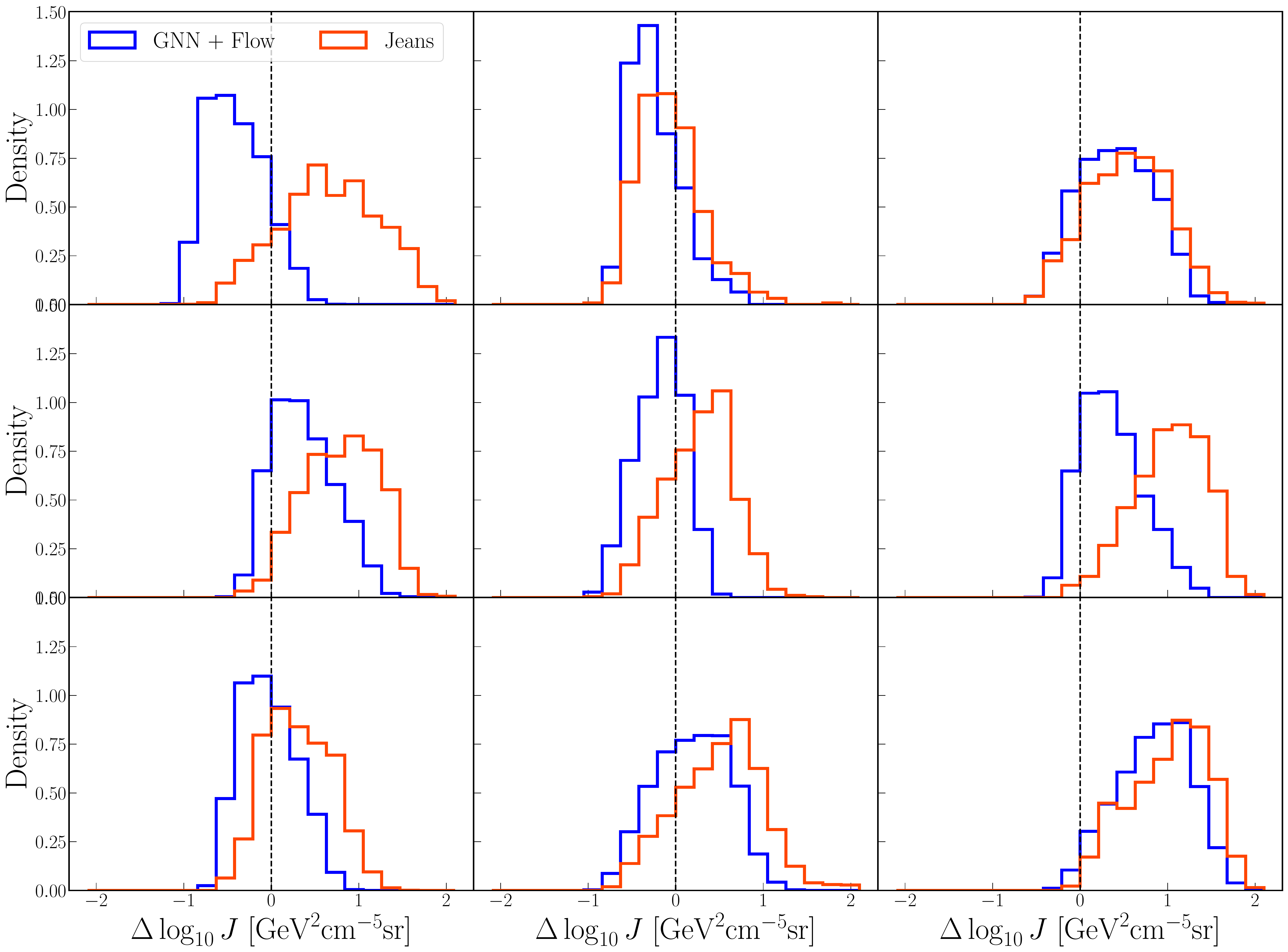}
    \caption{
    Posterior distributions of the $J$-factors predicted by the Jeans analysis (red) and by the GNN and normalizing flow (blue).
    Each panel show the $\Delta \log_{10} J = \log_{10} J_\mathrm{predict
    } - \log_{10} J_\mathrm{truth}$ posteriors of an example galaxy.
    The dashed black line represents the truth $J$-factor value (i.e. $\Delta \log_{10} J=0$).
    }
    \label{fig:j_factor_posteriors}
\end{figure}

As mentioned in the main text, we calculate the $J$-factors (normalized to distance of 100 kpc) for 100 galaxies randomly sampled from our test set.  
In this section, we plot the posterior distribution of the $J$-factor for a few examples for the Jeans analysis and our method in Fig.~\ref{fig:j_factor_posteriors}.
In all cases shown, the GNN and normalizing flow can predict the $J$-factor with a similar or higher accuracy compared to the Jeans analysis.

\subsection{Test of statistical coverage of the inferred posteriors}
\label{app:exp_cov_prob}

Simulation-based inference methods such as those employed in this work can be susceptible to producing overconfident posteriors~\cite{arxiv.2110.06581}. 
In this section, we examine the quality of the posterior distributions produced by our simulation-based inference pipeline following the prescription in Ref~.\cite{arxiv.2110.06581}.

Using the same notation as in the main text, let $\theta$ be the parameters of interest (i.e. the DM and stellar parameters) and $x$ be the observable (i.e. the graph representation of a galaxy constructed from its the stellar kinematics).
We denote our learned posterior density estimator as $\hat{p}(\theta | x)$.
For a confidence level $1 - \alpha$, the expected coverage probability is
\begin{equation}
    \mathbb{E}_{(\theta, x) \sim p(\theta, x)}\left[\mathds{1}_\Theta(\theta \in \Theta_{\hat{p}(\theta | x)} (1 - \alpha))\right],
\end{equation}
where $\Theta_{{\hat p(\theta | x)}}(1-\alpha)$ gives the $1 - \alpha$ highest posterior density interval (HDPI) of the estimator $\hat{p}(\theta | x)$ and $\mathds{1}_\Theta()$ is an indicator function mapping samples that fall within the HDPI to unity.
Given $N$ samples from the joint distribution $(\theta^\star, x) \sim p(\theta, x)$, the empirical expected coverage for the posterior estimator $\hat{p}(\theta | x)$ is defined as
\begin{equation}
    \frac{1}{N} \sum_{i=1}^{N} \mathds{1}_\Theta\left(\theta^\star \in \Theta_{\hat{p}(\theta | x)} (1 - \alpha)\right).
\end{equation}
The nominal expected coverage is the expected coverage in the case when $\hat p(\theta | x) = p(\theta | x)$ and equals to the confidence level $1 - \alpha$.
In general, we want our estimator to have an empirical expected coverage probability larger than or equal to the nominal expected coverage probability at all confidence levels. 
Such estimators will produce conservative posteriors, in contrast to overconfident posteriors which may spuriously exclude allowable regions of parameter space.
    
\begin{figure}[!b]
    \centering
    \includegraphics[width=0.5\linewidth]{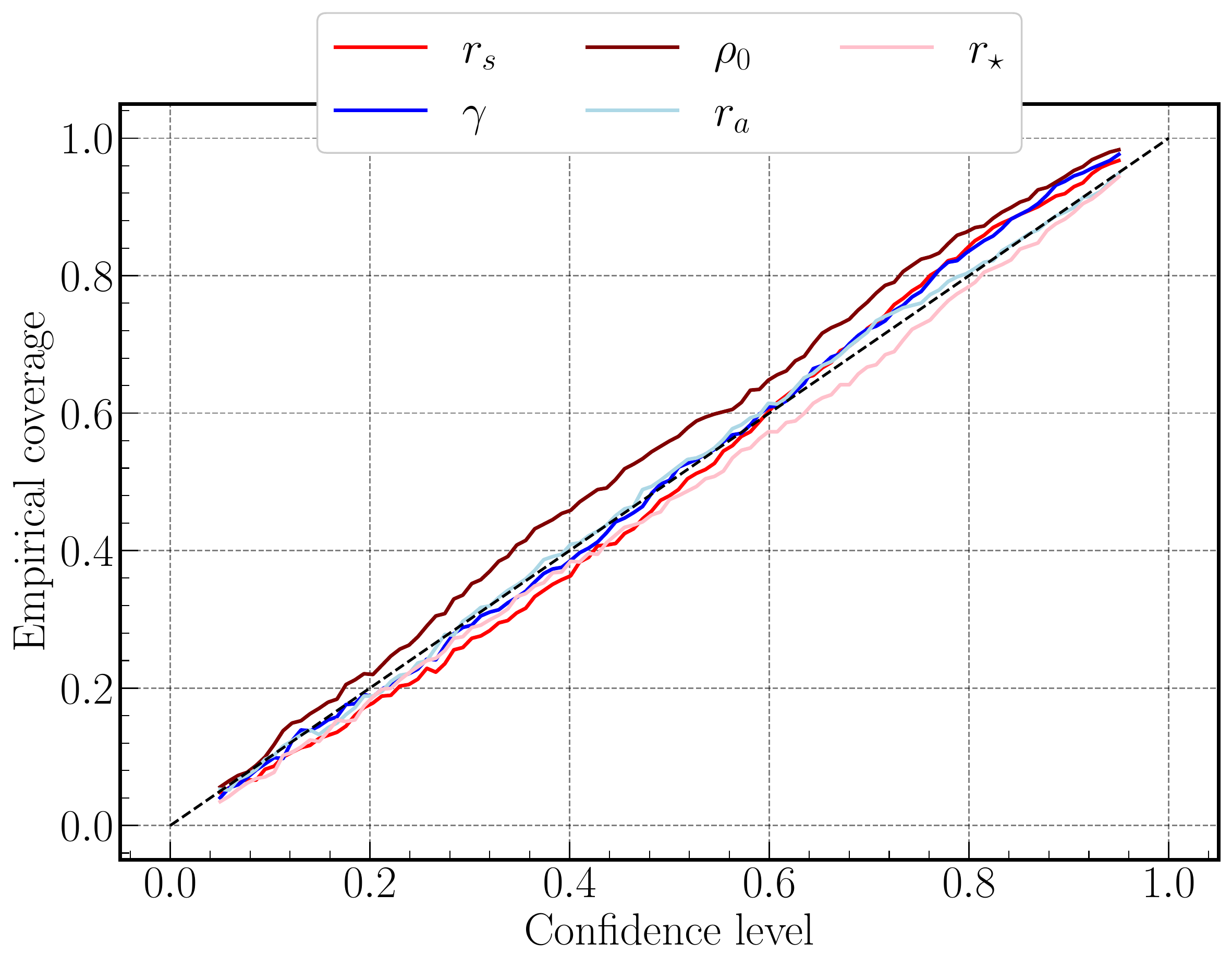}
    \caption{The expected coverage of the marginal DM and stellar parameters posteriors by the model presented in the main text ($\mu_\mathrm{stars}=100, \Delta v = 0.1 \, \si{km/s}$).
    If an estimator produces perfectly calibrated posteriors, then its empirical expected coverage probability is equal to the nominal expected coverage probability (dashed diagonal black line).
    The estimator is conservative (overconfident) if it produces an empirical expected coverage probability above (below) the diagonal line.
    }
    \label{fig:coverage}
\end{figure}

In Fig.~\ref{fig:coverage}, we plot the empirical expected coverage probability for the marginal posteriors as produced by the baseline model presented in the main text (i.e. $\mu_\mathrm{stars}=100$ and $\Delta v=0.1 \, \si{km/s}$) against the nominal confidence levels.
The dashed diagonal black line represents the nominal expected coverage probability. 
A conservative estimator will lie above the diagonal line, while an overconfident estimator will lie below it. 
In general, the posteriors produced by our model lie very close to the well-calibrated regime, although the posteriors for the stellar radius $r_\star$ are slightly overconfident.
We note that methodological improvement in calibration quality of posteriors produced using forward-modeling approaches is an ongoing, active area of research~\cite{dey2022calibrated}.

\subsection{Robustness to observational projection}
\label{app:projection}

Typically, only line-of-sight velocities and angular coordinates of tracer stars are observationally accessible, presenting the challenge of working with incomplete phase-space information when inferring the DM density profile. A test of our method is then its susceptibility to the specific direction from which a dwarf galaxy is viewed---its observational projection. Since different projections correspond to the same latent DM parameters, it is desirable for different projections to produce similar summary representations, and therefore similar posterior distributions. Since this is not explicitly baked into the network, approximate projective symmetry can be learned implicitly using the training sample.

\begin{figure}[!b]
    \centering
    \includegraphics[width=0.48\linewidth]{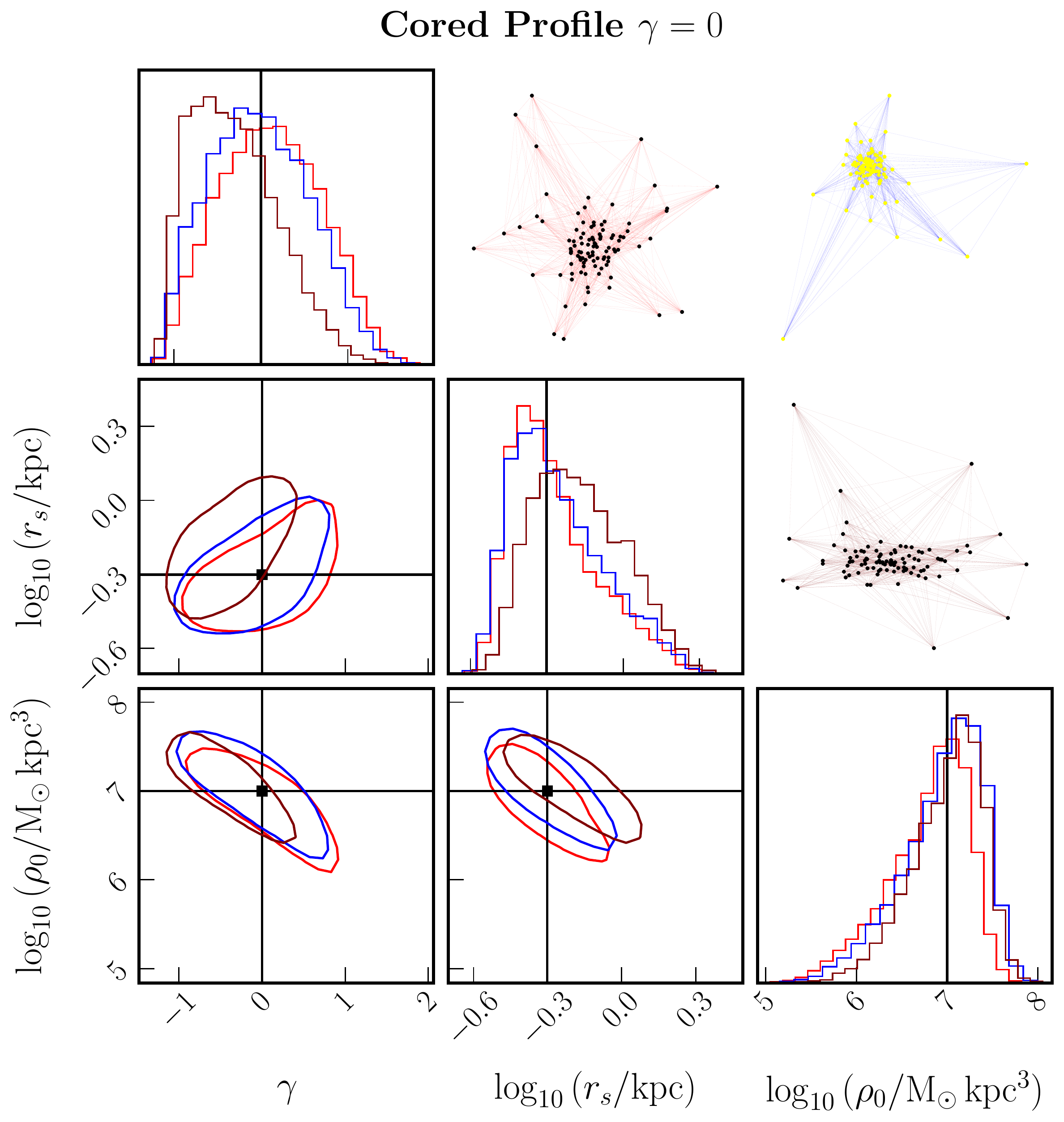}
    \includegraphics[width=0.48\linewidth]{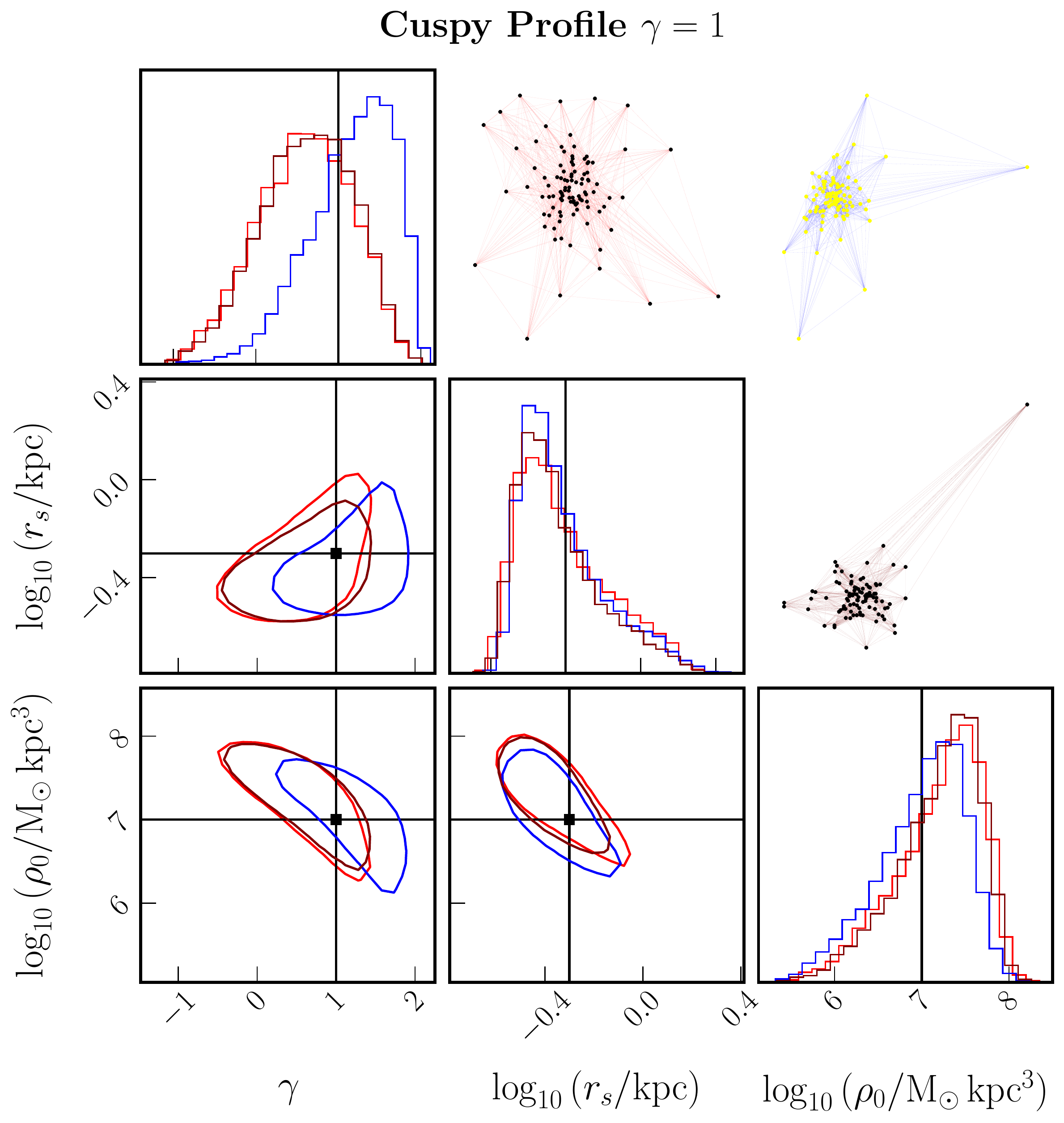}
    \caption{Example corner plots of the posterior DM parameters of two test galaxies, each with three orthogonal projections.
    The left (right) panel shows the posteriors for a galaxy with a cored (cuspy) DM profile.    
    The contour lines show the 68-\% containment region, with each color representing a different projection.
    For each projection, its graph representation is also shown (with the edge color matching the contour color) on the upper right corner.
    }
    \label{fig:projection}
\end{figure}

We take the galaxies with the same DM profiles presented in the main text in Fig.~\ref{fig:corner} and project them onto orthogonal planes. 
For each galaxy, we thus obtain 3 orthogonal projections (including the original projection).
In Fig.~\ref{fig:projection}, we show the 68\% contour line of the posterior DM parameters for three orthogonal projections of two galaxies (one with a cored DM profile and one with a cuspy DM profile) and the graph representations of the projections. 
For each projection, its graph representation matches the color of the contour line of the DM posteriors.
It can be seen that even though the graph representations may vary significantly between projections (e.g. the positions of each node may shift, forming new edge connections or breaking up old ones), the DM posteriors remain consistent. 
Note that we do not expect the DM posteriors to be identical, since the information given to the GNN is not the identical between projections. 

%% file: tables/priors.tex

\begin{table}[tb]
\small
\begin{center}
\begin{tabular}{C{4.7cm} | C{4cm} | C{3.5cm}}
\toprule
	 & \textbf{Parameter}  & \textbf{Prior distribution}\Tstrut\Bstrut	\\   
\Xhline{1\arrayrulewidth}
{\multirow{3}{*}{DM density profile}} & $\log_{10}(\rho_0 / (\si{M_\odot \, kpc^{-3}} )$ & $\mathcal U (5, 8)$ \Tstrut\Bstrut \\
& $\log_{10}(r_s / \si{kpc}))$ & $\mathcal U (-1, 0.7)$ \Tstrut\Bstrut  \\  
& $\gamma$ & $\mathcal U (-1, 2)$ \Tstrut\Bstrut  \\ 
\Xhline{1\arrayrulewidth}
{\multirow{2}{*}{Light profile (GNN + Flows)}} & $r_\star / r_s$ & $\mathcal U (0.2, 1)$ \Tstrut\Bstrut  \\ 
& $r_a / r_\star$ & $\mathcal U (0.5, 2)$ \Tstrut\Bstrut  \\ 
\Xhline{1\arrayrulewidth}
{\multirow{4}{*}{Light profile (Jeans analysis)}} & $\log_{10}(L / \mathrm L_\odot)$ & $\mathcal U (-2, 5)$ \Tstrut\Bstrut  \\ 
& $\log_{10}(r_\star / \si{kpc}) $ & $\mathcal U (-3, 3)$ \Tstrut\Bstrut  \\
& $r_a / r_\star$ & $\mathcal U (0.5, 2)$ \Tstrut\Bstrut  \\ 
& $\bar{v} / (\si{km \, s^{-1}}) $ & $\mathcal U (-100, 100)$ \Tstrut\Bstrut  \\
\botrule
\end{tabular}
\end{center}
\caption{Prior ranges for the DM and stellar parameter for the Jeans analysis and our method.}
\label{tab:priors}
\end{table}  

%% file: dsphs-gnn.bbl
\begin{thebibliography}{91}%
\makeatletter
\providecommand \@ifxundefined [1]{%
 \@ifx{#1\undefined}
}%
\providecommand \@ifnum [1]{%
 \ifnum #1\expandafter \@firstoftwo
 \else \expandafter \@secondoftwo
 \fi
}%
\providecommand \@ifx [1]{%
 \ifx #1\expandafter \@firstoftwo
 \else \expandafter \@secondoftwo
 \fi
}%
\providecommand \natexlab [1]{#1}%
\providecommand \enquote  [1]{``#1''}%
\providecommand \bibnamefont  [1]{#1}%
\providecommand \bibfnamefont [1]{#1}%
\providecommand \citenamefont [1]{#1}%
\providecommand \href@noop [0]{\@secondoftwo}%
\providecommand \href [0]{\begingroup \@sanitize@url \@href}%
\providecommand \@href[1]{\@@startlink{#1}\@@href}%
\providecommand \@@href[1]{\endgroup#1\@@endlink}%
\providecommand \@sanitize@url [0]{\catcode `\\12\catcode `\$12\catcode
  `\&12\catcode `\#12\catcode `\^12\catcode `\_12\catcode `\%12\relax}%
\providecommand \@@startlink[1]{}%
\providecommand \@@endlink[0]{}%
\providecommand \url  [0]{\begingroup\@sanitize@url \@url }%
\providecommand \@url [1]{\endgroup\@href {#1}{\urlprefix }}%
\providecommand \urlprefix  [0]{URL }%
\providecommand \Eprint [0]{\href }%
\providecommand \doibase [0]{https://doi.org/}%
\providecommand \selectlanguage [0]{\@gobble}%
\providecommand \bibinfo  [0]{\@secondoftwo}%
\providecommand \bibfield  [0]{\@secondoftwo}%
\providecommand \translation [1]{[#1]}%
\providecommand \BibitemOpen [0]{}%
\providecommand \bibitemStop [0]{}%
\providecommand \bibitemNoStop [0]{.\EOS\space}%
\providecommand \EOS [0]{\spacefactor3000\relax}%
\providecommand \BibitemShut  [1]{\csname bibitem#1\endcsname}%
\let\auto@bib@innerbib\@empty
\bibitem [{\citenamefont {{White}}\ and\ \citenamefont
  {{Rees}}(1978)}]{White&Rees_1978}%
  \BibitemOpen
  \bibfield  {author} {\bibinfo {author} {\bibfnamefont {S.~D.~M.}\
  \bibnamefont {{White}}}\ and\ \bibinfo {author} {\bibfnamefont {M.~J.}\
  \bibnamefont {{Rees}}},\ }\bibfield  {title} {\bibinfo {title} {{Core
  condensation in heavy halos: a two-stage theory for galaxy formation and
  clustering.}},\ }\href {https://doi.org/10.1093/mnras/183.3.341} {\bibfield
  {journal} {\bibinfo  {journal} {\mnras}\ }\textbf {\bibinfo {volume} {183}},\
  \bibinfo {pages} {341} (\bibinfo {year} {1978})}\BibitemShut {NoStop}%
\bibitem [{\citenamefont {{Pryor}}\ and\ \citenamefont
  {{Kormendy}}(1990)}]{1990AJ....100..127P}%
  \BibitemOpen
  \bibfield  {author} {\bibinfo {author} {\bibfnamefont {C.}~\bibnamefont
  {{Pryor}}}\ and\ \bibinfo {author} {\bibfnamefont {J.}~\bibnamefont
  {{Kormendy}}},\ }\bibfield  {title} {\bibinfo {title} {{The Dark Matter Halos
  of Draco and Ursa Minor}},\ }\href {https://doi.org/10.1086/115496}
  {\bibfield  {journal} {\bibinfo  {journal} {\aj}\ }\textbf {\bibinfo {volume}
  {100}},\ \bibinfo {pages} {127} (\bibinfo {year} {1990})}\BibitemShut
  {NoStop}%
\bibitem [{\citenamefont {Cirelli}\ \emph {et~al.}(2011)\citenamefont
  {Cirelli}, \citenamefont {Corcella}, \citenamefont {Hektor}, \citenamefont
  {Hütsi}, \citenamefont {Kadastik}, \citenamefont {Panci}, \citenamefont
  {Raidal}, \citenamefont {Sala},\ and\ \citenamefont
  {Strumia}}]{Cirelli_2011}%
  \BibitemOpen
  \bibfield  {author} {\bibinfo {author} {\bibfnamefont {M.}~\bibnamefont
  {Cirelli}}, \bibinfo {author} {\bibfnamefont {G.}~\bibnamefont {Corcella}},
  \bibinfo {author} {\bibfnamefont {A.}~\bibnamefont {Hektor}}, \bibinfo
  {author} {\bibfnamefont {G.}~\bibnamefont {Hütsi}}, \bibinfo {author}
  {\bibfnamefont {M.}~\bibnamefont {Kadastik}}, \bibinfo {author}
  {\bibfnamefont {P.}~\bibnamefont {Panci}}, \bibinfo {author} {\bibfnamefont
  {M.}~\bibnamefont {Raidal}}, \bibinfo {author} {\bibfnamefont
  {F.}~\bibnamefont {Sala}},\ and\ \bibinfo {author} {\bibfnamefont
  {A.}~\bibnamefont {Strumia}},\ }\bibfield  {title} {\bibinfo {title} {{PPPC}
  4 {DM} {ID}: a poor particle physicist cookbook for dark matter indirect
  detection},\ }\href {https://doi.org/10.1088/1475-7516/2011/03/051}
  {\bibfield  {journal} {\bibinfo  {journal} {Journal of Cosmology and
  Astroparticle Physics}\ }\textbf {\bibinfo {volume} {2011}}\bibinfo  {number}
  { (03)},\ \bibinfo {pages} {051}}\BibitemShut {NoStop}%
\bibitem [{\citenamefont {Ackermann}\ \emph {et~al.}(2015)\citenamefont
  {Ackermann} \emph {et~al.}}]{PhysRevLett.115.231301}%
  \BibitemOpen
\bibfield  {number} {  }\bibfield  {author} {\bibinfo {author} {\bibfnamefont
  {M.}~\bibnamefont {Ackermann}} \emph {et~al.} (\bibinfo {collaboration} {The
  Fermi-LAT Collaboration}),\ }\bibfield  {title} {\bibinfo {title} {Searching
  for dark matter annihilation from milky way dwarf spheroidal galaxies with
  six years of fermi large area telescope data},\ }\href
  {https://doi.org/10.1103/PhysRevLett.115.231301} {\bibfield  {journal}
  {\bibinfo  {journal} {Phys. Rev. Lett.}\ }\textbf {\bibinfo {volume} {115}},\
  \bibinfo {pages} {231301} (\bibinfo {year} {2015})}\BibitemShut {NoStop}%
\bibitem [{\citenamefont {Martinez}(2015)}]{10.1093/mnras/stv942}%
  \BibitemOpen
  \bibfield  {author} {\bibinfo {author} {\bibfnamefont {G.~D.}\ \bibnamefont
  {Martinez}},\ }\bibfield  {title} {\bibinfo {title} {{A robust determination
  of Milky Way satellite properties using hierarchical mass modelling}},\
  }\href {https://doi.org/10.1093/mnras/stv942} {\bibfield  {journal} {\bibinfo
   {journal} {Monthly Notices of the Royal Astronomical Society}\ }\textbf
  {\bibinfo {volume} {451}},\ \bibinfo {pages} {2524} (\bibinfo {year}
  {2015})},\ \Eprint
  {https://arxiv.org/abs/https://academic.oup.com/mnras/article-pdf/451/3/2524/4032739/stv942.pdf}
  {https://academic.oup.com/mnras/article-pdf/451/3/2524/4032739/stv942.pdf}
  \BibitemShut {NoStop}%
\bibitem [{\citenamefont {Geringer-Sameth}\ \emph
  {et~al.}(2015{\natexlab{a}})\citenamefont {Geringer-Sameth}, \citenamefont
  {Koushiappas},\ and\ \citenamefont {Walker}}]{Geringer_Sameth_2015}%
  \BibitemOpen
  \bibfield  {author} {\bibinfo {author} {\bibfnamefont {A.}~\bibnamefont
  {Geringer-Sameth}}, \bibinfo {author} {\bibfnamefont {S.~M.}\ \bibnamefont
  {Koushiappas}},\ and\ \bibinfo {author} {\bibfnamefont {M.}~\bibnamefont
  {Walker}},\ }\bibfield  {title} {\bibinfo {title} {Dwarf galaxy annihilation
  and decay emission profiles for dark matter experiments},\ }\href
  {https://doi.org/10.1088/0004-637x/801/2/74} {\bibfield  {journal} {\bibinfo
  {journal} {The Astrophysical Journal}\ }\textbf {\bibinfo {volume} {801}},\
  \bibinfo {pages} {74} (\bibinfo {year} {2015}{\natexlab{a}})}\BibitemShut
  {NoStop}%
\bibitem [{\citenamefont {Geringer-Sameth}\ and\ \citenamefont
  {Koushiappas}(2011)}]{PhysRevLett.107.241303}%
  \BibitemOpen
  \bibfield  {author} {\bibinfo {author} {\bibfnamefont {A.}~\bibnamefont
  {Geringer-Sameth}}\ and\ \bibinfo {author} {\bibfnamefont {S.~M.}\
  \bibnamefont {Koushiappas}},\ }\bibfield  {title} {\bibinfo {title}
  {Exclusion of canonical weakly interacting massive particles by joint
  analysis of milky way dwarf galaxies with data from the fermi gamma-ray space
  telescope},\ }\href {https://doi.org/10.1103/PhysRevLett.107.241303}
  {\bibfield  {journal} {\bibinfo  {journal} {Phys. Rev. Lett.}\ }\textbf
  {\bibinfo {volume} {107}},\ \bibinfo {pages} {241303} (\bibinfo {year}
  {2011})}\BibitemShut {NoStop}%
\bibitem [{\citenamefont {Ackermann}\ \emph {et~al.}(2011)\citenamefont
  {Ackermann} \emph {et~al.}}]{PhysRevLett.107.241302}%
  \BibitemOpen
  \bibfield  {author} {\bibinfo {author} {\bibfnamefont {M.}~\bibnamefont
  {Ackermann}} \emph {et~al.} (\bibinfo {collaboration} {The Fermi-LAT
  Collaboration}),\ }\bibfield  {title} {\bibinfo {title} {Constraining dark
  matter models from a combined analysis of milky way satellites with the fermi
  large area telescope},\ }\href
  {https://doi.org/10.1103/PhysRevLett.107.241302} {\bibfield  {journal}
  {\bibinfo  {journal} {Phys. Rev. Lett.}\ }\textbf {\bibinfo {volume} {107}},\
  \bibinfo {pages} {241302} (\bibinfo {year} {2011})}\BibitemShut {NoStop}%
\bibitem [{\citenamefont {Mazziotta}\ \emph {et~al.}(2012)\citenamefont
  {Mazziotta}, \citenamefont {Loparco}, \citenamefont {{de Palma}},\ and\
  \citenamefont {Giglietto}}]{MAZZIOTTA201226}%
  \BibitemOpen
  \bibfield  {author} {\bibinfo {author} {\bibfnamefont {M.}~\bibnamefont
  {Mazziotta}}, \bibinfo {author} {\bibfnamefont {F.}~\bibnamefont {Loparco}},
  \bibinfo {author} {\bibfnamefont {F.}~\bibnamefont {{de Palma}}},\ and\
  \bibinfo {author} {\bibfnamefont {N.}~\bibnamefont {Giglietto}},\ }\bibfield
  {title} {\bibinfo {title} {A model-independent analysis of the fermi large
  area telescope gamma-ray data from the milky way dwarf galaxies and halo to
  constrain dark matter scenarios},\ }\href
  {https://doi.org/https://doi.org/10.1016/j.astropartphys.2012.07.005}
  {\bibfield  {journal} {\bibinfo  {journal} {Astroparticle Physics}\ }\textbf
  {\bibinfo {volume} {37}},\ \bibinfo {pages} {26} (\bibinfo {year}
  {2012})}\BibitemShut {NoStop}%
\bibitem [{\citenamefont {Geringer-Sameth}\ \emph
  {et~al.}(2015{\natexlab{b}})\citenamefont {Geringer-Sameth}, \citenamefont
  {Koushiappas},\ and\ \citenamefont {Walker}}]{PhysRevD.91.083535}%
  \BibitemOpen
  \bibfield  {author} {\bibinfo {author} {\bibfnamefont {A.}~\bibnamefont
  {Geringer-Sameth}}, \bibinfo {author} {\bibfnamefont {S.~M.}\ \bibnamefont
  {Koushiappas}},\ and\ \bibinfo {author} {\bibfnamefont {M.~G.}\ \bibnamefont
  {Walker}},\ }\bibfield  {title} {\bibinfo {title} {Comprehensive search for
  dark matter annihilation in dwarf galaxies},\ }\href
  {https://doi.org/10.1103/PhysRevD.91.083535} {\bibfield  {journal} {\bibinfo
  {journal} {Phys. Rev. D}\ }\textbf {\bibinfo {volume} {91}},\ \bibinfo
  {pages} {083535} (\bibinfo {year} {2015}{\natexlab{b}})}\BibitemShut
  {NoStop}%
\bibitem [{\citenamefont {Ackermann}\ \emph {et~al.}(2014)\citenamefont
  {Ackermann} \emph {et~al.}}]{PhysRevD.89.042001}%
  \BibitemOpen
  \bibfield  {author} {\bibinfo {author} {\bibfnamefont {M.}~\bibnamefont
  {Ackermann}} \emph {et~al.} (\bibinfo {collaboration} {Fermi-LAT
  Collaboration}),\ }\bibfield  {title} {\bibinfo {title} {Dark matter
  constraints from observations of 25 milky way satellite galaxies with the
  fermi large area telescope},\ }\href
  {https://doi.org/10.1103/PhysRevD.89.042001} {\bibfield  {journal} {\bibinfo
  {journal} {Phys. Rev. D}\ }\textbf {\bibinfo {volume} {89}},\ \bibinfo
  {pages} {042001} (\bibinfo {year} {2014})}\BibitemShut {NoStop}%
\bibitem [{\citenamefont {Anderson}\ \emph {et~al.}(2015)\citenamefont
  {Anderson}, \citenamefont {Chiang}, \citenamefont {Cohen-Tanugi},
  \citenamefont {Conrad}, \citenamefont {Drlica-Wagner}, \citenamefont
  {Garde},\ and\ \citenamefont
  {Zimmer}}]{https://doi.org/10.48550/arxiv.1502.03081}%
  \BibitemOpen
  \bibfield  {author} {\bibinfo {author} {\bibfnamefont {B.}~\bibnamefont
  {Anderson}}, \bibinfo {author} {\bibfnamefont {J.}~\bibnamefont {Chiang}},
  \bibinfo {author} {\bibfnamefont {J.}~\bibnamefont {Cohen-Tanugi}}, \bibinfo
  {author} {\bibfnamefont {J.}~\bibnamefont {Conrad}}, \bibinfo {author}
  {\bibfnamefont {A.}~\bibnamefont {Drlica-Wagner}}, \bibinfo {author}
  {\bibfnamefont {M.~L.}\ \bibnamefont {Garde}},\ and\ \bibinfo {author}
  {\bibfnamefont {S.}~\bibnamefont {Zimmer}},\ }\href
  {https://doi.org/10.48550/ARXIV.1502.03081} {\bibinfo {title} {Using
  likelihood for combined data set analysis}} (\bibinfo {year}
  {2015})\BibitemShut {NoStop}%
\bibitem [{\citenamefont {{Navarro}}\ \emph {et~al.}(1996)\citenamefont
  {{Navarro}}, \citenamefont {{Eke}},\ and\ \citenamefont
  {{Frenk}}}]{1996MNRAS.283L..72N}%
  \BibitemOpen
  \bibfield  {author} {\bibinfo {author} {\bibfnamefont {J.~F.}\ \bibnamefont
  {{Navarro}}}, \bibinfo {author} {\bibfnamefont {V.~R.}\ \bibnamefont
  {{Eke}}},\ and\ \bibinfo {author} {\bibfnamefont {C.~S.}\ \bibnamefont
  {{Frenk}}},\ }\bibfield  {title} {\bibinfo {title} {{The cores of dwarf
  galaxy haloes}},\ }\href {https://doi.org/10.1093/mnras/283.3.L72} {\bibfield
   {journal} {\bibinfo  {journal} {MNRAS}\ }\textbf {\bibinfo {volume} {283}},\
  \bibinfo {pages} {L72} (\bibinfo {year} {1996})},\ \Eprint
  {https://arxiv.org/abs/astro-ph/9610187} {arXiv:astro-ph/9610187 [astro-ph]}
  \BibitemShut {NoStop}%
\bibitem [{\citenamefont {{Spekkens}}\ \emph {et~al.}(2005)\citenamefont
  {{Spekkens}}, \citenamefont {{Giovanelli}},\ and\ \citenamefont
  {{Haynes}}}]{2005AJ....129.2119S}%
  \BibitemOpen
  \bibfield  {author} {\bibinfo {author} {\bibfnamefont {K.}~\bibnamefont
  {{Spekkens}}}, \bibinfo {author} {\bibfnamefont {R.}~\bibnamefont
  {{Giovanelli}}},\ and\ \bibinfo {author} {\bibfnamefont {M.~P.}\ \bibnamefont
  {{Haynes}}},\ }\bibfield  {title} {\bibinfo {title} {{The Cusp/Core Problem
  in Galactic Halos: Long-Slit Spectra for a Large Dwarf Galaxy Sample}},\
  }\href {https://doi.org/10.1086/429592} {\bibfield  {journal} {\bibinfo
  {journal} {AJ}\ }\textbf {\bibinfo {volume} {129}},\ \bibinfo {pages} {2119}
  (\bibinfo {year} {2005})},\ \Eprint {https://arxiv.org/abs/astro-ph/0502166}
  {arXiv:astro-ph/0502166 [astro-ph]} \BibitemShut {NoStop}%
\bibitem [{\citenamefont {Navarro}\ \emph {et~al.}(1997)\citenamefont
  {Navarro}, \citenamefont {Frenk},\ and\ \citenamefont {White}}]{nfw1997}%
  \BibitemOpen
  \bibfield  {author} {\bibinfo {author} {\bibfnamefont {J.~F.}\ \bibnamefont
  {Navarro}}, \bibinfo {author} {\bibfnamefont {C.~S.}\ \bibnamefont {Frenk}},\
  and\ \bibinfo {author} {\bibfnamefont {S.~D.~M.}\ \bibnamefont {White}},\
  }\bibfield  {title} {\bibinfo {title} {A universal density profile from
  hierarchical clustering},\ }\href {https://doi.org/10.1086/304888} {\bibfield
   {journal} {\bibinfo  {journal} {ApJ}\ }\textbf {\bibinfo {volume} {490}},\
  \bibinfo {pages} {493} (\bibinfo {year} {1997})}\BibitemShut {NoStop}%
\bibitem [{\citenamefont {{Oh}}\ \emph {et~al.}(2011)\citenamefont {{Oh}},
  \citenamefont {{Brook}}, \citenamefont {{Governato}}, \citenamefont
  {{Brinks}}, \citenamefont {{Mayer}}, \citenamefont {{de Blok}}, \citenamefont
  {{Brooks}},\ and\ \citenamefont {{Walter}}}]{2011AJ....142...24O}%
  \BibitemOpen
  \bibfield  {author} {\bibinfo {author} {\bibfnamefont {S.-H.}\ \bibnamefont
  {{Oh}}}, \bibinfo {author} {\bibfnamefont {C.}~\bibnamefont {{Brook}}},
  \bibinfo {author} {\bibfnamefont {F.}~\bibnamefont {{Governato}}}, \bibinfo
  {author} {\bibfnamefont {E.}~\bibnamefont {{Brinks}}}, \bibinfo {author}
  {\bibfnamefont {L.}~\bibnamefont {{Mayer}}}, \bibinfo {author} {\bibfnamefont
  {W.~J.~G.}\ \bibnamefont {{de Blok}}}, \bibinfo {author} {\bibfnamefont
  {A.}~\bibnamefont {{Brooks}}},\ and\ \bibinfo {author} {\bibfnamefont
  {F.}~\bibnamefont {{Walter}}},\ }\bibfield  {title} {\bibinfo {title} {{The
  Central Slope of Dark Matter Cores in Dwarf Galaxies: Simulations versus
  THINGS}},\ }\href {https://doi.org/10.1088/0004-6256/142/1/24} {\bibfield
  {journal} {\bibinfo  {journal} {AJ}\ }\textbf {\bibinfo {volume} {142}},\
  \bibinfo {eid} {24} (\bibinfo {year} {2011})},\ \Eprint
  {https://arxiv.org/abs/1011.2777} {arXiv:1011.2777 [astro-ph.CO]}
  \BibitemShut {NoStop}%
\bibitem [{\citenamefont {{Oh}}\ \emph {et~al.}(2015)\citenamefont {{Oh}},
  \citenamefont {{Hunter}}, \citenamefont {{Brinks}}, \citenamefont
  {{Elmegreen}}, \citenamefont {{Schruba}}, \citenamefont {{Walter}},
  \citenamefont {{Rupen}}, \citenamefont {{Young}}, \citenamefont {{Simpson}},
  \citenamefont {{Johnson}}, \citenamefont {{Herrmann}}, \citenamefont
  {{Ficut-Vicas}}, \citenamefont {{Cigan}}, \citenamefont {{Heesen}},
  \citenamefont {{Ashley}},\ and\ \citenamefont
  {{Zhang}}}]{2015AJ....149..180O}%
  \BibitemOpen
  \bibfield  {author} {\bibinfo {author} {\bibfnamefont {S.-H.}\ \bibnamefont
  {{Oh}}}, \bibinfo {author} {\bibfnamefont {D.~A.}\ \bibnamefont {{Hunter}}},
  \bibinfo {author} {\bibfnamefont {E.}~\bibnamefont {{Brinks}}}, \bibinfo
  {author} {\bibfnamefont {B.~G.}\ \bibnamefont {{Elmegreen}}}, \bibinfo
  {author} {\bibfnamefont {A.}~\bibnamefont {{Schruba}}}, \bibinfo {author}
  {\bibfnamefont {F.}~\bibnamefont {{Walter}}}, \bibinfo {author}
  {\bibfnamefont {M.~P.}\ \bibnamefont {{Rupen}}}, \bibinfo {author}
  {\bibfnamefont {L.~M.}\ \bibnamefont {{Young}}}, \bibinfo {author}
  {\bibfnamefont {C.~E.}\ \bibnamefont {{Simpson}}}, \bibinfo {author}
  {\bibfnamefont {M.~C.}\ \bibnamefont {{Johnson}}}, \bibinfo {author}
  {\bibfnamefont {K.~A.}\ \bibnamefont {{Herrmann}}}, \bibinfo {author}
  {\bibfnamefont {D.}~\bibnamefont {{Ficut-Vicas}}}, \bibinfo {author}
  {\bibfnamefont {P.}~\bibnamefont {{Cigan}}}, \bibinfo {author} {\bibfnamefont
  {V.}~\bibnamefont {{Heesen}}}, \bibinfo {author} {\bibfnamefont
  {T.}~\bibnamefont {{Ashley}}},\ and\ \bibinfo {author} {\bibfnamefont
  {H.-X.}\ \bibnamefont {{Zhang}}},\ }\bibfield  {title} {\bibinfo {title}
  {{High-resolution Mass Models of Dwarf Galaxies from LITTLE THINGS}},\ }\href
  {https://doi.org/10.1088/0004-6256/149/6/180} {\bibfield  {journal} {\bibinfo
   {journal} {AJ}\ }\textbf {\bibinfo {volume} {149}},\ \bibinfo {eid} {180}
  (\bibinfo {year} {2015})},\ \Eprint {https://arxiv.org/abs/1502.01281}
  {arXiv:1502.01281 [astro-ph.GA]} \BibitemShut {NoStop}%
\bibitem [{\citenamefont {Bullock}\ and\ \citenamefont
  {Boylan-Kolchin}(2017)}]{Bullock:2017xww}%
  \BibitemOpen
  \bibfield  {author} {\bibinfo {author} {\bibfnamefont {J.~S.}\ \bibnamefont
  {Bullock}}\ and\ \bibinfo {author} {\bibfnamefont {M.}~\bibnamefont
  {Boylan-Kolchin}},\ }\bibfield  {title} {\bibinfo {title} {{Small-Scale
  Challenges to the $\Lambda$CDM Paradigm}},\ }\href
  {https://doi.org/10.1146/annurev-astro-091916-055313} {\bibfield  {journal}
  {\bibinfo  {journal} {Ann. Rev. Astron. Astrophys.}\ }\textbf {\bibinfo
  {volume} {55}},\ \bibinfo {pages} {343} (\bibinfo {year} {2017})},\ \Eprint
  {https://arxiv.org/abs/1707.04256} {arXiv:1707.04256 [astro-ph.CO]}
  \BibitemShut {NoStop}%
\bibitem [{\citenamefont {Navarro}\ \emph {et~al.}(1996)\citenamefont
  {Navarro}, \citenamefont {Eke},\ and\ \citenamefont
  {Frenk}}]{10.1093/mnras/283.3.L72}%
  \BibitemOpen
  \bibfield  {author} {\bibinfo {author} {\bibfnamefont {J.~F.}\ \bibnamefont
  {Navarro}}, \bibinfo {author} {\bibfnamefont {V.~R.}\ \bibnamefont {Eke}},\
  and\ \bibinfo {author} {\bibfnamefont {C.~S.}\ \bibnamefont {Frenk}},\
  }\bibfield  {title} {\bibinfo {title} {{The cores of dwarf galaxy haloes}},\
  }\href {https://doi.org/10.1093/mnras/283.3.L72} {\bibfield  {journal}
  {\bibinfo  {journal} {MNRAS}\ }\textbf {\bibinfo {volume} {283}},\ \bibinfo
  {pages} {L72} (\bibinfo {year} {1996})},\ \Eprint
  {https://arxiv.org/abs/https://academic.oup.com/mnras/article-pdf/283/3/L72/3305901/283-3-L72.pdf}
  {https://academic.oup.com/mnras/article-pdf/283/3/L72/3305901/283-3-L72.pdf}
  \BibitemShut {NoStop}%
\bibitem [{\citenamefont {{Read}}\ and\ \citenamefont
  {{Gilmore}}(2005)}]{2005MNRAS.356..107R}%
  \BibitemOpen
  \bibfield  {author} {\bibinfo {author} {\bibfnamefont {J.~I.}\ \bibnamefont
  {{Read}}}\ and\ \bibinfo {author} {\bibfnamefont {G.}~\bibnamefont
  {{Gilmore}}},\ }\bibfield  {title} {\bibinfo {title} {{Mass loss from dwarf
  spheroidal galaxies: the origins of shallow dark matter cores and exponential
  surface brightness profiles}},\ }\href
  {https://doi.org/10.1111/j.1365-2966.2004.08424.x} {\bibfield  {journal}
  {\bibinfo  {journal} {MNRAS}\ }\textbf {\bibinfo {volume} {356}},\ \bibinfo
  {pages} {107} (\bibinfo {year} {2005})},\ \Eprint
  {https://arxiv.org/abs/astro-ph/0409565} {arXiv:astro-ph/0409565 [astro-ph]}
  \BibitemShut {NoStop}%
\bibitem [{\citenamefont {{Mashchenko}}\ \emph {et~al.}(2006)\citenamefont
  {{Mashchenko}}, \citenamefont {{Couchman}},\ and\ \citenamefont
  {{Wadsley}}}]{2006Natur.442..539M}%
  \BibitemOpen
  \bibfield  {author} {\bibinfo {author} {\bibfnamefont {S.}~\bibnamefont
  {{Mashchenko}}}, \bibinfo {author} {\bibfnamefont {H.~M.~P.}\ \bibnamefont
  {{Couchman}}},\ and\ \bibinfo {author} {\bibfnamefont {J.}~\bibnamefont
  {{Wadsley}}},\ }\bibfield  {title} {\bibinfo {title} {{The removal of cusps
  from galaxy centres by stellar feedback in the early Universe}},\ }\href
  {https://doi.org/10.1038/nature04944} {\bibfield  {journal} {\bibinfo
  {journal} {Nature}\ }\textbf {\bibinfo {volume} {442}},\ \bibinfo {pages}
  {539} (\bibinfo {year} {2006})},\ \Eprint
  {https://arxiv.org/abs/astro-ph/0605672} {arXiv:astro-ph/0605672 [astro-ph]}
  \BibitemShut {NoStop}%
\bibitem [{\citenamefont {{Pontzen}}\ and\ \citenamefont
  {{Governato}}(2012)}]{2012MNRAS.421.3464P}%
  \BibitemOpen
  \bibfield  {author} {\bibinfo {author} {\bibfnamefont {A.}~\bibnamefont
  {{Pontzen}}}\ and\ \bibinfo {author} {\bibfnamefont {F.}~\bibnamefont
  {{Governato}}},\ }\bibfield  {title} {\bibinfo {title} {{How supernova
  feedback turns dark matter cusps into cores}},\ }\href
  {https://doi.org/10.1111/j.1365-2966.2012.20571.x} {\bibfield  {journal}
  {\bibinfo  {journal} {MNRAS}\ }\textbf {\bibinfo {volume} {421}},\ \bibinfo
  {pages} {3464} (\bibinfo {year} {2012})},\ \Eprint
  {https://arxiv.org/abs/1106.0499} {arXiv:1106.0499 [astro-ph.CO]}
  \BibitemShut {NoStop}%
\bibitem [{\citenamefont {Lovell}\ \emph {et~al.}(2012)\citenamefont {Lovell},
  \citenamefont {Eke}, \citenamefont {Frenk}, \citenamefont {Gao},
  \citenamefont {Jenkins}, \citenamefont {Theuns}, \citenamefont {Wang},
  \citenamefont {White}, \citenamefont {Boyarsky},\ and\ \citenamefont
  {Ruchayskiy}}]{10.1111/j.1365-2966.2011.20200.x}%
  \BibitemOpen
  \bibfield  {author} {\bibinfo {author} {\bibfnamefont {M.~R.}\ \bibnamefont
  {Lovell}}, \bibinfo {author} {\bibfnamefont {V.}~\bibnamefont {Eke}},
  \bibinfo {author} {\bibfnamefont {C.~S.}\ \bibnamefont {Frenk}}, \bibinfo
  {author} {\bibfnamefont {L.}~\bibnamefont {Gao}}, \bibinfo {author}
  {\bibfnamefont {A.}~\bibnamefont {Jenkins}}, \bibinfo {author} {\bibfnamefont
  {T.}~\bibnamefont {Theuns}}, \bibinfo {author} {\bibfnamefont
  {J.}~\bibnamefont {Wang}}, \bibinfo {author} {\bibfnamefont {S.~D.~M.}\
  \bibnamefont {White}}, \bibinfo {author} {\bibfnamefont {A.}~\bibnamefont
  {Boyarsky}},\ and\ \bibinfo {author} {\bibfnamefont {O.}~\bibnamefont
  {Ruchayskiy}},\ }\bibfield  {title} {\bibinfo {title} {{The haloes of bright
  satellite galaxies in a warm dark matter universe}},\ }\href
  {https://doi.org/10.1111/j.1365-2966.2011.20200.x} {\bibfield  {journal}
  {\bibinfo  {journal} {MNRAS}\ }\textbf {\bibinfo {volume} {420}},\ \bibinfo
  {pages} {2318} (\bibinfo {year} {2012})},\ \Eprint
  {https://arxiv.org/abs/https://academic.oup.com/mnras/article-pdf/420/3/2318/3020178/mnras0420-2318.pdf}
  {https://academic.oup.com/mnras/article-pdf/420/3/2318/3020178/mnras0420-2318.pdf}
  \BibitemShut {NoStop}%
\bibitem [{\citenamefont {Elbert}\ \emph {et~al.}(2015)\citenamefont {Elbert},
  \citenamefont {Bullock}, \citenamefont {Garrison-Kimmel}, \citenamefont
  {Rocha}, \citenamefont {O{\~n}orbe},\ and\ \citenamefont
  {Peter}}]{10.1093/mnras/stv1470}%
  \BibitemOpen
  \bibfield  {author} {\bibinfo {author} {\bibfnamefont {O.~D.}\ \bibnamefont
  {Elbert}}, \bibinfo {author} {\bibfnamefont {J.~S.}\ \bibnamefont {Bullock}},
  \bibinfo {author} {\bibfnamefont {S.}~\bibnamefont {Garrison-Kimmel}},
  \bibinfo {author} {\bibfnamefont {M.}~\bibnamefont {Rocha}}, \bibinfo
  {author} {\bibfnamefont {J.}~\bibnamefont {O{\~n}orbe}},\ and\ \bibinfo
  {author} {\bibfnamefont {A.~H.~G.}\ \bibnamefont {Peter}},\ }\bibfield
  {title} {\bibinfo {title} {{Core formation in dwarf haloes with
  self-interacting dark matter: no fine-tuning necessary}},\ }\href
  {https://doi.org/10.1093/mnras/stv1470} {\bibfield  {journal} {\bibinfo
  {journal} {MNRAS}\ }\textbf {\bibinfo {volume} {453}},\ \bibinfo {pages} {29}
  (\bibinfo {year} {2015})},\ \Eprint
  {https://arxiv.org/abs/https://academic.oup.com/mnras/article-pdf/453/1/29/4900820/stv1470.pdf}
  {https://academic.oup.com/mnras/article-pdf/453/1/29/4900820/stv1470.pdf}
  \BibitemShut {NoStop}%
\bibitem [{\citenamefont {{Burger}}\ \emph {et~al.}(2022)\citenamefont
  {{Burger}}, \citenamefont {{Zavala}}, \citenamefont {{Sales}}, \citenamefont
  {{Vogelsberger}}, \citenamefont {{Marinacci}},\ and\ \citenamefont
  {{Torrey}}}]{2022MNRAS.513.3458B}%
  \BibitemOpen
  \bibfield  {author} {\bibinfo {author} {\bibfnamefont {J.~D.}\ \bibnamefont
  {{Burger}}}, \bibinfo {author} {\bibfnamefont {J.}~\bibnamefont {{Zavala}}},
  \bibinfo {author} {\bibfnamefont {L.~V.}\ \bibnamefont {{Sales}}}, \bibinfo
  {author} {\bibfnamefont {M.}~\bibnamefont {{Vogelsberger}}}, \bibinfo
  {author} {\bibfnamefont {F.}~\bibnamefont {{Marinacci}}},\ and\ \bibinfo
  {author} {\bibfnamefont {P.}~\bibnamefont {{Torrey}}},\ }\bibfield  {title}
  {\bibinfo {title} {{Degeneracies between self-interacting dark matter and
  supernova feedback as cusp-core transformation mechanisms}},\ }\href
  {https://doi.org/10.1093/mnras/stac994} {\bibfield  {journal} {\bibinfo
  {journal} {\mnras}\ }\textbf {\bibinfo {volume} {513}},\ \bibinfo {pages}
  {3458} (\bibinfo {year} {2022})},\ \Eprint {https://arxiv.org/abs/2108.07358}
  {arXiv:2108.07358 [astro-ph.GA]} \BibitemShut {NoStop}%
\bibitem [{\citenamefont {{Jeans}}(1915)}]{1915MNRAS..76...70J}%
  \BibitemOpen
  \bibfield  {author} {\bibinfo {author} {\bibfnamefont {J.~H.}\ \bibnamefont
  {{Jeans}}},\ }\bibfield  {title} {\bibinfo {title} {{On the theory of
  star-streaming and the structure of the universe}},\ }\href
  {https://doi.org/10.1093/mnras/76.2.70} {\bibfield  {journal} {\bibinfo
  {journal} {MNRAS}\ }\textbf {\bibinfo {volume} {76}},\ \bibinfo {pages} {70}
  (\bibinfo {year} {1915})}\BibitemShut {NoStop}%
\bibitem [{\citenamefont {{Bonnivard}}\ \emph {et~al.}(2015)\citenamefont
  {{Bonnivard}}, \citenamefont {{Combet}}, \citenamefont {{Maurin}},\ and\
  \citenamefont {{Walker}}}]{2015MNRAS.446.3002B}%
  \BibitemOpen
  \bibfield  {author} {\bibinfo {author} {\bibfnamefont {V.}~\bibnamefont
  {{Bonnivard}}}, \bibinfo {author} {\bibfnamefont {C.}~\bibnamefont
  {{Combet}}}, \bibinfo {author} {\bibfnamefont {D.}~\bibnamefont {{Maurin}}},\
  and\ \bibinfo {author} {\bibfnamefont {M.~G.}\ \bibnamefont {{Walker}}},\
  }\bibfield  {title} {\bibinfo {title} {{Spherical Jeans analysis for dark
  matter indirect detection in dwarf spheroidal galaxies - impact of physical
  parameters and triaxiality}},\ }\href {https://doi.org/10.1093/mnras/stu2296}
  {\bibfield  {journal} {\bibinfo  {journal} {MNRAS}\ }\textbf {\bibinfo
  {volume} {446}},\ \bibinfo {pages} {3002} (\bibinfo {year} {2015})},\ \Eprint
  {https://arxiv.org/abs/1407.7822} {arXiv:1407.7822 [astro-ph.HE]}
  \BibitemShut {NoStop}%
\bibitem [{\citenamefont {{El-Badry}}\ \emph {et~al.}(2017)\citenamefont
  {{El-Badry}}, \citenamefont {{Wetzel}}, \citenamefont {{Geha}}, \citenamefont
  {{Quataert}}, \citenamefont {{Hopkins}}, \citenamefont {{Kere{\v{s}}}},
  \citenamefont {{Chan}},\ and\ \citenamefont
  {{Faucher-Gigu{\`e}re}}}]{2017ApJ...835..193E}%
  \BibitemOpen
  \bibfield  {author} {\bibinfo {author} {\bibfnamefont {K.}~\bibnamefont
  {{El-Badry}}}, \bibinfo {author} {\bibfnamefont {A.~R.}\ \bibnamefont
  {{Wetzel}}}, \bibinfo {author} {\bibfnamefont {M.}~\bibnamefont {{Geha}}},
  \bibinfo {author} {\bibfnamefont {E.}~\bibnamefont {{Quataert}}}, \bibinfo
  {author} {\bibfnamefont {P.~F.}\ \bibnamefont {{Hopkins}}}, \bibinfo {author}
  {\bibfnamefont {D.}~\bibnamefont {{Kere{\v{s}}}}}, \bibinfo {author}
  {\bibfnamefont {T.~K.}\ \bibnamefont {{Chan}}},\ and\ \bibinfo {author}
  {\bibfnamefont {C.-A.}\ \bibnamefont {{Faucher-Gigu{\`e}re}}},\ }\bibfield
  {title} {\bibinfo {title} {{When the Jeans Do Not Fit: How Stellar Feedback
  Drives Stellar Kinematics and Complicates Dynamical Modeling in Low-mass
  Galaxies}},\ }\href {https://doi.org/10.3847/1538-4357/835/2/193} {\bibfield
  {journal} {\bibinfo  {journal} {ApJ}\ }\textbf {\bibinfo {volume} {835}},\
  \bibinfo {eid} {193} (\bibinfo {year} {2017})},\ \Eprint
  {https://arxiv.org/abs/1610.04232} {arXiv:1610.04232 [astro-ph.GA]}
  \BibitemShut {NoStop}%
\bibitem [{\citenamefont {Genina}\ \emph {et~al.}(2020)\citenamefont {Genina},
  \citenamefont {Read}, \citenamefont {Frenk}, \citenamefont {Cole},
  \citenamefont {Ben{\'\i}tez-Llambay}, \citenamefont {Ludlow}, \citenamefont
  {Navarro}, \citenamefont {Oman},\ and\ \citenamefont {Robertson}}]{read2020}%
  \BibitemOpen
  \bibfield  {author} {\bibinfo {author} {\bibfnamefont {A.}~\bibnamefont
  {Genina}}, \bibinfo {author} {\bibfnamefont {J.~I.}\ \bibnamefont {Read}},
  \bibinfo {author} {\bibfnamefont {C.~S.}\ \bibnamefont {Frenk}}, \bibinfo
  {author} {\bibfnamefont {S.}~\bibnamefont {Cole}}, \bibinfo {author}
  {\bibfnamefont {A.}~\bibnamefont {Ben{\'\i}tez-Llambay}}, \bibinfo {author}
  {\bibfnamefont {A.~D.}\ \bibnamefont {Ludlow}}, \bibinfo {author}
  {\bibfnamefont {J.~F.}\ \bibnamefont {Navarro}}, \bibinfo {author}
  {\bibfnamefont {K.~A.}\ \bibnamefont {Oman}},\ and\ \bibinfo {author}
  {\bibfnamefont {A.}~\bibnamefont {Robertson}},\ }\bibfield  {title} {\bibinfo
  {title} {{To $\beta$ or not to $\beta$: can higher order Jeans analysis break
  the mass--anisotropy degeneracy in simulated dwarfs?}},\ }\href
  {https://doi.org/10.1093/mnras/staa2352} {\bibfield  {journal} {\bibinfo
  {journal} {Monthly Notices of the Royal Astronomical Society}\ }\textbf
  {\bibinfo {volume} {498}},\ \bibinfo {pages} {144} (\bibinfo {year}
  {2020})},\ \Eprint
  {https://arxiv.org/abs/https://academic.oup.com/mnras/article-pdf/498/1/144/33703538/staa2352.pdf}
  {https://academic.oup.com/mnras/article-pdf/498/1/144/33703538/staa2352.pdf}
  \BibitemShut {NoStop}%
\bibitem [{\citenamefont {Chang}\ and\ \citenamefont
  {Necib}(2021)}]{chang2021}%
  \BibitemOpen
  \bibfield  {author} {\bibinfo {author} {\bibfnamefont {L.~J.}\ \bibnamefont
  {Chang}}\ and\ \bibinfo {author} {\bibfnamefont {L.}~\bibnamefont {Necib}},\
  }\bibfield  {title} {\bibinfo {title} {{Dark matter density profiles in dwarf
  galaxies: linking Jeans modelling systematics and observation}},\ }\href
  {https://doi.org/10.1093/mnras/stab2440} {\bibfield  {journal} {\bibinfo
  {journal} {Monthly Notices of the Royal Astronomical Society}\ }\textbf
  {\bibinfo {volume} {507}},\ \bibinfo {pages} {4715} (\bibinfo {year}
  {2021})},\ \Eprint
  {https://arxiv.org/abs/https://academic.oup.com/mnras/article-pdf/507/4/4715/40384382/stab2440.pdf}
  {https://academic.oup.com/mnras/article-pdf/507/4/4715/40384382/stab2440.pdf}
  \BibitemShut {NoStop}%
\bibitem [{\citenamefont {{Helmi}}(2020)}]{2020ARA&A..58..205H}%
  \BibitemOpen
  \bibfield  {author} {\bibinfo {author} {\bibfnamefont {A.}~\bibnamefont
  {{Helmi}}},\ }\bibfield  {title} {\bibinfo {title} {{Streams, Substructures,
  and the Early History of the Milky Way}},\ }\href
  {https://doi.org/10.1146/annurev-astro-032620-021917} {\bibfield  {journal}
  {\bibinfo  {journal} {\araa}\ }\textbf {\bibinfo {volume} {58}},\ \bibinfo
  {pages} {205} (\bibinfo {year} {2020})},\ \Eprint
  {https://arxiv.org/abs/2002.04340} {arXiv:2002.04340 [astro-ph.GA]}
  \BibitemShut {NoStop}%
\bibitem [{\citenamefont {{Read}}\ \emph {et~al.}(2021)\citenamefont {{Read}},
  \citenamefont {{Mamon}}, \citenamefont {{Vasiliev}}, \citenamefont
  {{Watkins}}, \citenamefont {{Walker}}, \citenamefont {{Pe{\~n}arrubia}},
  \citenamefont {{Wilkinson}}, \citenamefont {{Dehnen}},\ and\ \citenamefont
  {{Das}}}]{2021MNRAS.501..978R}%
  \BibitemOpen
  \bibfield  {author} {\bibinfo {author} {\bibfnamefont {J.~I.}\ \bibnamefont
  {{Read}}}, \bibinfo {author} {\bibfnamefont {G.~A.}\ \bibnamefont {{Mamon}}},
  \bibinfo {author} {\bibfnamefont {E.}~\bibnamefont {{Vasiliev}}}, \bibinfo
  {author} {\bibfnamefont {L.~L.}\ \bibnamefont {{Watkins}}}, \bibinfo {author}
  {\bibfnamefont {M.~G.}\ \bibnamefont {{Walker}}}, \bibinfo {author}
  {\bibfnamefont {J.}~\bibnamefont {{Pe{\~n}arrubia}}}, \bibinfo {author}
  {\bibfnamefont {M.}~\bibnamefont {{Wilkinson}}}, \bibinfo {author}
  {\bibfnamefont {W.}~\bibnamefont {{Dehnen}}},\ and\ \bibinfo {author}
  {\bibfnamefont {P.}~\bibnamefont {{Das}}},\ }\bibfield  {title} {\bibinfo
  {title} {{Breaking beta: a comparison of mass modelling methods for spherical
  systems}},\ }\href {https://doi.org/10.1093/mnras/staa3663} {\bibfield
  {journal} {\bibinfo  {journal} {\mnras}\ }\textbf {\bibinfo {volume} {501}},\
  \bibinfo {pages} {978} (\bibinfo {year} {2021})},\ \Eprint
  {https://arxiv.org/abs/2011.09493} {arXiv:2011.09493 [astro-ph.GA]}
  \BibitemShut {NoStop}%
\bibitem [{\citenamefont {{Read}}\ and\ \citenamefont
  {{Steger}}(2017)}]{2017MNRAS.471.4541R}%
  \BibitemOpen
  \bibfield  {author} {\bibinfo {author} {\bibfnamefont {J.~I.}\ \bibnamefont
  {{Read}}}\ and\ \bibinfo {author} {\bibfnamefont {P.}~\bibnamefont
  {{Steger}}},\ }\bibfield  {title} {\bibinfo {title} {{How to break the
  density-anisotropy degeneracy in spherical stellar systems}},\ }\href
  {https://doi.org/10.1093/mnras/stx1798} {\bibfield  {journal} {\bibinfo
  {journal} {MNRAS}\ }\textbf {\bibinfo {volume} {471}},\ \bibinfo {pages}
  {4541} (\bibinfo {year} {2017})}\BibitemShut {NoStop}%
\bibitem [{\citenamefont {Read}\ \emph {et~al.}(2018)\citenamefont {Read},
  \citenamefont {Walker},\ and\ \citenamefont {Steger}}]{Read:2018pft}%
  \BibitemOpen
  \bibfield  {author} {\bibinfo {author} {\bibfnamefont {J.}~\bibnamefont
  {Read}}, \bibinfo {author} {\bibfnamefont {M.}~\bibnamefont {Walker}},\ and\
  \bibinfo {author} {\bibfnamefont {P.}~\bibnamefont {Steger}},\ }\bibfield
  {title} {\bibinfo {title} {{The case for a cold dark matter cusp in Draco}},\
  }\href {https://doi.org/10.1093/mnras/sty2286} {\bibfield  {journal}
  {\bibinfo  {journal} {MNRAS}\ }\textbf {\bibinfo {volume} {481}},\ \bibinfo
  {pages} {860} (\bibinfo {year} {2018})},\ \Eprint
  {https://arxiv.org/abs/1805.06934} {arXiv:1805.06934 [astro-ph.GA]}
  \BibitemShut {NoStop}%
\bibitem [{\citenamefont {{Walker}}\ and\ \citenamefont
  {{Pe{\~n}arrubia}}(2011)}]{2011ApJ...742...20W}%
  \BibitemOpen
  \bibfield  {author} {\bibinfo {author} {\bibfnamefont {M.~G.}\ \bibnamefont
  {{Walker}}}\ and\ \bibinfo {author} {\bibfnamefont {J.}~\bibnamefont
  {{Pe{\~n}arrubia}}},\ }\bibfield  {title} {\bibinfo {title} {{A Method for
  Measuring (Slopes of) the Mass Profiles of Dwarf Spheroidal Galaxies}},\
  }\href {https://doi.org/10.1088/0004-637X/742/1/20} {\bibfield  {journal}
  {\bibinfo  {journal} {ApJ}\ }\textbf {\bibinfo {volume} {742}},\ \bibinfo
  {eid} {20} (\bibinfo {year} {2011})},\ \Eprint
  {https://arxiv.org/abs/1108.2404} {arXiv:1108.2404} \BibitemShut {NoStop}%
\bibitem [{\citenamefont {{Amorisco}}\ and\ \citenamefont
  {{Evans}}(2012)}]{2012MNRAS.419..184A}%
  \BibitemOpen
  \bibfield  {author} {\bibinfo {author} {\bibfnamefont {N.~C.}\ \bibnamefont
  {{Amorisco}}}\ and\ \bibinfo {author} {\bibfnamefont {N.~W.}\ \bibnamefont
  {{Evans}}},\ }\bibfield  {title} {\bibinfo {title} {{Dark matter cores and
  cusps: the case of multiple stellar populations in dwarf spheroidals}},\
  }\href {https://doi.org/10.1111/j.1365-2966.2011.19684.x} {\bibfield
  {journal} {\bibinfo  {journal} {MNRAS}\ }\textbf {\bibinfo {volume} {419}},\
  \bibinfo {pages} {184} (\bibinfo {year} {2012})},\ \Eprint
  {https://arxiv.org/abs/1106.1062} {arXiv:1106.1062 [astro-ph.CO]}
  \BibitemShut {NoStop}%
\bibitem [{\citenamefont {{Zhu}}\ \emph {et~al.}(2016)\citenamefont {{Zhu}},
  \citenamefont {{van de Ven}}, \citenamefont {{Watkins}},\ and\ \citenamefont
  {{Posti}}}]{2016MNRAS.463.1117Z}%
  \BibitemOpen
  \bibfield  {author} {\bibinfo {author} {\bibfnamefont {L.}~\bibnamefont
  {{Zhu}}}, \bibinfo {author} {\bibfnamefont {G.}~\bibnamefont {{van de Ven}}},
  \bibinfo {author} {\bibfnamefont {L.~L.}\ \bibnamefont {{Watkins}}},\ and\
  \bibinfo {author} {\bibfnamefont {L.}~\bibnamefont {{Posti}}},\ }\bibfield
  {title} {\bibinfo {title} {{A discrete chemo-dynamical model of the dwarf
  spheroidal galaxy Sculptor: mass profile, velocity anisotropy and internal
  rotation}},\ }\href {https://doi.org/10.1093/mnras/stw2081} {\bibfield
  {journal} {\bibinfo  {journal} {\mnras}\ }\textbf {\bibinfo {volume} {463}},\
  \bibinfo {pages} {1117} (\bibinfo {year} {2016})},\ \Eprint
  {https://arxiv.org/abs/1608.08239} {arXiv:1608.08239 [astro-ph.GA]}
  \BibitemShut {NoStop}%
\bibitem [{\citenamefont {Strigari}\ \emph {et~al.}(2007)\citenamefont
  {Strigari}, \citenamefont {Bullock},\ and\ \citenamefont
  {Kaplinghat}}]{Strigari:2007vn}%
  \BibitemOpen
  \bibfield  {author} {\bibinfo {author} {\bibfnamefont {L.~E.}\ \bibnamefont
  {Strigari}}, \bibinfo {author} {\bibfnamefont {J.~S.}\ \bibnamefont
  {Bullock}},\ and\ \bibinfo {author} {\bibfnamefont {M.}~\bibnamefont
  {Kaplinghat}},\ }\bibfield  {title} {\bibinfo {title} {{Determining the
  Nature of Dark Matter with Astrometry}},\ }\href
  {https://doi.org/10.1086/512976} {\bibfield  {journal} {\bibinfo  {journal}
  {ApJ Lett.}\ }\textbf {\bibinfo {volume} {657}},\ \bibinfo {pages} {L1}
  (\bibinfo {year} {2007})},\ \Eprint {https://arxiv.org/abs/astro-ph/0701581}
  {arXiv:astro-ph/0701581} \BibitemShut {NoStop}%
\bibitem [{\citenamefont {Strigari}\ \emph {et~al.}(2018)\citenamefont
  {Strigari}, \citenamefont {Frenk},\ and\ \citenamefont
  {White}}]{Strigari:2018bcn}%
  \BibitemOpen
  \bibfield  {author} {\bibinfo {author} {\bibfnamefont {L.~E.}\ \bibnamefont
  {Strigari}}, \bibinfo {author} {\bibfnamefont {C.~S.}\ \bibnamefont
  {Frenk}},\ and\ \bibinfo {author} {\bibfnamefont {S.~D.}\ \bibnamefont
  {White}},\ }\bibfield  {title} {\bibinfo {title} {{Dynamical constraints on
  the dark matter distribution of the Sculptor dwarf spheroidal from stellar
  proper motions}},\ }\href {https://doi.org/10.3847/1538-4357/aac2d3}
  {\bibfield  {journal} {\bibinfo  {journal} {ApJ}\ }\textbf {\bibinfo {volume}
  {860}},\ \bibinfo {pages} {56} (\bibinfo {year} {2018})},\ \Eprint
  {https://arxiv.org/abs/1801.07343} {arXiv:1801.07343 [astro-ph.GA]}
  \BibitemShut {NoStop}%
\bibitem [{\citenamefont {{Mamon}}\ \emph {et~al.}(2013)\citenamefont
  {{Mamon}}, \citenamefont {{Biviano}},\ and\ \citenamefont
  {{Bou{\'e}}}}]{2013MNRAS.429.3079M}%
  \BibitemOpen
  \bibfield  {author} {\bibinfo {author} {\bibfnamefont {G.~A.}\ \bibnamefont
  {{Mamon}}}, \bibinfo {author} {\bibfnamefont {A.}~\bibnamefont {{Biviano}}},\
  and\ \bibinfo {author} {\bibfnamefont {G.}~\bibnamefont {{Bou{\'e}}}},\
  }\bibfield  {title} {\bibinfo {title} {{MAMPOSSt: Modelling Anisotropy and
  Mass Profiles of Observed Spherical Systems - I. Gaussian 3D velocities}},\
  }\href {https://doi.org/10.1093/mnras/sts565} {\bibfield  {journal} {\bibinfo
   {journal} {MNRAS}\ }\textbf {\bibinfo {volume} {429}},\ \bibinfo {pages}
  {3079} (\bibinfo {year} {2013})},\ \Eprint {https://arxiv.org/abs/1212.1455}
  {arXiv:1212.1455 [astro-ph.CO]} \BibitemShut {NoStop}%
\bibitem [{\citenamefont {Gelman}\ \emph {et~al.}(2003)\citenamefont {Gelman},
  \citenamefont {Carlin}, \citenamefont {Stern},\ and\ \citenamefont
  {Rubin}}]{gelman2003bayesian}%
  \BibitemOpen
  \bibfield  {author} {\bibinfo {author} {\bibfnamefont {A.}~\bibnamefont
  {Gelman}}, \bibinfo {author} {\bibfnamefont {J.}~\bibnamefont {Carlin}},
  \bibinfo {author} {\bibfnamefont {H.}~\bibnamefont {Stern}},\ and\ \bibinfo
  {author} {\bibfnamefont {D.}~\bibnamefont {Rubin}},\ }\href
  {https://books.google.com/books?id=BkGCmAEACAAJ} {\emph {\bibinfo {title}
  {Bayesian Data Analysis}}},\ Chapman \& Hall/CRC Texts in Statistical
  Science\ (\bibinfo  {publisher} {Chapman \& Hall/CRC},\ \bibinfo {year}
  {2003})\BibitemShut {NoStop}%
\bibitem [{\citenamefont {Smith}\ and\ \citenamefont
  {Gelfand}(1992)}]{doi:10.1080/00031305.1992.10475856}%
  \BibitemOpen
  \bibfield  {author} {\bibinfo {author} {\bibfnamefont {A.~F.~M.}\
  \bibnamefont {Smith}}\ and\ \bibinfo {author} {\bibfnamefont {A.~E.}\
  \bibnamefont {Gelfand}},\ }\bibfield  {title} {\bibinfo {title} {Bayesian
  statistics without tears: A sampling–resampling perspective},\ }\href
  {https://doi.org/10.1080/00031305.1992.10475856} {\bibfield  {journal}
  {\bibinfo  {journal} {The American Statistician}\ }\textbf {\bibinfo {volume}
  {46}},\ \bibinfo {pages} {84} (\bibinfo {year} {1992})},\ \Eprint
  {https://arxiv.org/abs/https://doi.org/10.1080/00031305.1992.10475856}
  {https://doi.org/10.1080/00031305.1992.10475856} \BibitemShut {NoStop}%
\bibitem [{\citenamefont {Rubin}(1988)}]{rubin1988using}%
  \BibitemOpen
  \bibfield  {author} {\bibinfo {author} {\bibfnamefont {D.~B.}\ \bibnamefont
  {Rubin}},\ }\bibfield  {title} {\bibinfo {title} {Using the sir algorithm to
  simulate posterior distributions},\ }\href@noop {} {\bibfield  {journal}
  {\bibinfo  {journal} {Bayesian statistics}\ }\textbf {\bibinfo {volume}
  {3}},\ \bibinfo {pages} {395} (\bibinfo {year} {1988})}\BibitemShut {NoStop}%
\bibitem [{\citenamefont {{Plummer}}(1911)}]{1911MNRAS..71..460P}%
  \BibitemOpen
  \bibfield  {author} {\bibinfo {author} {\bibfnamefont {H.~C.}\ \bibnamefont
  {{Plummer}}},\ }\bibfield  {title} {\bibinfo {title} {{On the problem of
  distribution in globular star clusters}},\ }\href
  {https://doi.org/10.1093/mnras/71.5.460} {\bibfield  {journal} {\bibinfo
  {journal} {MNRAS}\ }\textbf {\bibinfo {volume} {71}},\ \bibinfo {pages} {460}
  (\bibinfo {year} {1911})}\BibitemShut {NoStop}%
\bibitem [{\citenamefont {{Osipkov}}(1979)}]{1979PAZh....5...77O}%
  \BibitemOpen
  \bibfield  {author} {\bibinfo {author} {\bibfnamefont {L.~P.}\ \bibnamefont
  {{Osipkov}}},\ }\bibfield  {title} {\bibinfo {title} {{Spherical systems of
  gravitating bodies with an ellipsoidal velocity distribution}},\ }\href@noop
  {} {\bibfield  {journal} {\bibinfo  {journal} {Pisma v Astronomicheskii
  Zhurnal}\ }\textbf {\bibinfo {volume} {5}},\ \bibinfo {pages} {77} (\bibinfo
  {year} {1979})}\BibitemShut {NoStop}%
\bibitem [{\citenamefont {{Merritt}}(1985)}]{1985AJ.....90.1027M}%
  \BibitemOpen
  \bibfield  {author} {\bibinfo {author} {\bibfnamefont {D.}~\bibnamefont
  {{Merritt}}},\ }\bibfield  {title} {\bibinfo {title} {{Spherical stellar
  systems with spheroidal velocity distributions}},\ }\href
  {https://doi.org/10.1086/113810} {\bibfield  {journal} {\bibinfo  {journal}
  {AJ}\ }\textbf {\bibinfo {volume} {90}},\ \bibinfo {pages} {1027} (\bibinfo
  {year} {1985})}\BibitemShut {NoStop}%
\bibitem [{\citenamefont {{Simon}}\ and\ \citenamefont
  {{Geha}}(2007)}]{2007ApJ...670..313S}%
  \BibitemOpen
  \bibfield  {author} {\bibinfo {author} {\bibfnamefont {J.~D.}\ \bibnamefont
  {{Simon}}}\ and\ \bibinfo {author} {\bibfnamefont {M.}~\bibnamefont
  {{Geha}}},\ }\bibfield  {title} {\bibinfo {title} {{The Kinematics of the
  Ultra-faint Milky Way Satellites: Solving the Missing Satellite Problem}},\
  }\href {https://doi.org/10.1086/521816} {\bibfield  {journal} {\bibinfo
  {journal} {ApJ}\ }\textbf {\bibinfo {volume} {670}},\ \bibinfo {pages} {313}
  (\bibinfo {year} {2007})},\ \Eprint {https://arxiv.org/abs/0706.0516}
  {arXiv:0706.0516 [astro-ph]} \BibitemShut {NoStop}%
\bibitem [{\citenamefont {{Mateo}}\ \emph {et~al.}(2008)\citenamefont
  {{Mateo}}, \citenamefont {{Olszewski}},\ and\ \citenamefont
  {{Walker}}}]{2008ApJ...675..201M}%
  \BibitemOpen
  \bibfield  {author} {\bibinfo {author} {\bibfnamefont {M.}~\bibnamefont
  {{Mateo}}}, \bibinfo {author} {\bibfnamefont {E.~W.}\ \bibnamefont
  {{Olszewski}}},\ and\ \bibinfo {author} {\bibfnamefont {M.~G.}\ \bibnamefont
  {{Walker}}},\ }\bibfield  {title} {\bibinfo {title} {{The Velocity Dispersion
  Profile of the Remote Dwarf Spheroidal Galaxy Leo I: A Tidal Hit and Run?}},\
  }\href {https://doi.org/10.1086/522326} {\bibfield  {journal} {\bibinfo
  {journal} {ApJ}\ }\textbf {\bibinfo {volume} {675}},\ \bibinfo {pages} {201}
  (\bibinfo {year} {2008})},\ \Eprint {https://arxiv.org/abs/0708.1327}
  {arXiv:0708.1327 [astro-ph]} \BibitemShut {NoStop}%
\bibitem [{\citenamefont {{Walker}}\ \emph
  {et~al.}(2009{\natexlab{a}})\citenamefont {{Walker}}, \citenamefont
  {{Mateo}},\ and\ \citenamefont {{Olszewski}}}]{2009AJ....137.3100W}%
  \BibitemOpen
  \bibfield  {author} {\bibinfo {author} {\bibfnamefont {M.~G.}\ \bibnamefont
  {{Walker}}}, \bibinfo {author} {\bibfnamefont {M.}~\bibnamefont {{Mateo}}},\
  and\ \bibinfo {author} {\bibfnamefont {E.~W.}\ \bibnamefont {{Olszewski}}},\
  }\bibfield  {title} {\bibinfo {title} {{Stellar Velocities in the Carina,
  Fornax, Sculptor, and Sextans dSph Galaxies: Data From the Magellan/MMFS
  Survey}},\ }\href {https://doi.org/10.1088/0004-6256/137/2/3100} {\bibfield
  {journal} {\bibinfo  {journal} {ApJ}\ }\textbf {\bibinfo {volume} {137}},\
  \bibinfo {pages} {3100} (\bibinfo {year} {2009}{\natexlab{a}})},\ \Eprint
  {https://arxiv.org/abs/0811.0118} {arXiv:0811.0118 [astro-ph]} \BibitemShut
  {NoStop}%
\bibitem [{\citenamefont {{Walker}}\ \emph
  {et~al.}(2009{\natexlab{b}})\citenamefont {{Walker}}, \citenamefont
  {{Mateo}}, \citenamefont {{Olszewski}}, \citenamefont {{Pe{\~n}arrubia}},
  \citenamefont {{Evans}},\ and\ \citenamefont
  {{Gilmore}}}]{2009ApJ...704.1274W}%
  \BibitemOpen
  \bibfield  {author} {\bibinfo {author} {\bibfnamefont {M.~G.}\ \bibnamefont
  {{Walker}}}, \bibinfo {author} {\bibfnamefont {M.}~\bibnamefont {{Mateo}}},
  \bibinfo {author} {\bibfnamefont {E.~W.}\ \bibnamefont {{Olszewski}}},
  \bibinfo {author} {\bibfnamefont {J.}~\bibnamefont {{Pe{\~n}arrubia}}},
  \bibinfo {author} {\bibfnamefont {N.~W.}\ \bibnamefont {{Evans}}},\ and\
  \bibinfo {author} {\bibfnamefont {G.}~\bibnamefont {{Gilmore}}},\ }\bibfield
  {title} {\bibinfo {title} {{A Universal Mass Profile for Dwarf Spheroidal
  Galaxies?}},\ }\href {https://doi.org/10.1088/0004-637X/704/2/1274}
  {\bibfield  {journal} {\bibinfo  {journal} {ApJ}\ }\textbf {\bibinfo {volume}
  {704}},\ \bibinfo {pages} {1274} (\bibinfo {year} {2009}{\natexlab{b}})},\
  \Eprint {https://arxiv.org/abs/0906.0341} {arXiv:0906.0341 [astro-ph.CO]}
  \BibitemShut {NoStop}%
\bibitem [{\citenamefont {{Spencer}}\ \emph {et~al.}(2017)\citenamefont
  {{Spencer}}, \citenamefont {{Mateo}}, \citenamefont {{Walker}}, \citenamefont
  {{Olszewski}}, \citenamefont {{McConnachie}}, \citenamefont {{Kirby}},\ and\
  \citenamefont {{Koch}}}]{2017AJ....153..254S}%
  \BibitemOpen
  \bibfield  {author} {\bibinfo {author} {\bibfnamefont {M.~E.}\ \bibnamefont
  {{Spencer}}}, \bibinfo {author} {\bibfnamefont {M.}~\bibnamefont {{Mateo}}},
  \bibinfo {author} {\bibfnamefont {M.~G.}\ \bibnamefont {{Walker}}}, \bibinfo
  {author} {\bibfnamefont {E.~W.}\ \bibnamefont {{Olszewski}}}, \bibinfo
  {author} {\bibfnamefont {A.~W.}\ \bibnamefont {{McConnachie}}}, \bibinfo
  {author} {\bibfnamefont {E.~N.}\ \bibnamefont {{Kirby}}},\ and\ \bibinfo
  {author} {\bibfnamefont {A.}~\bibnamefont {{Koch}}},\ }\bibfield  {title}
  {\bibinfo {title} {{The Binary Fraction of Stars in Dwarf Galaxies: The Case
  of Leo II}},\ }\href {https://doi.org/10.3847/1538-3881/aa6d51} {\bibfield
  {journal} {\bibinfo  {journal} {AJ}\ }\textbf {\bibinfo {volume} {153}},\
  \bibinfo {eid} {254} (\bibinfo {year} {2017})},\ \Eprint
  {https://arxiv.org/abs/1706.04184} {arXiv:1706.04184 [astro-ph.GA]}
  \BibitemShut {NoStop}%
\bibitem [{\citenamefont {{Makinen}}\ \emph {et~al.}(2022)\citenamefont
  {{Makinen}}, \citenamefont {{Charnock}}, \citenamefont {{Lemos}},
  \citenamefont {{Porqueres}}, \citenamefont {{Heavens}},\ and\ \citenamefont
  {{Wandelt}}}]{2022arXiv220705202M}%
  \BibitemOpen
  \bibfield  {author} {\bibinfo {author} {\bibfnamefont {T.~L.}\ \bibnamefont
  {{Makinen}}}, \bibinfo {author} {\bibfnamefont {T.}~\bibnamefont
  {{Charnock}}}, \bibinfo {author} {\bibfnamefont {P.}~\bibnamefont {{Lemos}}},
  \bibinfo {author} {\bibfnamefont {N.}~\bibnamefont {{Porqueres}}}, \bibinfo
  {author} {\bibfnamefont {A.}~\bibnamefont {{Heavens}}},\ and\ \bibinfo
  {author} {\bibfnamefont {B.~D.}\ \bibnamefont {{Wandelt}}},\ }\bibfield
  {title} {\bibinfo {title} {{The Cosmic Graph: Optimal Information Extraction
  from Large-Scale Structure using Catalogues}},\ }\href@noop {} {\bibfield
  {journal} {\bibinfo  {journal} {arXiv e-prints}\ ,\ \bibinfo {eid}
  {arXiv:2207.05202}} (\bibinfo {year} {2022})},\ \Eprint
  {https://arxiv.org/abs/2207.05202} {arXiv:2207.05202 [astro-ph.CO]}
  \BibitemShut {NoStop}%
\bibitem [{\citenamefont {{Veli{\v{c}}kovi{\'c}}}\ \emph
  {et~al.}(2017)\citenamefont {{Veli{\v{c}}kovi{\'c}}}, \citenamefont
  {{Cucurull}}, \citenamefont {{Casanova}}, \citenamefont {{Romero}},
  \citenamefont {{Li{\`o}}},\ and\ \citenamefont
  {{Bengio}}}]{2017arXiv171010903V}%
  \BibitemOpen
  \bibfield  {author} {\bibinfo {author} {\bibfnamefont {P.}~\bibnamefont
  {{Veli{\v{c}}kovi{\'c}}}}, \bibinfo {author} {\bibfnamefont {G.}~\bibnamefont
  {{Cucurull}}}, \bibinfo {author} {\bibfnamefont {A.}~\bibnamefont
  {{Casanova}}}, \bibinfo {author} {\bibfnamefont {A.}~\bibnamefont
  {{Romero}}}, \bibinfo {author} {\bibfnamefont {P.}~\bibnamefont
  {{Li{\`o}}}},\ and\ \bibinfo {author} {\bibfnamefont {Y.}~\bibnamefont
  {{Bengio}}},\ }\bibfield  {title} {\bibinfo {title} {{Graph Attention
  Networks}},\ }\href@noop {} {\bibfield  {journal} {\bibinfo  {journal} {arXiv
  e-prints}\ ,\ \bibinfo {eid} {arXiv:1710.10903}} (\bibinfo {year} {2017})},\
  \Eprint {https://arxiv.org/abs/1710.10903} {arXiv:1710.10903 [stat.ML]}
  \BibitemShut {NoStop}%
\bibitem [{\citenamefont {{Defferrard}}\ \emph {et~al.}(2016)\citenamefont
  {{Defferrard}}, \citenamefont {{Bresson}},\ and\ \citenamefont
  {{Vandergheynst}}}]{2016arXiv160609375D}%
  \BibitemOpen
  \bibfield  {author} {\bibinfo {author} {\bibfnamefont {M.}~\bibnamefont
  {{Defferrard}}}, \bibinfo {author} {\bibfnamefont {X.}~\bibnamefont
  {{Bresson}}},\ and\ \bibinfo {author} {\bibfnamefont {P.}~\bibnamefont
  {{Vandergheynst}}},\ }\bibfield  {title} {\bibinfo {title} {{Convolutional
  Neural Networks on Graphs with Fast Localized Spectral Filtering}},\
  }\href@noop {} {\bibfield  {journal} {\bibinfo  {journal} {arXiv e-prints}\
  ,\ \bibinfo {eid} {arXiv:1606.09375}} (\bibinfo {year} {2016})},\ \Eprint
  {https://arxiv.org/abs/1606.09375} {arXiv:1606.09375 [cs.LG]} \BibitemShut
  {NoStop}%
\bibitem [{\citenamefont {Papamakarios}\ \emph {et~al.}(2019)\citenamefont
  {Papamakarios}, \citenamefont {Nalisnick}, \citenamefont {Rezende},
  \citenamefont {Mohamed},\ and\ \citenamefont
  {Lakshminarayanan}}]{papamakarios2019normalizing}%
  \BibitemOpen
  \bibfield  {author} {\bibinfo {author} {\bibfnamefont {G.}~\bibnamefont
  {Papamakarios}}, \bibinfo {author} {\bibfnamefont {E.}~\bibnamefont
  {Nalisnick}}, \bibinfo {author} {\bibfnamefont {D.~J.}\ \bibnamefont
  {Rezende}}, \bibinfo {author} {\bibfnamefont {S.}~\bibnamefont {Mohamed}},\
  and\ \bibinfo {author} {\bibfnamefont {B.}~\bibnamefont {Lakshminarayanan}},\
  }\bibfield  {title} {\bibinfo {title} {Normalizing flows for probabilistic
  modeling and inference},\ }\href@noop {} {\bibfield  {journal} {\bibinfo
  {journal} {Journal of Machine Learning Research}\ } (\bibinfo {year}
  {2019})},\ \Eprint {https://arxiv.org/abs/1912.02762} {arXiv:1912.02762
  [cs.LG]} \BibitemShut {NoStop}%
\bibitem [{\citenamefont {Rezende}\ and\ \citenamefont
  {Mohamed}(2015)}]{DBLP:conf/icml/RezendeM15}%
  \BibitemOpen
  \bibfield  {author} {\bibinfo {author} {\bibfnamefont {D.~J.}\ \bibnamefont
  {Rezende}}\ and\ \bibinfo {author} {\bibfnamefont {S.}~\bibnamefont
  {Mohamed}},\ }\bibfield  {title} {\bibinfo {title} {Variational inference
  with normalizing flows},\ }in\ \href
  {http://proceedings.mlr.press/v37/rezende15.html} {\emph {\bibinfo
  {booktitle} {Proceedings of the 32nd International Conference on Machine
  Learning, {ICML} 2015, Lille, France, 6-11 July 2015}}},\ \bibinfo {series}
  {{JMLR} Workshop and Conference Proceedings}, Vol.~\bibinfo {volume} {37},\
  \bibinfo {editor} {edited by\ \bibinfo {editor} {\bibfnamefont {F.~R.}\
  \bibnamefont {Bach}}\ and\ \bibinfo {editor} {\bibfnamefont {D.~M.}\
  \bibnamefont {Blei}}}\ (\bibinfo {year} {2015})\ pp.\ \bibinfo {pages}
  {1530--1538}\BibitemShut {NoStop}%
\bibitem [{\citenamefont {Papamakarios}\ \emph {et~al.}(2017)\citenamefont
  {Papamakarios}, \citenamefont {Pavlakou},\ and\ \citenamefont
  {Murray}}]{10.5555/3294771.3294994}%
  \BibitemOpen
  \bibfield  {author} {\bibinfo {author} {\bibfnamefont {G.}~\bibnamefont
  {Papamakarios}}, \bibinfo {author} {\bibfnamefont {T.}~\bibnamefont
  {Pavlakou}},\ and\ \bibinfo {author} {\bibfnamefont {I.}~\bibnamefont
  {Murray}},\ }\bibfield  {title} {\bibinfo {title} {Masked autoregressive flow
  for density estimation},\ }in\ \href
  {https://papers.nips.cc/paper/2017/hash/6c1da886822c67822bcf3679d04369fa-Abstract.html}
  {\emph {\bibinfo {booktitle} {Proceedings of the 31st International
  Conference on Neural Information Processing Systems}}},\ \bibinfo {series and
  number} {NIPS'17}\ (\bibinfo  {publisher} {Curran Associates Inc.},\ \bibinfo
  {address} {Red Hook, NY, USA},\ \bibinfo {year} {2017})\ pp.\ \bibinfo
  {pages} {2335--2344}\BibitemShut {NoStop}%
\bibitem [{\citenamefont {Germain}\ \emph {et~al.}(2015)\citenamefont
  {Germain}, \citenamefont {Gregor}, \citenamefont {Murray},\ and\
  \citenamefont {Larochelle}}]{DBLP:conf/icml/GermainGML15}%
  \BibitemOpen
  \bibfield  {author} {\bibinfo {author} {\bibfnamefont {M.}~\bibnamefont
  {Germain}}, \bibinfo {author} {\bibfnamefont {K.}~\bibnamefont {Gregor}},
  \bibinfo {author} {\bibfnamefont {I.}~\bibnamefont {Murray}},\ and\ \bibinfo
  {author} {\bibfnamefont {H.}~\bibnamefont {Larochelle}},\ }\bibfield  {title}
  {\bibinfo {title} {{MADE:} masked autoencoder for distribution estimation},\
  }in\ \href {http://proceedings.mlr.press/v37/germain15.html} {\emph {\bibinfo
  {booktitle} {Proceedings of the 32nd International Conference on Machine
  Learning, {ICML} 2015, Lille, France, 6-11 July 2015}}},\ \bibinfo {series}
  {{JMLR} Workshop and Conference Proceedings}, Vol.~\bibinfo {volume} {37},\
  \bibinfo {editor} {edited by\ \bibinfo {editor} {\bibfnamefont {F.~R.}\
  \bibnamefont {Bach}}\ and\ \bibinfo {editor} {\bibfnamefont {D.~M.}\
  \bibnamefont {Blei}}}\ (\bibinfo {year} {2015})\ pp.\ \bibinfo {pages}
  {881--889}\BibitemShut {NoStop}%
\bibitem [{\citenamefont {Cranmer}\ \emph {et~al.}(2020)\citenamefont
  {Cranmer}, \citenamefont {Brehmer},\ and\ \citenamefont
  {Louppe}}]{Cranmer:2019eaq}%
  \BibitemOpen
  \bibfield  {author} {\bibinfo {author} {\bibfnamefont {K.}~\bibnamefont
  {Cranmer}}, \bibinfo {author} {\bibfnamefont {J.}~\bibnamefont {Brehmer}},\
  and\ \bibinfo {author} {\bibfnamefont {G.}~\bibnamefont {Louppe}},\
  }\bibfield  {title} {\bibinfo {title} {{The frontier of simulation-based
  inference}},\ }\href {https://doi.org/10.1073/pnas.1912789117} {\bibfield
  {journal} {\bibinfo  {journal} {Proc. Nat. Acad. Sci.}\ }\textbf {\bibinfo
  {volume} {117}},\ \bibinfo {pages} {30055} (\bibinfo {year} {2020})},\
  \Eprint {https://arxiv.org/abs/1911.01429} {arXiv:1911.01429 [stat.ML]}
  \BibitemShut {NoStop}%
\bibitem [{\citenamefont {Cranmer}\ and\ \citenamefont
  {Louppe}(2016)}]{cranmer_kyle_2016_198541}%
  \BibitemOpen
  \bibfield  {author} {\bibinfo {author} {\bibfnamefont {K.}~\bibnamefont
  {Cranmer}}\ and\ \bibinfo {author} {\bibfnamefont {G.}~\bibnamefont
  {Louppe}},\ }\bibfield  {title} {\bibinfo {title} {{Unifying generative
  models and exact likelihood- free inference with conditional bijections}},\
  }\bibfield  {journal} {\bibinfo  {journal} {J. Brief Ideas}\ }\href
  {https://doi.org/10.5281/zenodo.198541} {10.5281/zenodo.198541} (\bibinfo
  {year} {2016})\BibitemShut {NoStop}%
\bibitem [{\citenamefont {Papamakarios}\ and\ \citenamefont
  {Murray}(2016)}]{10.5555/3157096.3157212}%
  \BibitemOpen
  \bibfield  {author} {\bibinfo {author} {\bibfnamefont {G.}~\bibnamefont
  {Papamakarios}}\ and\ \bibinfo {author} {\bibfnamefont {I.}~\bibnamefont
  {Murray}},\ }\bibfield  {title} {\bibinfo {title} {Fast $\epsilon$-free
  inference of simulation models with bayesian conditional density
  estimation},\ }in\ \href
  {https://proceedings.neurips.cc/paper/2016/hash/6aca97005c68f1206823815f66102863-Abstract.html}
  {\emph {\bibinfo {booktitle} {Proceedings of the 30th International
  Conference on Neural Information Processing Systems}}},\ \bibinfo {series and
  number} {NIPS'16}\ (\bibinfo  {publisher} {Curran Associates Inc.},\ \bibinfo
  {address} {Red Hook, NY, USA},\ \bibinfo {year} {2016})\ pp.\ \bibinfo
  {pages} {1036--1044},\ \Eprint {https://arxiv.org/abs/1605.06376}
  {arXiv:1605.06376 [stat.ML]} \BibitemShut {NoStop}%
\bibitem [{\citenamefont {Kingma}\ and\ \citenamefont
  {Ba}(2014)}]{kingma2014adam}%
  \BibitemOpen
  \bibfield  {author} {\bibinfo {author} {\bibfnamefont {D.~P.}\ \bibnamefont
  {Kingma}}\ and\ \bibinfo {author} {\bibfnamefont {J.}~\bibnamefont {Ba}},\
  }\bibfield  {title} {\bibinfo {title} {Adam: A method for stochastic
  optimization},\ }\href@noop {} {\bibfield  {journal} {\bibinfo  {journal}
  {arXiv preprint arXiv:1412.6980}\ } (\bibinfo {year} {2014})}\BibitemShut
  {NoStop}%
\bibitem [{\citenamefont {Loshchilov}\ and\ \citenamefont
  {Hutter}(2019)}]{adamw2019}%
  \BibitemOpen
  \bibfield  {author} {\bibinfo {author} {\bibfnamefont {I.}~\bibnamefont
  {Loshchilov}}\ and\ \bibinfo {author} {\bibfnamefont {F.}~\bibnamefont
  {Hutter}},\ }\bibfield  {title} {\bibinfo {title} {Decoupled weight decay
  regularization},\ }in\ \href {https://openreview.net/forum?id=Bkg6RiCqY7}
  {\emph {\bibinfo {booktitle} {International Conference on Learning
  Representations}}}\ (\bibinfo {year} {2019})\BibitemShut {NoStop}%
\bibitem [{\citenamefont {{Skilling}}(2004)}]{2004AIPC..735..395S}%
  \BibitemOpen
  \bibfield  {author} {\bibinfo {author} {\bibfnamefont {J.}~\bibnamefont
  {{Skilling}}},\ }\bibfield  {title} {\bibinfo {title} {{Nested Sampling}},\
  }in\ \href {https://doi.org/10.1063/1.1835238} {\emph {\bibinfo {booktitle}
  {Bayesian Inference and Maximum Entropy Methods in Science and Engineering:
  24th International Workshop on Bayesian Inference and Maximum Entropy Methods
  in Science and Engineering}}},\ \bibinfo {series} {American Institute of
  Physics Conference Series}, Vol.\ \bibinfo {volume} {735},\ \bibinfo {editor}
  {edited by\ \bibinfo {editor} {\bibfnamefont {R.}~\bibnamefont {{Fischer}}},
  \bibinfo {editor} {\bibfnamefont {R.}~\bibnamefont {{Preuss}}},\ and\
  \bibinfo {editor} {\bibfnamefont {U.~V.}\ \bibnamefont {{Toussaint}}}}\
  (\bibinfo {year} {2004})\ pp.\ \bibinfo {pages} {395--405}\BibitemShut
  {NoStop}%
\bibitem [{\citenamefont {Skilling}(2006)}]{skilling2006}%
  \BibitemOpen
  \bibfield  {author} {\bibinfo {author} {\bibfnamefont {J.}~\bibnamefont
  {Skilling}},\ }\bibfield  {title} {\bibinfo {title} {Nested sampling for
  general bayesian computation},\ }\href {https://doi.org/10.1214/06-BA127}
  {\bibfield  {journal} {\bibinfo  {journal} {Bayesian Anal.}\ }\textbf
  {\bibinfo {volume} {1}},\ \bibinfo {pages} {833} (\bibinfo {year}
  {2006})}\BibitemShut {NoStop}%
\bibitem [{\citenamefont {{Speagle}}(2020)}]{dynesty}%
  \BibitemOpen
  \bibfield  {author} {\bibinfo {author} {\bibfnamefont {J.~S.}\ \bibnamefont
  {{Speagle}}},\ }\bibfield  {title} {\bibinfo {title} {{DYNESTY: a dynamic
  nested sampling package for estimating Bayesian posteriors and evidences}},\
  }\href {https://doi.org/10.1093/mnras/staa278} {\bibfield  {journal}
  {\bibinfo  {journal} {\mnras}\ }\textbf {\bibinfo {volume} {493}},\ \bibinfo
  {pages} {3132} (\bibinfo {year} {2020})},\ \Eprint
  {https://arxiv.org/abs/1904.02180} {arXiv:1904.02180 [astro-ph.IM]}
  \BibitemShut {NoStop}%
\bibitem [{\citenamefont {Nielsen}(2019)}]{e21050485}%
  \BibitemOpen
  \bibfield  {author} {\bibinfo {author} {\bibfnamefont {F.}~\bibnamefont
  {Nielsen}},\ }\bibfield  {title} {\bibinfo {title} {On the jensen–shannon
  symmetrization of distances relying on abstract means},\ }\bibfield
  {journal} {\bibinfo  {journal} {Entropy}\ }\textbf {\bibinfo {volume} {21}},\
  \href {https://doi.org/10.3390/e21050485} {10.3390/e21050485} (\bibinfo
  {year} {2019})\BibitemShut {NoStop}%
\bibitem [{\citenamefont {Manning}\ and\ \citenamefont
  {Sch{\"u}tze}(1999)}]{manning99foundations}%
  \BibitemOpen
  \bibfield  {author} {\bibinfo {author} {\bibfnamefont {C.~D.}\ \bibnamefont
  {Manning}}\ and\ \bibinfo {author} {\bibfnamefont {H.}~\bibnamefont
  {Sch{\"u}tze}},\ }\href {http://nlp.stanford.edu/fsnlp/} {\emph {\bibinfo
  {title} {Foundations of Statistical Natural Language Processing}}}\ (\bibinfo
   {publisher} {The {MIT} Press},\ \bibinfo {address} {Cambridge,
  Massachusetts},\ \bibinfo {year} {1999})\BibitemShut {NoStop}%
\bibitem [{\citenamefont {{Garrison-Kimmel}}\ \emph {et~al.}(2019)\citenamefont
  {{Garrison-Kimmel}}, \citenamefont {{Hopkins}}, \citenamefont {{Wetzel}},
  \citenamefont {{Bullock}}, \citenamefont {{Boylan-Kolchin}}, \citenamefont
  {{Kere{\v{s}}}}, \citenamefont {{Faucher-Gigu{\`e}re}}, \citenamefont
  {{El-Badry}}, \citenamefont {{Lamberts}}, \citenamefont {{Quataert}},\ and\
  \citenamefont {{Sanderson}}}]{2019MNRAS.487.1380G}%
  \BibitemOpen
  \bibfield  {author} {\bibinfo {author} {\bibfnamefont {S.}~\bibnamefont
  {{Garrison-Kimmel}}}, \bibinfo {author} {\bibfnamefont {P.~F.}\ \bibnamefont
  {{Hopkins}}}, \bibinfo {author} {\bibfnamefont {A.}~\bibnamefont {{Wetzel}}},
  \bibinfo {author} {\bibfnamefont {J.~S.}\ \bibnamefont {{Bullock}}}, \bibinfo
  {author} {\bibfnamefont {M.}~\bibnamefont {{Boylan-Kolchin}}}, \bibinfo
  {author} {\bibfnamefont {D.}~\bibnamefont {{Kere{\v{s}}}}}, \bibinfo {author}
  {\bibfnamefont {C.-A.}\ \bibnamefont {{Faucher-Gigu{\`e}re}}}, \bibinfo
  {author} {\bibfnamefont {K.}~\bibnamefont {{El-Badry}}}, \bibinfo {author}
  {\bibfnamefont {A.}~\bibnamefont {{Lamberts}}}, \bibinfo {author}
  {\bibfnamefont {E.}~\bibnamefont {{Quataert}}},\ and\ \bibinfo {author}
  {\bibfnamefont {R.}~\bibnamefont {{Sanderson}}},\ }\bibfield  {title}
  {\bibinfo {title} {{The Local Group on FIRE: dwarf galaxy populations across
  a suite of hydrodynamic simulations}},\ }\href
  {https://doi.org/10.1093/mnras/stz1317} {\bibfield  {journal} {\bibinfo
  {journal} {\mnras}\ }\textbf {\bibinfo {volume} {487}},\ \bibinfo {pages}
  {1380} (\bibinfo {year} {2019})},\ \Eprint {https://arxiv.org/abs/1806.04143}
  {arXiv:1806.04143 [astro-ph.GA]} \BibitemShut {NoStop}%
\bibitem [{\citenamefont {Brown}\ \emph {et~al.}(2021)\citenamefont {Brown},
  \citenamefont {Buitrago}, \citenamefont {Hanna}, \citenamefont {Sanielevici},
  \citenamefont {Scibek},\ and\ \citenamefont {Nystrom}}]{bridges2}%
  \BibitemOpen
  \bibfield  {author} {\bibinfo {author} {\bibfnamefont {S.~T.}\ \bibnamefont
  {Brown}}, \bibinfo {author} {\bibfnamefont {P.}~\bibnamefont {Buitrago}},
  \bibinfo {author} {\bibfnamefont {E.}~\bibnamefont {Hanna}}, \bibinfo
  {author} {\bibfnamefont {S.}~\bibnamefont {Sanielevici}}, \bibinfo {author}
  {\bibfnamefont {R.}~\bibnamefont {Scibek}},\ and\ \bibinfo {author}
  {\bibfnamefont {N.~A.}\ \bibnamefont {Nystrom}},\ }\bibfield  {title}
  {\bibinfo {title} {Bridges-2: A platform for rapidly-evolving and data
  intensive research},\ }in\ \href {https://doi.org/10.1145/3437359.3465593}
  {\emph {\bibinfo {booktitle} {Practice and Experience in Advanced Research
  Computing}}},\ \bibinfo {series and number} {PEARC '21}\ (\bibinfo
  {publisher} {Association for Computing Machinery},\ \bibinfo {address} {New
  York, NY, USA},\ \bibinfo {year} {2021})\BibitemShut {NoStop}%
\bibitem [{\citenamefont {Towns}\ \emph {et~al.}(2014)\citenamefont {Towns},
  \citenamefont {Cockerill}, \citenamefont {Dahan}, \citenamefont {Foster},
  \citenamefont {Gaither}, \citenamefont {Grimshaw}, \citenamefont {Hazlewood},
  \citenamefont {Lathrop}, \citenamefont {Lifka}, \citenamefont {Peterson},
  \citenamefont {Roskies}, \citenamefont {Scott},\ and\ \citenamefont
  {Wilkins-Diehr}}]{xsede}%
  \BibitemOpen
  \bibfield  {author} {\bibinfo {author} {\bibfnamefont {J.}~\bibnamefont
  {Towns}}, \bibinfo {author} {\bibfnamefont {T.}~\bibnamefont {Cockerill}},
  \bibinfo {author} {\bibfnamefont {M.}~\bibnamefont {Dahan}}, \bibinfo
  {author} {\bibfnamefont {I.}~\bibnamefont {Foster}}, \bibinfo {author}
  {\bibfnamefont {K.}~\bibnamefont {Gaither}}, \bibinfo {author} {\bibfnamefont
  {A.}~\bibnamefont {Grimshaw}}, \bibinfo {author} {\bibfnamefont
  {V.}~\bibnamefont {Hazlewood}}, \bibinfo {author} {\bibfnamefont
  {S.}~\bibnamefont {Lathrop}}, \bibinfo {author} {\bibfnamefont
  {D.}~\bibnamefont {Lifka}}, \bibinfo {author} {\bibfnamefont {G.~D.}\
  \bibnamefont {Peterson}}, \bibinfo {author} {\bibfnamefont {R.}~\bibnamefont
  {Roskies}}, \bibinfo {author} {\bibfnamefont {J.}~\bibnamefont {Scott}},\
  and\ \bibinfo {author} {\bibfnamefont {N.}~\bibnamefont {Wilkins-Diehr}},\
  }\bibfield  {title} {\bibinfo {title} {Xsede: Accelerating scientific
  discovery},\ }\href {https://doi.org/10.1109/MCSE.2014.80} {\bibfield
  {journal} {\bibinfo  {journal} {Computing in Science; Engineering}\ }\textbf
  {\bibinfo {volume} {16}},\ \bibinfo {pages} {62} (\bibinfo {year}
  {2014})}\BibitemShut {NoStop}%
\bibitem [{\citenamefont {{Ashton}}\ \emph {et~al.}(2019)\citenamefont
  {{Ashton}} \emph {et~al.}}]{bilby}%
  \BibitemOpen
  \bibfield  {author} {\bibinfo {author} {\bibfnamefont {G.}~\bibnamefont
  {{Ashton}}} \emph {et~al.},\ }\bibfield  {title} {\bibinfo {title} {{BILBY: A
  User-friendly Bayesian Inference Library for Gravitational-wave Astronomy}},\
  }\href {https://doi.org/10.3847/1538-4365/ab06fc} {\bibfield  {journal}
  {\bibinfo  {journal} {\apjs}\ }\textbf {\bibinfo {volume} {241}},\ \bibinfo
  {eid} {27} (\bibinfo {year} {2019})},\ \Eprint
  {https://arxiv.org/abs/1811.02042} {arXiv:1811.02042 [astro-ph.IM]}
  \BibitemShut {NoStop}%
\bibitem [{\citenamefont {{Perez}}\ and\ \citenamefont
  {{Granger}}(2007)}]{PER-GRA:2007}%
  \BibitemOpen
  \bibfield  {author} {\bibinfo {author} {\bibfnamefont {F.}~\bibnamefont
  {{Perez}}}\ and\ \bibinfo {author} {\bibfnamefont {B.~E.}\ \bibnamefont
  {{Granger}}},\ }\bibfield  {title} {\bibinfo {title} {{IPython: A System for
  Interactive Scientific Computing}},\ }\href
  {https://doi.org/10.1109/MCSE.2007.53} {\bibfield  {journal} {\bibinfo
  {journal} {Computing in Science and Engineering}\ }\textbf {\bibinfo {volume}
  {9}},\ \bibinfo {pages} {21} (\bibinfo {year} {2007})}\BibitemShut {NoStop}%
\bibitem [{\citenamefont {Kluyver}\ \emph {et~al.}(2016)\citenamefont {Kluyver}
  \emph {et~al.}}]{Kluyver2016JupyterN}%
  \BibitemOpen
  \bibfield  {author} {\bibinfo {author} {\bibfnamefont {T.}~\bibnamefont
  {Kluyver}} \emph {et~al.},\ }\bibfield  {title} {\bibinfo {title} {Jupyter
  notebooks - a publishing format for reproducible computational workflows},\
  }in\ \href@noop {} {\emph {\bibinfo {booktitle} {ELPUB}}}\ (\bibinfo {year}
  {2016})\BibitemShut {NoStop}%
\bibitem [{\citenamefont {Hunter}(2007)}]{Hunter:2007}%
  \BibitemOpen
  \bibfield  {author} {\bibinfo {author} {\bibfnamefont {J.~D.}\ \bibnamefont
  {Hunter}},\ }\bibfield  {title} {\bibinfo {title} {{Matplotlib: A 2D graphics
  environment}},\ }\href@noop {} {\bibfield  {journal} {\bibinfo  {journal}
  {Computing In Science \& Engineering}\ }\textbf {\bibinfo {volume} {9}},\
  \bibinfo {pages} {90} (\bibinfo {year} {2007})}\BibitemShut {NoStop}%
\bibitem [{\citenamefont {Durkan}\ \emph {et~al.}(2020)\citenamefont {Durkan},
  \citenamefont {Bekasov}, \citenamefont {Murray},\ and\ \citenamefont
  {Papamakarios}}]{nflows}%
  \BibitemOpen
  \bibfield  {author} {\bibinfo {author} {\bibfnamefont {C.}~\bibnamefont
  {Durkan}}, \bibinfo {author} {\bibfnamefont {A.}~\bibnamefont {Bekasov}},
  \bibinfo {author} {\bibfnamefont {I.}~\bibnamefont {Murray}},\ and\ \bibinfo
  {author} {\bibfnamefont {G.}~\bibnamefont {Papamakarios}},\ }\href
  {https://doi.org/10.5281/zenodo.4296287} {\bibinfo {title} {{nflows}:
  normalizing flows in {PyTorch}}} (\bibinfo {year} {2020})\BibitemShut
  {NoStop}%
\bibitem [{\citenamefont {Harris}\ \emph {et~al.}(2020)\citenamefont {Harris},
  \citenamefont {Millman}, \citenamefont {Van Der~Walt}, \citenamefont
  {Gommers}, \citenamefont {Virtanen}, \citenamefont {Cournapeau},
  \citenamefont {Wieser}, \citenamefont {Taylor}, \citenamefont {Berg},
  \citenamefont {Smith} \emph {et~al.}}]{harris2020array}%
  \BibitemOpen
  \bibfield  {author} {\bibinfo {author} {\bibfnamefont {C.~R.}\ \bibnamefont
  {Harris}}, \bibinfo {author} {\bibfnamefont {K.~J.}\ \bibnamefont {Millman}},
  \bibinfo {author} {\bibfnamefont {S.~J.}\ \bibnamefont {Van Der~Walt}},
  \bibinfo {author} {\bibfnamefont {R.}~\bibnamefont {Gommers}}, \bibinfo
  {author} {\bibfnamefont {P.}~\bibnamefont {Virtanen}}, \bibinfo {author}
  {\bibfnamefont {D.}~\bibnamefont {Cournapeau}}, \bibinfo {author}
  {\bibfnamefont {E.}~\bibnamefont {Wieser}}, \bibinfo {author} {\bibfnamefont
  {J.}~\bibnamefont {Taylor}}, \bibinfo {author} {\bibfnamefont
  {S.}~\bibnamefont {Berg}}, \bibinfo {author} {\bibfnamefont {N.~J.}\
  \bibnamefont {Smith}}, \emph {et~al.},\ }\bibfield  {title} {\bibinfo {title}
  {Array programming with numpy},\ }\href@noop {} {\bibfield  {journal}
  {\bibinfo  {journal} {Nature}\ }\textbf {\bibinfo {volume} {585}},\ \bibinfo
  {pages} {357} (\bibinfo {year} {2020})}\BibitemShut {NoStop}%
\bibitem [{\citenamefont {Paszke}\ \emph {et~al.}(2019)\citenamefont {Paszke}
  \emph {et~al.}}]{NEURIPS2019_9015}%
  \BibitemOpen
  \bibfield  {author} {\bibinfo {author} {\bibfnamefont {A.}~\bibnamefont
  {Paszke}} \emph {et~al.},\ }\bibfield  {title} {\bibinfo {title} {Pytorch: An
  imperative style, high-performance deep learning library},\ }in\ \href
  {http://papers.neurips.cc/paper/9015-pytorch-an-imperative-style-high-performance-deep-learning-library.pdf}
  {\emph {\bibinfo {booktitle} {Advances in Neural Information Processing
  Systems 32}}},\ \bibinfo {editor} {edited by\ \bibinfo {editor}
  {\bibfnamefont {H.}~\bibnamefont {Wallach}}, \bibinfo {editor} {\bibfnamefont
  {H.}~\bibnamefont {Larochelle}}, \bibinfo {editor} {\bibfnamefont
  {A.}~\bibnamefont {Beygelzimer}}, \bibinfo {editor} {\bibfnamefont
  {F.}~\bibnamefont {d\textquotesingle Alch\'{e}-Buc}}, \bibinfo {editor}
  {\bibfnamefont {E.}~\bibnamefont {Fox}},\ and\ \bibinfo {editor}
  {\bibfnamefont {R.}~\bibnamefont {Garnett}}}\ (\bibinfo  {publisher} {Curran
  Associates, Inc.},\ \bibinfo {year} {2019})\ pp.\ \bibinfo {pages}
  {8024--8035}\BibitemShut {NoStop}%
\bibitem [{\citenamefont {Fey}\ and\ \citenamefont
  {Lenssen}(2019)}]{Fey/Lenssen/2019}%
  \BibitemOpen
  \bibfield  {author} {\bibinfo {author} {\bibfnamefont {M.}~\bibnamefont
  {Fey}}\ and\ \bibinfo {author} {\bibfnamefont {J.~E.}\ \bibnamefont
  {Lenssen}},\ }\bibfield  {title} {\bibinfo {title} {Fast graph representation
  learning with {PyTorch Geometric}},\ }in\ \href@noop {} {\emph {\bibinfo
  {booktitle} {ICLR Workshop on Representation Learning on Graphs and
  Manifolds}}}\ (\bibinfo {year} {2019})\ \Eprint
  {https://arxiv.org/abs/1903.02428} {arXiv:1903.02428 [cs.LG]} \BibitemShut
  {NoStop}%
\bibitem [{\citenamefont {Falcon}\ \emph {et~al.}(2020)\citenamefont {Falcon}
  \emph {et~al.}}]{william_falcon_2020_3828935}%
  \BibitemOpen
  \bibfield  {author} {\bibinfo {author} {\bibfnamefont {W.}~\bibnamefont
  {Falcon}} \emph {et~al.},\ }\href {https://doi.org/10.5281/zenodo.3828935}
  {\bibinfo {title} {Pytorchlightning/pytorch-lightning: 0.7.6 release}}
  (\bibinfo {year} {2020})\BibitemShut {NoStop}%
\bibitem [{\citenamefont {{Virtanen}}\ \emph {et~al.}(2020)\citenamefont
  {{Virtanen}} \emph {et~al.}}]{2020SciPy-NMeth}%
  \BibitemOpen
  \bibfield  {author} {\bibinfo {author} {\bibfnamefont {P.}~\bibnamefont
  {{Virtanen}}} \emph {et~al.},\ }\bibfield  {title} {\bibinfo {title} {{SciPy
  1.0: Fundamental Algorithms for Scientific Computing in Python}},\ }\bibfield
   {journal} {\bibinfo  {journal} {Nature Methods}\ }\href
  {https://doi.org/https://doi.org/10.1038/s41592-019-0686-2}
  {https://doi.org/10.1038/s41592-019-0686-2} (\bibinfo {year}
  {2020})\BibitemShut {NoStop}%
\bibitem [{\citenamefont {{Binney}}\ and\ \citenamefont
  {{Mamon}}(1982)}]{1982MNRAS.200..361B}%
  \BibitemOpen
  \bibfield  {author} {\bibinfo {author} {\bibfnamefont {J.}~\bibnamefont
  {{Binney}}}\ and\ \bibinfo {author} {\bibfnamefont {G.~A.}\ \bibnamefont
  {{Mamon}}},\ }\bibfield  {title} {\bibinfo {title} {{M/L and velocity
  anisotropy from observations of spherical galaxies, or must M87 have a
  massive black hole}},\ }\href {https://doi.org/10.1093/mnras/200.2.361}
  {\bibfield  {journal} {\bibinfo  {journal} {MNRAS}\ }\textbf {\bibinfo
  {volume} {200}},\ \bibinfo {pages} {361} (\bibinfo {year}
  {1982})}\BibitemShut {NoStop}%
\bibitem [{\citenamefont {Binney}\ and\ \citenamefont {Tremaine}(2008)}]{BT2}%
  \BibitemOpen
  \bibfield  {author} {\bibinfo {author} {\bibfnamefont {J.}~\bibnamefont
  {Binney}}\ and\ \bibinfo {author} {\bibfnamefont {S.}~\bibnamefont
  {Tremaine}},\ }\href@noop {} {\emph {\bibinfo {title} {Galactic Dynamics}}},\
  \bibinfo {edition} {2nd}\ ed.,\ Princeton Series in Astrophysics\ (\bibinfo
  {publisher} {Princeton University Press},\ \bibinfo {year}
  {2008})\BibitemShut {NoStop}%
\bibitem [{\citenamefont {Barlow}(2004)}]{Barlow:2004wg}%
  \BibitemOpen
  \bibfield  {author} {\bibinfo {author} {\bibfnamefont {R.}~\bibnamefont
  {Barlow}},\ }\bibfield  {title} {\bibinfo {title} {{Asymmetric statistical
  errors}},\ }in\ \href@noop {} {\emph {\bibinfo {booktitle} {{Statistical
  Problems in Particle Physics, Astrophysics and Cosmology}}}}\ (\bibinfo
  {year} {2004})\ pp.\ \bibinfo {pages} {56--59},\ \Eprint
  {https://arxiv.org/abs/physics/0406120} {arXiv:physics/0406120} \BibitemShut
  {NoStop}%
\bibitem [{\citenamefont {{Strigari}}\ \emph {et~al.}(2008)\citenamefont
  {{Strigari}}, \citenamefont {{Koushiappas}}, \citenamefont {{Bullock}},
  \citenamefont {{Kaplinghat}}, \citenamefont {{Simon}}, \citenamefont
  {{Geha}},\ and\ \citenamefont {{Willman}}}]{2008ApJ...678..614S}%
  \BibitemOpen
  \bibfield  {author} {\bibinfo {author} {\bibfnamefont {L.~E.}\ \bibnamefont
  {{Strigari}}}, \bibinfo {author} {\bibfnamefont {S.~M.}\ \bibnamefont
  {{Koushiappas}}}, \bibinfo {author} {\bibfnamefont {J.~S.}\ \bibnamefont
  {{Bullock}}}, \bibinfo {author} {\bibfnamefont {M.}~\bibnamefont
  {{Kaplinghat}}}, \bibinfo {author} {\bibfnamefont {J.~D.}\ \bibnamefont
  {{Simon}}}, \bibinfo {author} {\bibfnamefont {M.}~\bibnamefont {{Geha}}},\
  and\ \bibinfo {author} {\bibfnamefont {B.}~\bibnamefont {{Willman}}},\
  }\bibfield  {title} {\bibinfo {title} {{The Most Dark-Matter-dominated
  Galaxies: Predicted Gamma-Ray Signals from the Faintest Milky Way Dwarfs}},\
  }\href {https://doi.org/10.1086/529488} {\bibfield  {journal} {\bibinfo
  {journal} {ApJ}\ }\textbf {\bibinfo {volume} {678}},\ \bibinfo {pages} {614}
  (\bibinfo {year} {2008})},\ \Eprint {https://arxiv.org/abs/0709.1510}
  {arXiv:0709.1510 [astro-ph]} \BibitemShut {NoStop}%
\bibitem [{\citenamefont {Lisanti}\ \emph {et~al.}(2018)\citenamefont
  {Lisanti}, \citenamefont {Mishra-Sharma}, \citenamefont {Rodd}, \citenamefont
  {Safdi},\ and\ \citenamefont {Wechsler}}]{Lisanti:2017qoz}%
  \BibitemOpen
  \bibfield  {author} {\bibinfo {author} {\bibfnamefont {M.}~\bibnamefont
  {Lisanti}}, \bibinfo {author} {\bibfnamefont {S.}~\bibnamefont
  {Mishra-Sharma}}, \bibinfo {author} {\bibfnamefont {N.~L.}\ \bibnamefont
  {Rodd}}, \bibinfo {author} {\bibfnamefont {B.~R.}\ \bibnamefont {Safdi}},\
  and\ \bibinfo {author} {\bibfnamefont {R.~H.}\ \bibnamefont {Wechsler}},\
  }\bibfield  {title} {\bibinfo {title} {{Mapping Extragalactic Dark Matter
  Annihilation with Galaxy Surveys: A Systematic Study of Stacked Group
  Searches}},\ }\href {https://doi.org/10.1103/PhysRevD.97.063005} {\bibfield
  {journal} {\bibinfo  {journal} {Phys. Rev. D}\ }\textbf {\bibinfo {volume}
  {97}},\ \bibinfo {pages} {063005} (\bibinfo {year} {2018})},\ \Eprint
  {https://arxiv.org/abs/1709.00416} {arXiv:1709.00416 [astro-ph.CO]}
  \BibitemShut {NoStop}%
\bibitem [{\citenamefont {Veli{\v{c}}kovi{\'c}}\ \emph
  {et~al.}(2017)\citenamefont {Veli{\v{c}}kovi{\'c}}, \citenamefont {Cucurull},
  \citenamefont {Casanova}, \citenamefont {Romero}, \citenamefont {Lio},\ and\
  \citenamefont {Bengio}}]{velivckovic2017graph}%
  \BibitemOpen
  \bibfield  {author} {\bibinfo {author} {\bibfnamefont {P.}~\bibnamefont
  {Veli{\v{c}}kovi{\'c}}}, \bibinfo {author} {\bibfnamefont {G.}~\bibnamefont
  {Cucurull}}, \bibinfo {author} {\bibfnamefont {A.}~\bibnamefont {Casanova}},
  \bibinfo {author} {\bibfnamefont {A.}~\bibnamefont {Romero}}, \bibinfo
  {author} {\bibfnamefont {P.}~\bibnamefont {Lio}},\ and\ \bibinfo {author}
  {\bibfnamefont {Y.}~\bibnamefont {Bengio}},\ }\bibfield  {title} {\bibinfo
  {title} {Graph attention networks},\ }\href@noop {} {\bibfield  {journal}
  {\bibinfo  {journal} {arXiv preprint arXiv:1710.10903}\ } (\bibinfo {year}
  {2017})}\BibitemShut {NoStop}%
\bibitem [{\citenamefont {Kipf}\ and\ \citenamefont
  {Welling}(2016)}]{kipf2016semi}%
  \BibitemOpen
  \bibfield  {author} {\bibinfo {author} {\bibfnamefont {T.~N.}\ \bibnamefont
  {Kipf}}\ and\ \bibinfo {author} {\bibfnamefont {M.}~\bibnamefont {Welling}},\
  }\bibfield  {title} {\bibinfo {title} {Semi-supervised classification with
  graph convolutional networks},\ }\href@noop {} {\bibfield  {journal}
  {\bibinfo  {journal} {arXiv preprint arXiv:1609.02907}\ } (\bibinfo {year}
  {2016})}\BibitemShut {NoStop}%
\bibitem [{\citenamefont {Kullback}\ and\ \citenamefont {Leibler}(1951)}]{kld}%
  \BibitemOpen
  \bibfield  {author} {\bibinfo {author} {\bibfnamefont {S.}~\bibnamefont
  {Kullback}}\ and\ \bibinfo {author} {\bibfnamefont {R.~A.}\ \bibnamefont
  {Leibler}},\ }\bibfield  {title} {\bibinfo {title} {{On Information and
  Sufficiency}},\ }\href {https://doi.org/10.1214/aoms/1177729694} {\bibfield
  {journal} {\bibinfo  {journal} {The Annals of Mathematical Statistics}\
  }\textbf {\bibinfo {volume} {22}},\ \bibinfo {pages} {79 } (\bibinfo {year}
  {1951})}\BibitemShut {NoStop}%
\bibitem [{\citenamefont {Hermans}\ \emph {et~al.}(2021)\citenamefont
  {Hermans}, \citenamefont {Delaunoy}, \citenamefont {Rozet}, \citenamefont
  {Wehenkel},\ and\ \citenamefont {Louppe}}]{arxiv.2110.06581}%
  \BibitemOpen
  \bibfield  {author} {\bibinfo {author} {\bibfnamefont {J.}~\bibnamefont
  {Hermans}}, \bibinfo {author} {\bibfnamefont {A.}~\bibnamefont {Delaunoy}},
  \bibinfo {author} {\bibfnamefont {F.}~\bibnamefont {Rozet}}, \bibinfo
  {author} {\bibfnamefont {A.}~\bibnamefont {Wehenkel}},\ and\ \bibinfo
  {author} {\bibfnamefont {G.}~\bibnamefont {Louppe}},\ }\href
  {https://doi.org/10.48550/ARXIV.2110.06581} {\bibinfo {title} {Averting a
  crisis in simulation-based inference}} (\bibinfo {year} {2021})\BibitemShut
  {NoStop}%
\bibitem [{\citenamefont {Dey}\ \emph {et~al.}(2022)\citenamefont {Dey},
  \citenamefont {Zhao}, \citenamefont {Newman}, \citenamefont {Andrews},
  \citenamefont {Izbicki},\ and\ \citenamefont {Lee}}]{dey2022calibrated}%
  \BibitemOpen
  \bibfield  {author} {\bibinfo {author} {\bibfnamefont {B.}~\bibnamefont
  {Dey}}, \bibinfo {author} {\bibfnamefont {D.}~\bibnamefont {Zhao}}, \bibinfo
  {author} {\bibfnamefont {J.~A.}\ \bibnamefont {Newman}}, \bibinfo {author}
  {\bibfnamefont {B.~H.}\ \bibnamefont {Andrews}}, \bibinfo {author}
  {\bibfnamefont {R.}~\bibnamefont {Izbicki}},\ and\ \bibinfo {author}
  {\bibfnamefont {A.~B.}\ \bibnamefont {Lee}},\ }\bibfield  {title} {\bibinfo
  {title} {Calibrated predictive distributions via diagnostics for conditional
  coverage},\ }\href@noop {} {\bibfield  {journal} {\bibinfo  {journal} {arXiv
  preprint arXiv:2205.14568}\ } (\bibinfo {year} {2022})}\BibitemShut {NoStop}%
\end{thebibliography}%
